\documentclass[]{article}
\usepackage{amsthm}
\usepackage{amsmath}
\usepackage[colorlinks,citecolor=blue,urlcolor=blue,filecolor=blue,backref=page]{hyperref}
\usepackage{url}
\usepackage{subfig}
\usepackage{ mathrsfs }
\usepackage{algorithm}
\usepackage{algcompatible}

\usepackage{multirow}
\usepackage{booktabs}
\usepackage{amssymb}
\usepackage{tikz}
\usetikzlibrary{bayesnet}
\usepackage{framed}
\usepackage{rotating}
\usepackage{pdflscape}
\usepackage{arydshln}
\usepackage{threeparttable}
\usepackage[export]{adjustbox}

\newenvironment{changemargin}[2]{%
	\begin{list}{}{%
			\setlength{\topsep}{0pt}%
			\setlength{\leftmargin}{#1}%
			\setlength{\rightmargin}{#2}%
			\setlength{\listparindent}{\parindent}%
			\setlength{\itemindent}{\parindent}%
			\setlength{\parsep}{\parskip}%
		}%
		\item[]}{
	\end{list}}

\usepackage{arydshln}
\newcommand\footnoteref[1]{\protected@xdef\@thefnmark{\ref{#1}}\@footnotemark}
\usepackage{threeparttable}
\usepackage[authoryear,round]{natbib}
\usepackage[margin=1.31in]{geometry}
\usepackage{scrextend}

\providecommand{\keywords}[1]{\textbf{\textit{Keywords:}} #1}

\numberwithin{equation}{section}
\theoremstyle{plain}

\usepackage{xr}
\newcommand*\sfref[1]{%
	Supplementary Figure \ref{#1}}
\newcommand*\salgoref[1]{%
	Supplementary Algorithm \ref{#1}}
\newcommand*\stableref[1]{%
	Supplementary Table \ref{#1}}

\def\code#1{\texttt{#1}}
\newcounter{nextsec}

\newcommand\nextsubsection{%
	\setcounter{nextsec}{\expandafter\parsesub\thesubsection\relax}%
	\stepcounter{nextsec}%
	\thesection.\thenextsec%
}

\def\parsesub#1.#2\relax{#2}
\def\parsesubsub#1.#2.#3\relax{#3}

\DeclareMathOperator{\vect}{vec}
\DeclareMathOperator{\reshape}{\mathrm{reshape}}

\newcommand{\OES}{\text{OES}}
\newcommand{\ESS}{\text{ESS}}
\newcommand{\CPU}{\text{CPUtime}}

\newcommand{\beginsupplement}{%
	\setcounter{table}{0}
	\renewcommand{\thetable}{S\arabic{table}}%
	\setcounter{figure}{0}
	\renewcommand{\thefigure}{S\arabic{figure}}%
}

\usepackage{titlesec}
\titlespacing\section{0pt}{12pt plus 4pt minus 2pt}{0pt plus 2pt minus 2pt}
\titlespacing\subsection{0pt}{12pt plus 4pt minus 2pt}{0pt plus 2pt minus 2pt}
\titlespacing\subsubsection{0pt}{12pt plus 4pt minus 2pt}{0pt plus 2pt minus 2pt}

\title{\textbf{Posterior Inference for Sparse Hierarchical Non-stationary Models}}
\author{Karla Monterrubio-G\'omez\thanks{University of Warwick, UK}, Lassi Roininen\thanks{LUT University, Finland} \thanks{These authors contributed equally to this work.}, Sara Wade\thanks{University of Edinburgh, UK} \footnotemark[3], Theodoros Damoulas\footnotemark[1],\\ and Mark Girolami\thanks{University of Cambridge, UK, Alan Turing Institute, UK}}
\date{}

\begin{document}
\maketitle
	
	\begin{abstract}
	Gaussian processes are valuable tools for non-parametric modelling, where typically an assumption of stationarity is employed.  While removing this assumption can improve prediction, fitting such models is challenging. 
In this work, hierarchical models are constructed based on Gaussian Markov random fields with stochastic spatially varying parameters. Importantly, this allows for non-stationarity while also addressing the computational burden through a sparse banded representation of the precision matrix.
In this setting, efficient Markov chain Monte Carlo (MCMC) sampling is challenging due to the strong coupling a posteriori of the parameters and hyperparameters. We develop and compare three adaptive MCMC schemes and make use of banded matrix operations for faster inference. 
Furthermore, a novel extension to multi-dimensional settings is proposed through an additive structure that retains the flexibility and scalability of the model, while also inheriting interpretability from the additive approach. 
A thorough assessment of the 
efficiency and accuracy of the methods in nonstationary settings is presented for both simulated experiments and a computer emulation problem.
	\end{abstract}

\keywords{Gaussian Process; 	Multilevel models; Gaussian Markov random fields; MCMC; SPDE}

\section{Introduction}
Gaussian processes are frequently utilised in constructing powerful nonparametric models, which are appealing due to their analytical properties. 
The flexibility and nonparametric nature of these models make them appropriate and useful in a wide range of applications. Gaussian process (GP) priors have been used in geostatistics \citep{matheron} under the name of Kriging. They are also common in other applications; for instance, in atmospheric sciences \citep{berrocal2010}, 
biology \citep{bat} 
and inverse problems \citep{kaipio2006statistical}.

A large amount of research on GPs and their applications has focused on models where an assumption of stationarity for the process of interest is made. \citet{heaton2017methods} provides a complete review and comparison of available methods under this assumption. Nevertheless, this assumption is rarely realistic in practice and as a consequence, several approaches to introduce non-stationarity have been proposed \citep[e.g.][]{anderes2008estimating, gramacy2012bayesian,kim2005, montagna2016,sampson}. 
Although comparative evaluations show that removing the stationary assumption improves predictive accuracy \citep{fouedjio2016generalized, gramacy2012bayesian, neto2014}, fitting such non-stationary models has proven to be challenging. 
This, combined with the well-known computational constraints of GP models, arising from storing covariance matrices, solving linear systems and computing determinants, poses important questions on how to efficiently perform Bayesian inference in non-stationary problems.

The stochastic partial differential equation (SPDE) approach introduced by \citet{lindgren2011an} employs Gaussian Markov random fields (GMRFs) to ameliorate the computational burden of working with GPs and incorporates a non-stationary framework through spatially varying parameters that are modelled as a linear combination of basis functions. 
Similarly, \citet{paciorek2006} proposed a family of closed-form non-stationary covariance functions with spatially varying parameters modelled by a second latent GP prior. While recognised as a flexible construction, doing inference in a fully Bayesian framework becomes impractical due to the computational demands of such models. Moreover, standard Markov Chain Monte Carlo  (MCMC) procedures require careful parameter tuning, exhibit mixing difficulties and require long runs to reach convergence \citep{neto2014, paciorek2006}.

In this paper, we extend the SPDE formulation of non-stationary GPs considered by \citet{roininen2016hyperpriors}. This model is analogous to SPDE-based constructions in spatial interpolation \citep{Fuglstad2015,FUGLSTAD2015505,yue2014}, and to the non-stationary framework proposed by \citet{paciorek2006}, where the spatially varying parameters are modelled as random objects. We incorporate and account for uncertainty in the measurement noise variance and hyperprior parameters and consider two
hyperpriors for the spatially varying length-scale to account for different smoothness assumptions.

The hierarchical structure of these models, that we refer to as 2-level GPs, introduces strong dependencies and hence efficient sampling from the posterior distribution is problematic. To address this, we introduce and offer a comparative evaluation of three MCMC sampling schemes. 
The first corresponds to an adaptive Metropolis-within-Gibbs scheme. 
The second employs elliptical slice sampling (ELL-SS) combined with re-parametrisations for decoupling the prior, hyperprior, and hyperparameters. The third is a marginal sampler with ELL-SS for a re-parametrised length-scale process. 
The developed methodology results in a non-stationary hierarchical construction that retains the flexibility of the model introduced by \citet{paciorek2006} but is computationally more efficient, due to the sparse and banded structure of the finite-dimensional approximation of the precision matrix. 

The 2-level models studied here naturally extend to multiple levels to construct the deep GP models of \citet{dunlop2017deep}. Deep GPs have received increased interest in literature and proposals differ in how the layers are combined \citep[e.g.] []{blomqvist2018deep,damianou2013, dunlop2017deep, hegde2018deep}. However, the key challenges, preventing wide-spread use of Deep GPs, include developing interpretable constructions that lack degeneracy \citep{duvenaud2014} and efficient and scalable inference, despite the highly coupled layers and computational expense of GPs. The hierarchical construction considered here provides an interpretable structure for nonstationary problems, as well as a sparse framework to address the computational burden, providing a promising route to deeper constructions.

Finally, extensions of the 2-level GPs to multi-dimensional settings are important and necessary in many applications.  Existing approaches for two-dimensional settings are based on heavily parametrised models using spectral decompositions \cite{neto2014,paciorek2006, risser2017local}, basis function representations \cite{katzfuss2013bayesian}, or an isotropic assumption \cite{heinonen2016non,roininen2016hyperpriors}. 
Instead, we propose a novel extension based on additive GPs \citep{duvenaud2011additive}, that decomposes the function of interest in terms of low-dimensional functions, which are modelled as separable non-stationary processes. Important advantages include increased intrepretability and robustness to curse of dimensionality, while inheriting the appealing flexibility of 2-level GPs. The additive structure permits scalability, by taking advantage of the sparse banded precision matrices, low-dimensional representation, and efficient Kroneacker algebra for the separable interaction terms. Moreover, it can capture long-range structures in the data. The choice of interaction terms may be application driven, and hyperpriors can be employed to determine their importance. In this case, the MCMC schemes can be extended through a Gibbs sampling framework. 
This extension provide an  efficient method for data-dense problems in low dimensions but also enables  using the construction for multidimensional (nD) problems with relatively sparse data, similar to \citep{Volodina2018}.



The paper is organised as follows. We start by summarising related work in Section \ref{RelatedWork}. 
In Section \ref{Themodel}, we present the sparse non-stationary hierarchical model for one-dimensional problems 
and describe the proposed sampling schemes in Section \ref{Algorithms}. 
Section~\ref{Simulated_examples2D} 
extends the model to multi-dimensional settings, while retaining the computational benefits and flexibility. The experiments in Section \ref{Examples} provide a complete empirical evaluation, with a study of the discretisation and sample size effects and performance for different signal types, as well as a comparison with alternative GP models.
Finally, Section~\ref{NASA_data} applies the methodology to a computer emulation problem for a NASA rocket booster vehicle.
\section{Related work and background} \label{RelatedWork}

We begin with a review of Gaussian process models, providing a connection between the non-stationary GPs of \citet{paciorek2006} and the SPDE formulation in \citet{lindgren2011an} and \citet{roininen2016hyperpriors}.

\subsection{Gaussian process models}\label{GPformulation}
Let us denote by $\mathbf{y} \in \mathbb{R}^m$ noisy realisations of an unknown random process $\{z(\mathbf{x}), \mathbf{x} \in \mathbb{R}^d  \}$. 
A standard GP regression model assumes
\begin{equation} \label{eqn:GPformulation}
y_i=z(\mathbf{x}_i)+\varepsilon_i,
\end{equation}
where $\varepsilon_i$ is zero-mean Gaussian noise with variance $\sigma^2_\varepsilon$ and $z(\cdot)$ a Gaussian process. More precisely, the model can be written in a hierarchical form,
\begin{equation}
\begin{split}
y_i &\sim \mathcal{N} (z(\mathbf{x}_i),\sigma^2_\varepsilon), \quad i=1,\dots,m,  \\
z(\cdot)&\sim \mathcal{GP}\left(0,C_{\boldsymbol{\phi}}(\cdot, \cdot)\right),  \\
(\boldsymbol{\phi},\sigma^2_\varepsilon) &\sim \pi(\boldsymbol{\phi})\pi(\sigma^2_\varepsilon),  
\end{split}
\label{eq:onelevel} 
\end{equation}
where $C_{ \boldsymbol{\phi}}(\cdot, \cdot)$ is a covariance function parametrised by $\boldsymbol{\phi}$ and must define a valid covariance matrix (symmetric and positive semi-definite). The covariance function encodes important properties of the process, such as its variation and smoothness.  Stationary covariance functions only depend on the inputs $(\mathbf{x}_i, \mathbf{x}_j)$ through $|\mathbf{x}_i- \mathbf{x}_j|$ and are most often the default choice. Typical covariance functions include the stationary squared exponential (SE),
\begin{equation}
C^{\text{\scalebox{.6}{S}}}(\mathbf{x}_i,\mathbf{x}_j)= \tau^2 \exp \left( -\frac{\|\mathbf{x}_i-\mathbf{x}_j\|^2}{2 \lambda^2} \right),
\label{eq:Sq}
\end{equation} 
and the stationary Mat\'ern family, formulated as 
\begin{equation} 
C^{\text{\tiny{S}}}(\mathbf{x}_i,\mathbf{x}_j) = \tau^2 \frac{2^{1-\nu}}{\Gamma(\nu)}\left(\frac{\|\mathbf{x}_i-\mathbf{x}_j \|}{\lambda}\right)^\nu K_\nu\left(\frac{\|\mathbf{x}_i-\mathbf{x}_j \|}{\lambda}\right),
\label{eq:Maternstat}
\end{equation}
where $\Gamma(\cdot)$ is the gamma-function, $\nu>0$ is the smoothness parameter, $\lambda>0$ is the length-scale, $\tau^2>0$ is the magnitude or variance parameter, and $K_\nu$ denotes the modified Bessel function of the second kind of order $\nu$. 

However, the translation-invariance assumption of stationary covariance functions may be inappropriate for certain applications where the process is spatially dependent, such as, for problems in environmental, geospatial and urban sciences. In these cases, a non-stationary formulation of the model is desirable. 
\citet{paciorek2006} introduced a family of non-stationary covariance functions,
\begin{equation*}
C^{\text{\tiny{NS}}}(\mathbf{x}_i,\mathbf{x}_j)= \frac{\tau^2 |\Sigma(\mathbf{x}_i)|^{\frac{1}{4}}|\Sigma(\mathbf{x}_j)|^{\frac{1}{4}}}{|(\Sigma(\mathbf{x}_i)+\Sigma(\mathbf{x}_j))/2|^{\frac{1}{2}}} R\left(\sqrt{ Q_{ij}}\right), 
\end{equation*}
where $R(\cdot)$ is a stationary correlation function on $\mathbb{R}$; $\Sigma(\cdot)$ is a $d \times d$ spatially varying covariance matrix, referred to as a kernel matrix, which describes local anisotropies; and \[Q_{ij}= \left(\mathbf{x}_i-\mathbf{x}_j\right)^{\text{\tiny{T}}} \left((\Sigma(\mathbf{x}_i)+\Sigma(\mathbf{x}_j))/2\right)^{-1}\left(\mathbf{x}_i-\mathbf{x}_j\right).\]
The non-stationary version of the Mat\'ern covariance function is therefore,
\begin{equation}
C^{\text{\tiny{NS}}}(\mathbf{x}_i,\mathbf{x}_j)=\frac{\tau^2 2^{1-\nu} |\Sigma(\mathbf{x}_i)|^{\frac{1}{4}}|\Sigma(\mathbf{x}_j)|^{\frac{1}{4}}}{\Gamma(\nu)|(\Sigma(\mathbf{x}_i)+\Sigma(\mathbf{x}_j))/2|^{\frac{1}{2}}}\left(\sqrt{ Q_{ij}}\right)^\nu {K}_\nu\left(\sqrt{Q_{ij}}\right), \label{eq:NsMatern}
\end{equation}
with hyperparameters $ \boldsymbol{\phi}= \{\Sigma(\cdot), \nu, \tau^2 \}$. When employing this type of non-stationary covariance function 
in equation \eqref{eq:onelevel}, we are required to infer the kernel matrices at every location where the process was observed. \citet{paciorek2006} modelled the kernel matrices as a continuous-parameter random process by utilising its spectral decomposition. Nonetheless, this approach results in computationally expensive inference \citep[Section 5.1]{paciorek2006} even for one-dimensional problems. As a consequence, alternative approaches to model the spatially varying parameters have been proposed \citep{lang, neto2014,  risser2016nonstationary}.  

We note that for one-dimensional problems, the kernel matrices, $\Sigma(\cdot)$, are reduced to scalars, which we denote as $\ell(\cdot)$. In this setting, when modelling the spatially varying length-scale with a GP, the hierarchical formulation of the model is
\begin{equation}
\begin{split}
y_i &\sim \mathcal{N} (z(x_i),\sigma^2_\varepsilon), \quad i=1,\dots,m,  \\
z(\cdot)&\sim \mathcal{GP}\left(0,C^{\text{\tiny{NS}}}_{\boldsymbol{\phi}}(\cdot, \cdot)\right), \\
\log \ell(\cdot)&\sim \mathcal{GP}\left(\mu_{\ell},C^{\text{\tiny{S}}}_{\boldsymbol{\varphi}}(\cdot, \cdot)\right),  \\
( {\tau^2}, \boldsymbol{\varphi}, \sigma^2_\varepsilon, \mu_{\ell}) &\sim  \pi({\tau^2}) \pi( \boldsymbol{\varphi})\pi(\sigma^2_\varepsilon)   \pi({\mu_{\ell}}) ,   \label{eq:twolevel} 
\end{split}
\end{equation}
where $C_{\boldsymbol{\phi}}^{\text{\tiny{NS}}}(\cdot, \cdot)$ is as in equation \eqref{eq:NsMatern} and $C_{\boldsymbol{\varphi}}^{\text{\tiny{S}}}(\cdot, \cdot)$ is a stationary covariance function with parameters $\boldsymbol{\varphi}$. We note that the prior for the spatially varying length-scale is assigned over a transformed parameter, defined as $u(\cdot) := \log \ell(\cdot)$, with $\mu_\ell$ representing the a priori constant mean of the log length-scale process. 

Efficient sampling from the posterior 
is challenging and the computational burden introduced by the spatially varying parameter is noticeable even in one-dimensional problems \citep{heinonen2016non,paciorek2006}. These difficulties arise from different sources. First, the computational complexity inherited from dense covariance matrices makes the model unsuitable for large datasets. Second, the latent processes and hyperparameters tend to be strongly coupled, leaving vanilla MCMC schemes inefficient. Finally, as in a stationary formulation, the model is sensitive to the choice of hyperparameters, ${\boldsymbol{\varphi}}$, and therefore these must be inferred \citep{neto2014}.

\subsection{SPDE formulation of Mat\'ern fields} \label{SPDE}  
\citet{lindgren2011an} showed that Gaussian Markov random fields can be presented equivalently as stochastic partial differential equations.
By fixing $\nu=2-d/2$, a GP with stationary Mat\'ern covariance  \eqref{eq:Maternstat} and a Markov property can be defined 
through
\begin{equation} \label{eqn:statSPDE}
\left(1 - \lambda^2 \Delta \right){z} = \tau \sqrt{\lambda^d} {w},
\end{equation}
where   \(\Delta:=\sum_{k=1}^d\partial^2/\partial x_{k}^2\) is the Laplace operator, $w$ is white noise on $\mathbb{R}^d$, and  \( \mathrm{Var}(w) = \Gamma(\nu+d/2)(4\pi)^{d/2}/\Gamma(\nu).\)

Analogous to the construction of \citet{paciorek2006} for non-stationary covariance functions with spatially varying length-scales, \citet{roininen2016hyperpriors} derive an SPDE formulation for non-stationary Mat\'ern fields,
\begin{equation} \label{eqn:NonstatSPDE}
\left(1 - \ell(\cdot)^2 \Delta \right){z} = \tau \sqrt{\ell(\cdot)^d}{w},
\end{equation} 
where $\ell(\cdot)$ is a spatially varying length-scale, that is modelled as a log-transformed continuous-parameter GP in the hyperprior in equation \eqref{eq:twolevel}. 
An alternative formulation was proposed by \citet[Section 3.2]{lindgren2011an}, where spatially varying parameters were modelled through a basis function representation. Such a choice gives computational advantages, through a lower dimensional parameter space. 
However, this requires selecting the number of basis functions, and the ability to flexibly recover changes in the length-scale strongly depends on this choice. 

A finite-dimensional approximation of our continuous-parameter model \eqref{eqn:NonstatSPDE} 
can be written in vector-matrix format as $L(\boldsymbol{\ell})\mathbf{z}=\mathbf{w},$ 
where $L(\boldsymbol{\ell})$ is a  sparse matrix depending on $\ell_j:= \ell(jh)$, with $h$ denoting the discretisation step in a chosen finite difference approximation. 
This model is constructed in such a way that the finite-dimensional approximation 
converges to the continuous-parameter model \eqref{eqn:NonstatSPDE} in the discretisation limit $h\rightarrow 0$ (for proofs, see \citet{roininen2016hyperpriors}). 
This property guarantees that irrespective of the choice of $h$, the posteriors, and hence also the estimators, on different meshes, that are dense enough,
are essentially the same.

The SPDE formulation in \eqref{eqn:statSPDE} considers periodic boundary conditions, which can lead to undesirable effects in the edges of the estimators. In order to correct a possible boundary effect, one can add points around the boundary. 
This domain extension offers also a possible benefit in the sparse structure of $L(\boldsymbol{\ell})$. By construction, the matrix $L(\boldsymbol{\ell})$ is a cyclic tridiagonal matrix, and while Sherman-Morrison formula can be applied to solve this type of systems efficiently (e.g. \citet{seiler1989numerical}), we can simply neglect the matrix elements in the corners once we have applied domain extension and take advantage of the resulting tridiagonal structure. 

We note that employing a GP to model $\ell (\cdot)$ results in a similar construction to that discussed in
Section \ref{GPformulation}. In the next sections, we extend the work of \citet{roininen2016hyperpriors}, by including inference of the measurement noise variance and the length-scale hyperparameter. Additionally, we explore different hyperprior models, discuss MCMC algorithms to do inference with these types of models, and present an efficient way to extend the model to higher dimensions.

\section{Sparse non-stationary hierarchical models} \label{Themodel}
The GP formulation in equation \eqref{eqn:GPformulation} can be rephrased through
\begin{equation} \label{eqn:LS}
\mathbf{y}=\mathcal{A}z+ \boldsymbol{\varepsilon} \approx A \mathbf{z}+ \boldsymbol{\varepsilon},
\end{equation}
where $\mathcal{A}$ represents a linear mapping from some function space to a finite-dimensional space $\mathbb{R}^m$ and $ \boldsymbol{\varepsilon} \in \mathbb{R}^{m} $ is assumed to be zero-mean Gaussian noise with variance $\sigma^2_\varepsilon{I_m}$, which is independent of $z$. For computational reasons, we discretise this equation, such that $\mathcal{A}z \approx A \mathbf{z} $, obtaining the right hand side of equation \eqref{eqn:LS},
where $A \in \mathbb{R}^{m \times n}$ is a known matrix and $\mathbf{z} \in \mathbb{R}^{n}$ with $\mathbf{z} \sim \mathcal{N}( 0, C^{\text{\tiny{NS}}}_{\boldsymbol{\phi}})$.  In this case, through the matrix $A$, we are able to define the grid resolution of the latent fields. In particular, for more rough processes, we may be interested in finer resolutions, while for smooth functions, a sparse grid may be sufficient to obtain an accurate representation.

Our aim is to decompose the inverse covariance matrix $({C^{\text{\tiny{NS}}}_{\mathbf{u}}})^{-1} := Q_{\mathbf{u} } = {L(\mathbf{u} )}^{\text{\tiny{T}}}{L(\mathbf{u} )}$, where ${L(\mathbf{u} )}$ is a sparse matrix that depends on the log length-scale parameters $\mathbf{u}=\log(\boldsymbol{\ell})  $. The required decomposition can be achieved employing the SPDE approach from Section~\ref{SPDE}. 
An explicit hierarchical formulation of the model is
\begin{equation}
\begin{split}
\mathbf{y} \mid \mathbf{z}, \sigma^2_{\varepsilon}&\sim \mathcal{N} (A \mathbf{z},\sigma^2_\varepsilon {I}_m), \\
\mathbf{z} \mid \mathbf{u} &\sim \mathcal{N}\left( 0,Q_{\mathbf{u}}^{-1}\right),  \\
\mathbf{u} \mid \lambda&\sim\mathcal{N}\left( \boldsymbol{\mu}_{\ell},C_{\lambda}\right), \\
(\sigma^2_\varepsilon , \lambda) &\sim \pi(\sigma^2_\varepsilon)\pi(\lambda),
\end{split} \label{eqn:hierarchicalSPDE}
\end{equation}
where  $\boldsymbol{\mu}_{\ell}$ denotes the  $n$-dimensional vector with all elements equal to $\mu_\ell$. As both the length-scale and magnitude parameters cannot be estimated consistently \citep{zhang2004inconsistent}, we use the observe data to set the magnitude and mean of both the stationary and non-stationary processes to improve identifiability, with full details provided in 
the Supplementary Material.
The key component of the model is $Q_{\mathbf{u}}$, the inverse covariance of the GMRF 
employed to represent the non-stationary GP. 
This precision matrix depends on $\mathbf{u}$, which is assumed to be a constant-mean GP 
that describes the spatially varying log length-scale, and $\lambda$ denotes the length-scale parameter of the covariance function that describes the properties of the log length-scale process. 
A plate diagram of this model is given in Figure \ref{fig:DAGS} (left).

\begin{figure}[!h]
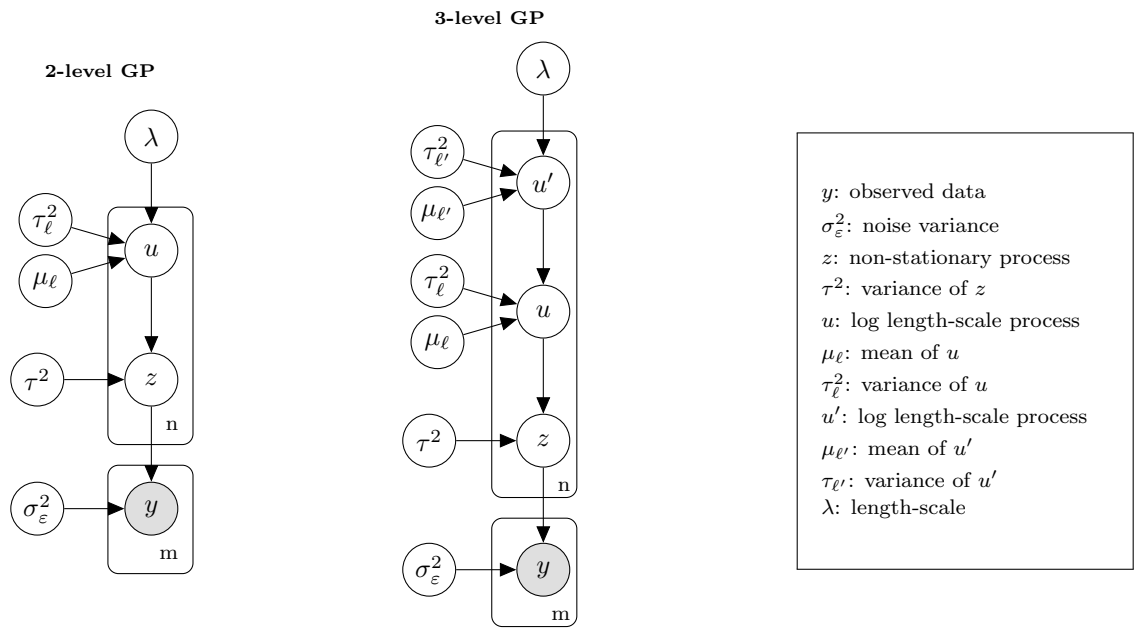

	\centering
	\scriptsize
	\begin{minipage}[c]{0.16\textwidth} 
		\centering 	{	\textbf{2-level GP}}\\	
		\vspace{4mm}	
		\tikz{ 
			\node[obs] (y) {$y$};
			\node[latent, above= of y] (z){$z$};
			\node[latent, above=of z] (u) {$u$};
			\node[latent, above=8mm of u] (lambda) {$\lambda$};
			\node[latent, below left=-1mm and 9mm of u] (muell) {$\mu_{\ell}$};
			\node[latent, left=8mm of y] (noise) {$\sigma^2_\varepsilon$}; 
			\node[latent,  left=8mm of z] (tau) {$\tau^2$} ; %
			\node[latent,  above left=-1mm and 9mm of u] (tau2) {$\tau_{\ell}^2$};
			\plate[inner sep=2mm, xshift=-0.12mm, yshift=0.12mm] {plate1} {(y)} {m}; %
			\plate[inner sep=2mm, xshift=-0.12mm, yshift=0.12mm] {plate1} { (z) (u)} {n}; %
			\edge {z} {y} ; %
			\edge {u} {z} ;
			\edge {tau2} {u};
			\edge {muell} {u};
			\edge {lambda} {u} ;
			\edge {noise} {y};
			\edge {tau} {z}}  
	\end{minipage}\hfill
	\begin{minipage}[c]{0.16\textwidth}  
		\centering	{\textbf{3-level GP}}  \\ \vspace{2mm}			   
		\tikz{ 
			\node[obs] (y) {$y$};
			\node[latent, above=of y] (z) {$z$};
			\node[latent, above=of z] (u) {$u$};
			\node[latent, above=of u] (up) {$u^{\prime}$};
			\node[latent, below left=-1mm and 9mm of u] (muell) {$\mu_{\ell}$};
			\node[latent, above=8mm of up] (lambda) {$\lambda$};
			\node[latent,  below left=-1mm and 9mm of up] (muellp) {$\mu_{\ell^{\prime}}$};
			\node[latent, left=8mm of y] (noise) {$\sigma^2_\varepsilon$}; 
			\node[latent,  left=8mm of z] (tau) {$\tau^2$} ; %
			\node[latent,  above left=-1mm and 9mm of u] (tau2) {$\tau_{\ell}^2$} ;
			\node[latent,   above left=-1mm and 9mm of up] (tau3) {${\tau^2_{\ell^\prime}}$} ;
			\plate[inner sep=2mm, xshift=-0.12cm, yshift=0.12cm] {plate1} {(y)} {m}; %
			\plate[inner sep=2mm, xshift=-0.12cm, yshift=0.12cm] {plate1} { (z) (u) (up)} {n}; %
			\edge {z} {y} ; %
			\edge {u} {z} ;
			\edge {up} {u};
			\edge {lambda} {up}
			\edge {noise} {y};
			\edge {tau} {z};
			\edge {tau2} {u};
			\edge {muell} {u};
			\edge {muellp} {up};
			\edge {tau3} {up};
		}  
	\end{minipage} \hfill
	\begin{minipage}{0.3\textwidth} 
		\vspace{5mm}
		\hspace{10mm}
		\begin{framed}
			\begin{flushleft}
				{\footnotesize{ 
						$y$: observed data\\ \vspace{1.1mm}
						$\sigma^2_{\varepsilon}$: noise variance\\ \vspace{.9mm}
						$z$: non-stationary process\\ \vspace{.9mm}
						$\tau^2$: variance of $z$\\ \vspace{.9mm}
						$u$: log length-scale process\\ \vspace{.9mm}
						$\mu_{\ell}$: mean of $u$\\ \vspace{.9mm}
						$\tau_{\ell}^2$: variance of $u$\\ \vspace{.9mm}
						$u^{\prime}$: log length-scale process\\ \vspace{.9mm}
						$\mu_{\ell^{\prime}}$: mean of $u^{\prime}$\\ \vspace{.9mm}
						$\tau_{\ell^{\prime}}$: variance of $u^{\prime}$\\ \vspace{.9mm}
						$\lambda$: length-scale}}
			\end{flushleft}
		\end{framed}	
	\end{minipage}
	\caption{Plate diagram for a non-stationary hierarchical model.}
\end{figure}	\label{fig:DAGS}

In the following, we discuss different types of hyperpriors for $\mathbf{u}$. Notice that we are free to assign an inhomogeneous Mat\'ern field for the log length-scale process, introducing more flexibility to the model. A graphical representation of this type of 3-level construction is given to the right of Figure \ref{fig:DAGS}. For simplicity, we focus on the 2-level case, when the parameters of the log length-scale process are restricted to be constant along the input space. 

\paragraph{AR(1) hyperprior.} 
%
A hyperprior with sample paths smoother than white noise is needed, otherwise different discretisations of $z$ may affect the posterior estimates \citep{roininen2016hyperpriors}. One such process is the Ornstein-Uhlenbeck, a member of the stationary Mat\'ern family (equation~\eqref{eq:Maternstat}), with exponential covariance function obtained by setting $\nu=1/2$.
The Ornstein-Uhlenbeck has non-differentiable sample paths, allowing quick changes in the behaviour of the log length-scale process. 
It is the continuous-time counterpart of the first-order autoregressive model AR(1) given by $u_j=\beta u_{j-1}+e_j$ and $e_j\sim \mathcal{N}(0, \sigma^2)$,
where 
$u_j$ is on an uniform lattice $t_j:=jh$, $j \in \mathbb{Z}$ with discretisation step $h$. Without a proof, we note that the AR(1) has an exponential autocovariance for all $\beta>0$ except for $\beta=1$ which corresponds to Gaussian random walk, i.e.\ Brownian motion.
While the stable AR(1) requires  that $\beta<1$, this is not a necessary condition here, as our goal is in forming covariance matrices. 
Let us denote by $a_0:=1/\sigma$ and $a_1:=\beta/\sigma$. 
Then, we can construct the inverse of the exponential covariance matrix $ ({C^{\text{\tiny{S}}}_{\lambda}})^{-1}:= Q_{\lambda} = {L(\lambda)}^{\text{\tiny{T}}}{L(\lambda)}$, where ${L(\lambda)}$ is a sparse matrix that depends on $\lambda$ and $\tau_{\ell}$. More precisely, $L(\lambda)$ is a banded matrix, with nonzero elements only on the main diagonal given by $(a_0,\ldots,a_0,1)$ and the  first diagonal above this given by $(a_1,\ldots,a_1)$.
The coefficients 
are defined as $$a_0=( \sqrt{h/ \lambda}+ \sqrt{h/ \lambda+4\lambda/ h})/ \tau_{\ell}\sqrt{8} \text{ and } a_1=( \sqrt{h/ \lambda}- \sqrt{h/ \lambda+4\lambda/h})/  \tau_{\ell}\sqrt{8}.$$
Hence, we have a sparse representation for the hyperprior precision matrix, and 
the banded structure in \( L(\lambda)\) offers important computational advantages when evaluating \( \mathcal{N}( \mathbf{u} \mid  \boldsymbol{\mu}_{\ell} ,Q^{-1}_{{\lambda}}) \), as the required determinant computations, matrix multiplications, and system of equations can be significantly simplified.

\paragraph{SE hyperprior.}
In contrast to the AR(1) hyperprior, we have the squared exponential hyperprior (equation \eqref{eq:Sq}) for $C_{\lambda}$. This covariance function, also referred to as the radial basis function (RBF), is recovered when $\nu \rightarrow \infty $ in the stationary Mat\'ern covariance of equation~\eqref{eq:Maternstat}. Sample paths from a SE are infinitely differentiable and consequently very smooth.  Therefore, when employing a SE hyperprior for the length-scale process, we introduce strong prior smoothness assumptions on how the correlation of the non-stationary process changes with distance. 
We note that for the SE hyperprior, the precision matrix is dense and 
therefore, comes at an increased computational cost.

\section{Inference for one-dimensional problems} \label{Algorithms}
In order to efficiently draw samples from the posterior distributions of interest, we explore three MCMC sampling approaches. The first draws samples from the multidimensional vector $\mathbf{u}$ through an adaptive Metropolis-within-Gibbs algorithm. The second employs ancillary augmentation \citep{yu2011center} over $\mathbf{z}$ and $\mathbf{u}$ and uses elliptical slice sampling \citep[ELL-SS,][]{murray2010elliptical} over the re-parametrised log length-scale process. The third integrates out the non-stationary process, resulting in a marginal sampler that draws from $\mathbf{u}$ by combining ancillary augmentation and ELL-SS to break the correlation between $\mathbf{u}$ and $\lambda$.

\subsection{Metropolis-within-Gibbs (MWG)}
This sampling scheme is inspired by that proposed in \citet{roininen2016hyperpriors} and additionally incorporates adaptive random walks \citep{adaptive} for the noise variance, length-scale hyperparameter, and log length-scale process. 
The procedure is detailed in~\salgoref{alg:MWG}.

The MWG framework updates the log length-scale process at each location individually and, regardless of the hyperprior employed, offers computational gains due to the fact that when proposing a single element of the log length-scale process $u_k^*$, for $k=1,\ldots,n$, the $\log$-ratio of the prior density of $\mathbf{z}$ used in the acceptance probability simplifies to 
\begin{align*}
\log \left( \frac{\mathcal{N}(\mathbf{z} \mid 0 ,Q^{-1}_{\mathbf{u}^*})}{\mathcal{N}(\mathbf{z} \mid 0 ,Q^{-1}_{\mathbf{u}})}\right)&= \log \det ({L} (\mathbf{u}^*){L} (\mathbf{u})^{-1})\\
&-\frac{1}{2} \mathbf{z}^{\text{\tiny{T}}} \left({L} (\mathbf{u}^*)^{\text{\tiny{T}}}	{L} (\mathbf{u}^*)-{L} (\mathbf{u})^{\text{\tiny{T}}}	{L} (\mathbf{u})\right)\mathbf{z}.
\end{align*}
Here $\mathbf{u}^*$ is the proposed log length-scale vector, obtained by updating the $k$th element of $\mathbf{u}$ to $u_k^*$, and
combined with pentadiagonal form of the precision matrix, resulting from multiplication of tridiagonal matrices $Q_{\mathbf{u}}={L} (\mathbf{u})^{\text{\tiny{T}}}	{L} (\mathbf{u})$, the computational complexity of the quadratic term in the $\log$-ratio is reduced from $O(n^2)$ to $O(1)$. 
Moreover, the $\log$-determinant  can be computed through numerically stable and inexpensive operations; for details, see \citet[Section 6]{roininen2016hyperpriors}. 
Similarly, the $\log$-ratio of the prior density of $\mathbf{u}$ 
simplifies to 
\begin{align*}
\log \left(\frac{\mathcal{N}(\mathbf{u}^* \mid \boldsymbol{\mu}_\ell ,C_{\lambda})}{\mathcal{N}(\mathbf{u} \mid \boldsymbol{\mu}_\ell ,C_{\lambda})} \right)
&=  - \frac{1}{2} \left( [(u_k^*)^2-u_k^2]Q_{\lambda\, k,k}+ \sum_{j \neq k}[u_k^*-u_k]u_jQ_{\lambda\, k,j} \right),
\end{align*}
where $Q_{\lambda\, k,j}$ denotes the $(k,j)$ element of the matrix $Q_{\lambda}$. Further computational gains are possible when we utilise the AR(1) hyperprior, as the tridiagonal form $Q_{\lambda}=L(\lambda)^{\text{\tiny{T}}}L(\lambda)$, resulting from the sparse AR(1) construction of $L(\lambda)$, reduces this operation from $O(n)$ to $O(1)$.

Additionally, when proposing a new hyperparameter $\lambda^*$, we must evaluate 
\begin{align*}
\log \left(\frac{\mathcal{N}(\mathbf{u} \mid \boldsymbol{\mu}_\ell ,C_{\lambda^*})}{\mathcal{N}(\mathbf{u} \mid \boldsymbol{\mu}_\ell ,C_{\lambda})} \right)=\frac{1}{2} \log \det (Q_{\lambda^*}Q^{-1}_{\lambda}) - \frac{1}{2} (\mathbf{u}-\boldsymbol{\mu}_{\ell})^{\text{\tiny{T}}} (Q_{\lambda}-Q_{\lambda^*})(\mathbf{u}-\boldsymbol{\mu}_{\ell}).
\end{align*}
For the SE hyperprior, this requires the inversion of a dense $n \times n$ matrix, while the tridiagonal form of $Q_{\lambda}$ for the AR(1) hyperprior makes this considerably cheaper by reducing the computational complexity of this $\log$-ratio term from $O(n^3)$ to $O(n)$. In addition, our simulation studies show that this algorithm does not perform well when the hyperprior for $u(\cdot)$ has strong smoothness assumptions, such as those induced by employing a SE covariance function. This flaw motives us to explore alternative algorithms.

\subsection{Whitened elliptical slice sampling (w-ELL-SS)} \label{wellss}
Elliptical slice sampling is a state-of-the-art MCMC algorithm 
for latent Gaussian models \citep{murray2010elliptical}.
Here, we combine this sampling algorithm with ancillary augmentation or {\it{whitening}} \citep{yu2011center}, which represents a computationally cheap and effective strategy to break the correlation between the prior and its corresponding hyperparameters \citep{filippone2013comparative,murray2010slice}.

We can equivalently define the unknown function as \( \mathbf{z}= L(\mathbf{u})^{-1}\boldsymbol{\xi}\)  with \(\boldsymbol{\xi} \sim \mathcal{N}(0, I_n) \) and the log length-scale vector as \( \mathbf{u}=R_{\lambda} \boldsymbol{\zeta} + \boldsymbol{\mu}_{\ell}\) with \(\boldsymbol{\zeta} \sim \mathcal{N}(0, I_n) \). For the AR(1) hyperprior, $R_{\lambda}:=L(\lambda)^{-1}$; 
whereas, for the SE hyperprior, we define $R_{\lambda}$ to be the lower-triangular Cholesky factor of $C_{\lambda}$. 
Re-parametrising in terms of the whitened parameters $\boldsymbol{\xi}$ and $\boldsymbol{\zeta}$, results in the joint posterior
\begin{align*}
&\pi( \boldsymbol{\zeta} , \boldsymbol{\xi}, \lambda, \sigma^2_{\varepsilon} \mid \mathbf{y} ) \\
&\quad \propto \mathcal{N}(\mathbf{y} \mid A {L(R_{\lambda} \boldsymbol{\zeta} + \boldsymbol{\mu}_{\ell} )}^{-1}\boldsymbol{\xi}, \sigma^2_{\varepsilon}I_m) \mathcal{N}( \boldsymbol{\xi} \mid 0, I_n) \mathcal{N}( \boldsymbol{\zeta} \mid 0, I_n)\pi(\lambda)\pi(\sigma^2_{\varepsilon}).
\end{align*} 
The sampling method is described in~\salgoref{alg:W_ESS}. As opposed to the MWG, the log length scales $\mathbf{u}$ are updated jointly through the whitened parameter $\boldsymbol{\zeta}$. In this case, the likelihood can be evaluated as a product of univariate Gaussian distributions, after computing $\mathbf{u}=R_{\lambda} \boldsymbol{\zeta} + \boldsymbol{\mu}_{\ell}$ and solving $L(\mathbf{u}) \mathbf{z}= \boldsymbol{\xi}$.  Regardless of the hyperprior employed, the latter system of equations $L(\mathbf{u}) \mathbf{z}= \boldsymbol{\xi}$ can be solved in $O(n)$ operations by taking advantage of the tridiagonal structure of $L(\mathbf{u})$ \citep{rue2005gaussian}. The former system of equations $\mathbf{u}=R_{\lambda} \boldsymbol{\zeta} + \boldsymbol{\mu}_{\ell}$ requires matrix multiplication, 
resulting in $O(n^2)$ operations; however, for the AR(1) hyperprior, we can equivalently solve $L(\lambda)(\mathbf{u}-\boldsymbol{\mu}_{\ell})=\boldsymbol{\zeta}$ and make use of the banded form of $L(\lambda)$ to reduce this to $O(n)$ operations.

Thus, while MWG requires looping over the elements of the $n$-dimensional log length-scale vector, with each operation costing $O(1)$ operations for the AR(1) hyperprior and $O(n)$ operations for the SE hyperprior, the w-ELL-SS instead updates this vector jointly through $O(n)$ for the AR(1) hyperprior and $O(n^2)$ operations for the SE hyperprior. However, as ELL-SS is a rejection free sampling method, each iteration may require several likelihood evaluations, mitigating any gain in computation time of this scheme. 

%

\subsection{ Marginal elliptical slice sampling (m-ELL-SS)} \label{mellss}

In simulation studies, we found that integrating out the unknown function $\mathbf{z}$ significantly improves the mixing of $\mathbf{u}$ and its hyperparameters. 
The $\log$ marginal likelihood of the data corresponds to
\begin{equation} 
\log \pi(\mathbf{y} \mid  \mathbf{u}, \lambda, \sigma^2_{\varepsilon}) = - \frac{m}{2} \log(2\pi) -\frac{1}{2} \log \det (\Psi)- \frac{1}{2} \mathbf{y}^{\text{\tiny{T}}} \Psi^{-1}\mathbf{y}, \label{eq:marglike}
\end{equation} 
where $\Psi= AQ^{-1}_{\mathbf{u}}A^{\text{\tiny{T}}}+\sigma^2_{\varepsilon}I_m$. 
Again, we use whitening to decouple $\mathbf{u}$ and $\lambda$, with the re-parametrisation \( \boldsymbol{\zeta}= R_{\lambda}^{-1}( \mathbf{u}-\boldsymbol{\mu}_{\ell})\)
and $R_{\lambda}=L(\lambda)^{-1}$ for the AR(1) hyperprior or $R_{\lambda}=\text{chol}(C_{\lambda})$ for the SE hyperprior. The posterior  is 
\begin{equation*}
\pi( \boldsymbol{\zeta}, \lambda, \sigma_\varepsilon^2 \mid \mathbf{y} ) \propto \mathcal{N}(\mathbf{y} \mid 0, AQ^{-1}_{R_{\lambda} \boldsymbol{\zeta} +\boldsymbol{\mu}_{\ell}}A^{\text{\tiny{T}}}+\sigma^2_{\varepsilon}I_n) \mathcal{N}( \boldsymbol{\zeta} \mid 0, I_m)\pi(\lambda)\pi(\sigma_\varepsilon^2).
\end{equation*} 
The sampling scheme is detailed in~\salgoref{alg:M_ESS}. Again, the log length scales $\mathbf{u}$ are updated jointly through the whitened parameter $\boldsymbol{\zeta}$. This requires first computing $\mathbf{u}=R_{\lambda} \boldsymbol{\zeta} + \boldsymbol{\mu}_{\ell}$, an $O(n)$ operation for the AR(1) hyperprior and $O(n^2)$ operation for the SE hyperprior. However, in comparison with the w-ELL-SS, which proceeds by solving $L(\mathbf{u}) \mathbf{z}= \boldsymbol{\xi}$ and simply taking the product of univariate Gaussians in $O(n)$ operations, we must evaluate the marginal likelihood in \eqref{eq:marglike}.

When computing the marginal likelihood, we emphasise that the required 
calculations for $\Psi$ can be computed employing the Woodbury identity; $$\Psi^{-1}=\sigma^{-2}_{\varepsilon} \left(I_m - A \left( {L(\mathbf{u})}^{\text{\tiny{T}}}{L(\mathbf{u})} + \sigma^{-2}_{\varepsilon}A^{\text{\tiny{T}}}A \right)^{-1}  A^{\text{\tiny{T}}}\right).$$
While this identity also requires a matrix inversion, note that \( {L(\mathbf{u})}^{\text{\tiny{T}}}{L(\mathbf{u})} + \sigma^{-2}_{\varepsilon}A^{\text{\tiny{T}}}A\) is also banded and therefore computations are considerably cheaper. Indeed, the quadratic term in the marginal likelihood \eqref{eq:marglike} is
$$\sigma^{-2}_{\varepsilon} \left(\mathbf{y}^{\text{\tiny{T}}} \mathbf{y}  - \mathbf{y}^{\text{\tiny{T}}}A \left( {L(\mathbf{u})}^{\text{\tiny{T}}}{L(\mathbf{u})} + \sigma^{-2}_{\varepsilon}A^{\text{\tiny{T}}}A \right)^{-1}  A^{\text{\tiny{T}}}\mathbf{y} \right),$$
with the most expensive operation of order $O(n)$. Specifically, the first term $\mathbf{y}^{\text{\tiny{T}}} \mathbf{y}$ can be computed  in $O(m)$ operations, while the second term can be efficiently computed by breaking it into three separate operations. First, we set $\boldsymbol{\varsigma}=A^{\text{\tiny{T}}}\mathbf{y}$, with computational complexity reduced from $O(nm)$ to $O(n)$ through sparsity in $A$. Next, we solve $ ({L(\mathbf{u})}^{\text{\tiny{T}}}{L(\mathbf{u})} + \sigma^{-2}_{\varepsilon}A^{\text{\tiny{T}}}A) \boldsymbol{\varrho}=\boldsymbol{\varsigma}$ in $O(n)$ operations due to the banded form of the matrix. 
Finally, we compute $ \boldsymbol{\varsigma}^{\text{\tiny{T}}}\boldsymbol{\varrho}$, with a cost of $O(n)$ operations. Computing the determinant, on the other hand, is more expensive with the dominant term costing $O(m^3)$ or $O(nm)$, whichever is greater. Specifically, we must first solve $ ({L(\mathbf{u})}^{\text{\tiny{T}}}{L(\mathbf{u})} + \sigma^{-2}_{\varepsilon}A^{\text{\tiny{T}}}A)B=A^{\text{\tiny{T}}}$, with complexity $O(nm)$, and then compute $AB$, with reduced complexity $O(nm)$ due to sparsity in $A$. Finally, the determinant of the $m \times m$ matrix $\Psi^{-1}$ is computed. 

In addition, when proposing new values for the noise variance $\sigma^2_{\varepsilon}$ or the length scale $\lambda$, we must recompute the marginal likelihood \eqref{eq:marglike}, as opposed to evaluating the product of $m$ univariate Gaussians for the w-ELL-SS scheme, increasing the cost of these steps as well. However, in the marginal scheme, in contrast to both MWG and w-ELL-SS,  sampling of $\mathbf{z}$ is no longer required. We also note the computational gains of the AR(1) over the SE hyperprior deteriorate when the determinant  evaluation dominates this computation, i.e. when $m^3>n^2$.

The increased computational cost of the marginal scheme comes with improved mixing, and this trade-off is examined in the simulation studies of Section \ref{Comparative_eval}.  In contrast to MWG, this scheme performs well regardless of the hyperprior employed.



\section{Extensions for $D$-dimensional problems}\label{Simulated_examples2D}

To extend the model from Section~\ref{Themodel} to higher dimensional settings, while maintaining its computational benefits, a novel construction is proposed utilising additive Gaussian process models \citep[AGP,][]{duvenaud2011additive}. First, the model is presented, followed by a description of the extended inference procedure.

\subsection{Sparse non-stationary additive models}\label{sec:2dmodels}
Additive regression models decompose the regression function into main effects and interactions. Linear regression is a classic example, and nonparametric additive models \citep{Friedman81,buja1989} provide increased flexibility, while retaining interpretability and robustness to the input dimension, when compared with general nonparameteric surfaces. The additive GP formulation results from considering the sum and product of covariance functions, two operations for constructing valid covariance functions in $D$-dimensions. This provides a flexible and interpretable model for the unknown function to include main first-order terms up to $D$-order interaction terms, assumed to be separable across dimensions. 

In the additive GP, the choice between low-order and high-order terms  represents a trade-off between between interpretability and accuracy. On one hand, by including only first-order terms, the model can capture long-range structures and has increased intrepretability. On the other, including only a $D$-order separable function increases flexibility and complexity. 
\citet{duvenaud2011additive} include all iteration terms and develop a maximum marginal likelihood approach to determine the importance of each term. Additionally, they develop an efficient algorithm, despite the exponential number of terms, through parametrisations that limit the number of hyperparameters. Interestingly, their experiments show that typically only a few orders of interactions are important. Alternatively, the choice of terms in the additive GP may be application driven; more recently, this is the approach taken in \citet{lonGP} for longitudinal biomedical data. Another interesting direction in \citet{gilboa2015scaling} constructs projected additive GPs through first-order functions of linear projections of the inputs.

For notational simplicity, in the following, we focus on the 2-dimensional setting, including both the main and interaction terms for generality. The model construction and inference can be applied to $D$-dimensional settings, through appropriate choice of the terms to include in the additive formulation. 
In two-dimensional problems, the discretisation is based on a complete $n_1 \times n_2$ grid, with the noisy realisations modelled through
\begin{equation*} 
\mathbf{y}=A_1\mathbf{z}_1+ A_2\mathbf{z}_2+A_3\mathbf{z}_3+\boldsymbol{\varepsilon},
\end{equation*}
where $A_1 \in \mathbb{R}^{m \times n_1}$, $A_2 \in \mathbb{R}^{m \times n_2}$ and $A_3 \in \mathbb{R}^{m \times (n_1 n_2)}$ are known matrices. We assume $z_1(\cdot)$ and $z_2(\cdot)$ are independent one-dimensional non-stationary processes, while $z_3(\cdot)$ is a two-dimensional, separable non-stationary process. Thus, $\mathbf{z}_r  \in \mathbb{R}^{n_r}$ denotes the vector formed by the first-order non-stationary processes at the $n_r$ locations in dimension $r=1,2$, while $\mathbf{z}_3  \in \mathbb{R}^{n_1  n_2}$ collects  the second-order non-stationary process at all locations on the complete $n_1 \times n_2$ grid.  

\begin{figure}[!t]
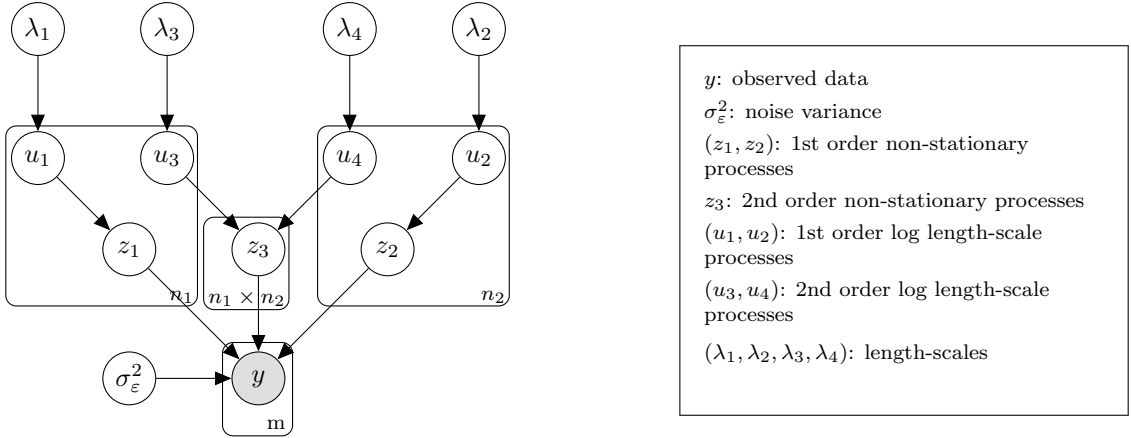

	\centering
	\begin{minipage}{0.38\textwidth}\raggedright
		\centering 	
		\scriptsize
		\tikz{ 
			\node[obs] (y) {$y$};
			\node[latent, above=of y] (z3) {$z_3$};
			\node[latent, left=of z3] (z1) {$z_1$};
			\node[latent, right=of z3] (z2) {$z_2$};
			\node[latent, above left =of z1] (u1) {$u_1$};
			\node[latent, above right=of z2] (u2) {$u_2$};
			\node[latent, above left=of z3] (u3) {$u_3$};
			\node[latent, above right=of z3] (u4) {$u_4$};
			\node[latent, above=of u1] (lambda1) {$\lambda_1$};
			\node[latent, above=of u2] (lambda2) {$\lambda_2$};
			\node[latent, above=of u3] (lambda3) {$\lambda_3$};
			\node[latent, above=of u4] (lambda4) {$\lambda_4$};
			\node[latent, left=of y] (noise) {$\sigma^2_\varepsilon$}; 
			\plate[inner sep=1mm, xshift=-0.12mm, yshift=0.12mm] {plate1} {(y)} {m}; %
			\plate[inner sep=0.5mm, xshift=-0.12mm, yshift=0.12mm] {plate1} { (z1) (u1) (u3)} {$n_1$}; %
			\plate[inner sep=0.5mm, xshift=-0.12mm, yshift=0.12mm] {plate1} { (z2) (u2) (u4)} {$n_2$}; %
			\plate[inner sep=.5mm, xshift=-0.10mm, yshift=0.12mm] {plate1} { (z3)} {$n_1 \times n_2$}; %
			\edge {z1} {y} ; %
			\edge {z2} {y} ; %
			\edge {z3} {y} ; %
			\edge {u1} {z1} ;
			\edge {u2} {z2} ;
			\edge {u3} {z3} ;
			\edge {u4} {z3} ;
			\edge {lambda1} {u1} ;
			\edge {lambda2} {u2} ;
			\edge {lambda3} {u3} ;
			\edge {lambda4} {u4} ;
			\edge {noise} {y}}
	\end{minipage}\hfill
	\begin{minipage}{0.4\textwidth}
		\vspace{3mm}
		\begin{framed}
			\begin{flushleft}
				{\footnotesize{ 
						$y$: observed data\\ \vspace{1.1mm}
						$\sigma^2_{\varepsilon}$: noise variance\\ \vspace{.9mm}
						$(z_1,z_2)$: 1st order non-stationary processes\\ \vspace{.9mm}
						$z_3$: 2nd order non-stationary processes\\ \vspace{.9mm}
						$(u_1,u_2)$: 1st order log length-scale processes \\ \vspace{.9mm}
						$(u_3,u_4)$: 2nd order log length-scale processes \\ \vspace{.9mm}
						$(\lambda_1,\lambda_2, \lambda_3,\lambda_4)$: length-scales}}
			\end{flushleft}
		\end{framed}	
	\end{minipage}
	\caption{Plate diagram for a non-stationary 2-level additive GP model.}
	\label{fig:DAGS2d}
\end{figure}

The hierarchical structure of the model (depicted in Figure \ref{fig:DAGS2d}) is
\begin{equation}
\begin{split}
\mathbf{y} \mid \{\mathbf{z}_r\}_{r=1}^3,\sigma^2_{\varepsilon}&\sim \mathcal{N} (A_1\mathbf{z}_1+ A_2\mathbf{z}_2+A_3\mathbf{z}_3,\sigma^2_\varepsilon {I}_m), \\
\mathbf{z}_r \mid \mathbf{u} _r&\sim \mathcal{N}\left(0,C_{\mathbf{u}_r}^{\text{\tiny{NS}}}\right), \quad r=1,2, \\
& \mathbf{z}_3 \mid \mathbf{u} _3,\mathbf{u} _4\sim \mathcal{N}\left(0,C_{\mathbf{u}_3,\mathbf{u}_4}^{\text{\tiny{NS}}}\right),\\
\mathbf{u}_s\mid \lambda_s&\sim\mathcal{N}\left(\boldsymbol{\mu}_{\ell_s},C_{\lambda_s}^{\text{\tiny{S}}}\right), \quad s=1,2,3,4, \\
(\sigma^2_\varepsilon , \boldsymbol{\lambda}) &\sim  \pi(\sigma^2_\varepsilon)\pi(\lambda_1)\pi(\lambda_2)\pi(\lambda_3)\pi(\lambda_4),
\end{split} \label{eqn:hierarchicalSPDE_2_d}
\end{equation}
with $\boldsymbol{\lambda}=(\lambda_1,\ldots, \lambda_4)$. In equation~\eqref{eqn:hierarchicalSPDE_2_d}, we have four one-dimensional length-scale processes: two describing the correlation changes in each direction independently and two incorporating that information in a two-dimensional process, through a separable assumption $C_{\mathbf{u}_3,\mathbf{u}_4}^{\text{\tiny{NS}}}(\mathbf{x}_i,\mathbf{x}_j)=C_{\mathbf{u}_3}^{\text{\tiny{NS}}}(x_{i,1},x_{j,1})C_{\mathbf{u}_4}^{\text{\tiny{NS}}}(x_{i,2},x_{j,2}).$ A visualisation of the non-stationary additive covariance function is provided in \sfref{fig:kernels}.



Because the AGP is based on one-dimensional kernels, we can directly apply the methodology discussed in Section~\ref{Themodel} for any of the hyperpriors studied. Instead, a direct extension of the SPDE model to two-dimensional settings will not allow us to employ the AR(1) hyperprior and benefit from its computational advantages. This is because a two-dimensional exponential covariance does not have a valid Markov representation. Furthermore, the additive and hierarchical structure of the model in equation~\eqref{eqn:hierarchicalSPDE_2_d} favours interpretability about the behaviour of the correlation in each dimension. 

\subsection{Inference for additive non-stationary models}\label{sec:2dinference}

The posterior for the additive non-stationary model in equation~\eqref{eqn:hierarchicalSPDE_2_d} is
\begin{equation*}
\begin{split}
\pi(\{\mathbf{z}_r\}_{r=1}^3, \{ \mathbf{u}_s,\lambda_s\}_{s=1}^4, \sigma_{\varepsilon}^2 \mid \mathbf{y}) \propto \mathcal{N}(\mathbf{y} \mid A_1\mathbf{z}_1+ A_2\mathbf{z}_2+A_3\mathbf{z}_3,\sigma^2_{\varepsilon}{I}_m) \\ \mathcal{N}( \mathbf{z}_1 \mid 0 ,Q^{-1}_{\mathbf{u}_1}) \mathcal{N}( \mathbf{z}_2 \mid 0 ,Q^{-1}_{\mathbf{u}_2})   \mathcal{N}( \mathbf{z}_3 \mid 0 ,Q^{-1}_{\mathbf{u}_{3},\mathbf{u}_{4}}) \\ \mathcal{N}( \mathbf{u}_1 \mid {\boldsymbol{\mu}_{{\ell_1}}} ,C_{\lambda_1})\cdots \mathcal{N}( \mathbf{u}_4 \mid {\boldsymbol{\mu}_{\ell_4}} ,C_{\lambda_4}) \pi(\lambda_1)\cdots \pi(\lambda_4) \pi(\sigma^2_{\varepsilon}), 
\end{split}
\end{equation*}
with $Q^{-1}_{\mathbf{u}_{3},\mathbf{u}_{4}}$ being a separable covariance matrix, defined as $ Q^{-1}_{\mathbf{u}_{3,4}}:= Q^{-1}_{\mathbf{u}_{3}} \otimes Q^{-1}_{\mathbf{u}_{4}}$, where $\otimes$ denotes the Kronecker product. The three inference schemes described in Section~\ref{Algorithms} can be appropriately extended through a blocked Gibbs sampler, that updates the three blocks of parameters $(\mathbf{z}_1,\mathbf{u}_1,\lambda_1)$; $(\mathbf{z}_2,\mathbf{u}_2,\lambda_2)$; and $(\mathbf{z}_3,\mathbf{u}_3,\mathbf{u}_4,\lambda_3,\lambda_4)$ from their full conditional distributions. Following from 
the one-dimensional synthetic experiments of Section~\ref{Simulated_examples}, we focus on the marginal sampler of Section~\ref{mellss}. We will refer to it as the block marginal elliptical slice sampler (Block-m-ELL-SS); in this case, although we are not integrating out the processes $\{\mathbf{z}_r\}_{r=1}^3$, we use the marginal likelihood to sample the length-scale process and corresponding length-scale hyperparameters in each block. 
For instance, when sampling the block $(\mathbf{z}_1,\mathbf{u}_1,\lambda_1)$, the full conditional factorises as
\begin{equation*}
\pi ( \mathbf{z}_1,  \boldsymbol{\zeta}_1,\lambda_1 \mid \mathbf{y}, \sigma_{\varepsilon}^2 ,\mathbf{z}_2, \mathbf{z}_3) = \pi ( \boldsymbol{\zeta}_1,\lambda_1 \mid \mathbf{y}, \sigma_{\varepsilon}^2 ,\mathbf{z}_2, \mathbf{z}_3) \pi ( \mathbf{z}_1 \mid \boldsymbol{\zeta}_1,\lambda_1, \mathbf{y}, \sigma_{\varepsilon}^2 ,\mathbf{z}_2, \mathbf{z}_3),
\end{equation*}
with $\boldsymbol{\zeta}_1=R_{\lambda_1}^{-1}( \mathbf{u}_1-\boldsymbol{\mu}_{\ell_1})$ denoting the whitened parameter. 
Thus, we first sample from the block marginal $\pi ( \boldsymbol{\zeta}_1,\lambda_1 \mid \mathbf{y}, \sigma_{\varepsilon}^2 ,\mathbf{z}_2, \mathbf{z}_3)$ utilising the steps described in Section~\ref{mellss}, with 
the marginal likelihood replaced by  $\mathcal{N} ( \mathbf{y}- A_2\mathbf{z}_2- A_3\mathbf{z}_3 |  0, A_1Q_{\mathbf{u}_1}^{-1}A_1^{\text{\tiny{T}}}+ {\sigma^2_{\varepsilon}}{I_m} )$. The algorithm is detailed in \salgoref{alg:M_ESS_2d}. 
For efficiency in evaluating the block marginal likelihood obtained from integration of $\mathbf{z}_r$, $r=1,2$, the matrix determinant lemma \citep{harville1997matrix} must be employed to avoid computing the determinant of an $m \times m$ matrix  and instead evaluate the  determinant of three small matrices.

When an interaction term is employed in the model, the algorithm requires samples from the posterior of $\mathbf{z}_3$, which is a Gaussian distribution with mean
$\boldsymbol{\mu}_{z_3}= \sigma^{-2}_{\varepsilon} \Sigma_{z_3} A_3^{\text{\tiny{T}}} ( \mathbf{y}- A_1\mathbf{z}_1- A_2\mathbf{z}_2) $ and variance $\Sigma_{z_3}= (Q_{\mathbf{u}_{3}} \otimes Q_{\mathbf{u}_{4}}+ \sigma^{-2}_{\varepsilon}  A_3^{\text{\tiny{T}}} A_3)^{-1} $. These posterior moment computations need the inversion of an $n_1n_2 \times n_1 n_2$ matrix and cannot exploit the Kronecker structure because of the second summand in $\Sigma_{z_3}$.  To overcome this, we utilise the efficient method of \citet[Section 2.2]{gilboa2015scaling}, based on eigendecompositions and matrix-vector multiplications for Kronecker matrices.
This procedure applies to the case when $A_3^{\text{\tiny{T}}} A_3 =I_{n_1n_2}$; this constraint requires the data to be observed on the complete grid (not necessarily equidistant), but can easily be relaxed for incomplete grids and domain extensions with an additional Gibbs step to sample the missing observations. Specifically, we make use of the identity
\begin{equation}\label{eq:identity1}
\begin{split}
\Sigma_{z_3}&= \left(Q_{\mathbf{u}_{3}} \otimes Q_{\mathbf{u}_{4}}+ \sigma^{-2}_{\varepsilon} I_{n_1n_2} \right)^{-1}\\
&= E_3\otimes E_4 (\Lambda_3 \otimes \Lambda_4+ \sigma^{-2}_{\varepsilon} I_{n_1n_2} ) ^{-1}E_3^{\text{\tiny{T}}}\otimes E_4^{\text{\tiny{T}}},
\end{split}
\end{equation}
where $Q_{\mathbf{u}_{3}}= E_{3} \Lambda_3 E_{3}^{\text{\tiny{T}}}$  and $Q_{\mathbf{u}_{4}}= E_{4} \Lambda_4 E_{4}^{\text{\tiny{T}}}$, with $E_3$ and $E_4$ denoting the eigenvectors matrices and $\Lambda_3$ and $\Lambda_4$ denoting the diagonal matrices of eigenvalues of $Q_{\mathbf{u}_{3}}$ and $Q_{\mathbf{u}_{4}}$, respectively. The second key identity is
\begin{equation}\label{eq:identity2}
(E_3\otimes E_4)\boldsymbol{\alpha}=\vect [(E_3[E_4 \reshape(  \boldsymbol{\alpha},n_2,n_1)]^{\text{\tiny{T}}})^\text{\tiny{T}}],
\end{equation}
where the operator 
$\reshape(b,p,q)$ returns a $p \times q$ matrix whose elements are taken from the vector $b$, and 
$\vect(M)$ denotes the vectorisation of a matrix $M$.


Thus, to efficiently compute the posterior mean, $\boldsymbol{\mu}_{z_3}$, we follow three steps:
\begin{equation*}
\begin{split}
\boldsymbol{\alpha} &= \vect   \left[ \left(  E_3^{\text{\tiny{T}}} [E_4^{\text{\tiny{T}}} \reshape(  \mathbf{\tilde{y}}, n_2, n_1) ] ^{\text{\tiny{T}}}  \right) ^{\text{\tiny{T}}} \right], \\
\boldsymbol{\alpha} &=  (\Lambda_3 \otimes \Lambda_4+ \sigma^{-2}_{\varepsilon}  I_{n_1n_2} ) ^{-1}\boldsymbol{\alpha},\\
\boldsymbol{\mu}_{z_3} &=  \sigma^{-2}_{\varepsilon}\vect \left[ \left(  E_3[E_4 \reshape(\boldsymbol{\alpha}, n_2, n_1) ] ^{\text{\tiny{T}}}  \right) ^{\text{\tiny{T}}} \right],
\end{split}
\end{equation*}
where $\mathbf{\tilde{y}} := \mathbf{y}- A_1\mathbf{z}_1- A_2\mathbf{z}_2$. Note that $(\Lambda_3 \otimes \Lambda_4+ \sigma^{-2}_{\varepsilon}  I_{n_1n_2} )$ is diagonal and therefore easy to invert. 
A posterior sample of $\mathbf{z}_3$ is then obtained by sampling $\boldsymbol{\eta} \sim \mathcal{N}( 0 ,I_{n_1n_2})$ and setting $\mathbf{z}_3=\boldsymbol{\mu}_{z_3}+ E_3\otimes E_4 (\Lambda_3 \otimes \Lambda_4+ \sigma^{-2}_{\varepsilon} I_{n_1n_2}) ^{-1/2}\boldsymbol{\eta}$, where for the latter operation, we again make use of the second identity \eqref{eq:identity2} and the diagonal form of $(\Lambda_3 \otimes \Lambda_4+ \sigma^{-2}_{\varepsilon}  I_{n_1n_2} )$. 

The last critical computation is the evaluation of the block marginal likelihood $\mathcal{N}(\mathbf{\tilde{y}} \mid  0 ,  Q_{\mathbf{u}_{3}}^{-1} \otimes Q_{\mathbf{u}_{4}}^{-1}+\sigma^{2}_{\varepsilon}I_{n_1n_2})$, which is required  to sample $ (\boldsymbol{\zeta}_3,\boldsymbol{\zeta}_4)$ and the corresponding hyperparameters, $\lambda_3$ and $\lambda_4$. First, the quadratic term can be calculated efficiently following the approach employed for the posterior mean. 
Next,  for the $\log$ determinant computation, one can use again the eigendecomposition; namely,  
\begin{equation*}
\begin{split}
&\log \det \left( Q_{\mathbf{u}_{3}}^{-1} \otimes Q_{\mathbf{u}_{4}}^{-1}+ {\sigma^2_{\varepsilon}}{I_{n_1n_2}} \right)^{-1} \\
&\quad  \quad= \log \det \left(   E_3\otimes E_4 (\Lambda_3^{-1} \otimes \Lambda_4^{-1}+ \sigma^{2}_{\varepsilon} {I_{n_1n_2}}) ^{-1}E_3^{\text{\tiny{T}}}\otimes E_4^{\text{\tiny{T}}}\right)\\
&\quad  \quad= -\log \det \left(   \Lambda_3^{-1} \otimes \Lambda_4^{-1}+ \sigma^{2}_{\varepsilon} {I_{n_1n_2}}\right),
\end{split}
\end{equation*}
where $  \Lambda_3^{-1} \otimes \Lambda_4^{-1}+ \sigma^{2}_{\varepsilon} {I_{n_1n_2}}$ is a diagonal matrix, whose $\log$ determinant is straightforward to calculate. 
We emphasize the required terms can also be efficiently computed for higher-order interactions through $D$-dimensional versions of the two key identities \eqref{eq:identity1} and \eqref{eq:identity2} in \citet{gilboa2015scaling}.

\section{Experiments} \label{Examples}
We apply the sparse non-stationary hierarchical methodology to three simulated 1-dimensional interpolation experiments and a two-dimensional synthetic example.  First, the one-dimensional experiments study the effects of the discretisation and sample size on the efficiency of the algorithms presented in Section~\ref{Algorithms} under two extreme hyperpriors. In addition, the experiments show that our model can recover different signal types, while also providing information on the correlation structure. 
Second, a two-dimensional synthetic experiment demonstrates how the model can be extended to higher dimensions utilising an AGP model.
Finally, in Section \ref{Comparative_eval}, we present a comparative evaluation on the performance of 2-level GP models against two other methods: a stationary GP model and a Bayesian treed GP \citep[TGP,][]{gramacy2007tgp} model, a popular approach for dealing with non-stationarity.

\subsection{One-dimensional synthetic data } \label{Simulated_examples}
We consider three simulated datasets with different signal types. The first example (\sfref{fig:simulated_data}a) is a function with smooth parts and edges and is also piecewise constant.
The second synthetic dataset (\sfref{fig:simulated_data}b) is a damped sine wave function with smooth decaying oscillations. 
The third example corresponds to the {\textit{Bumps}} (\sfref{fig:simulated_data}c) function employed by \citet{donoho1995adapting}, which depicts a signal with pronounced spikes and constant parts. 
In the first dataset, we investigate, empirically, posterior consistency of the estimates with respect to the discretisation scheme. The second experiment explores the performance of the sampling schemes for increased sample size and measurement noise. The last example examines 
emphasises the importance of the prior choice.

\subsubsection*{Experiment 1: Smooth-piecewise constant function}
\begin{figure}[!t]   
	\begin{center}
		\subfloat[{\tiny{$\boldsymbol{\ell}$, $n=85$}}]{\includegraphics[scale=.35]{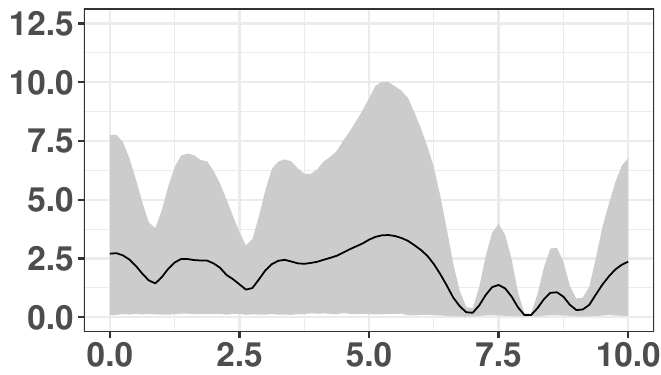} \label{subfig:MWGa}}
		\subfloat[{\tiny{$\boldsymbol{\ell}$, $n=169$}}]{\includegraphics[scale=.35]{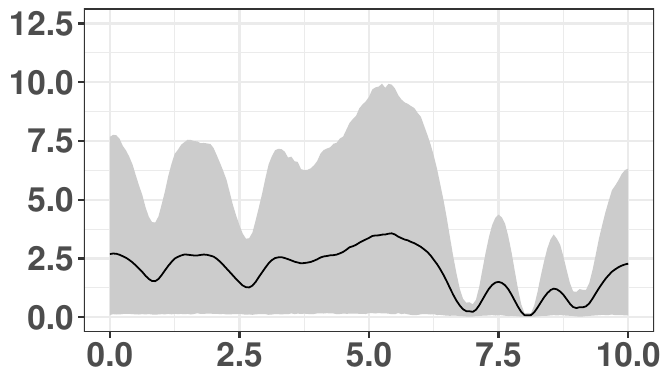}\label{subfig:MWGb}}
		\subfloat[{\tiny{$\boldsymbol{\ell}$, $n=253$}}]{\includegraphics[scale=.35]{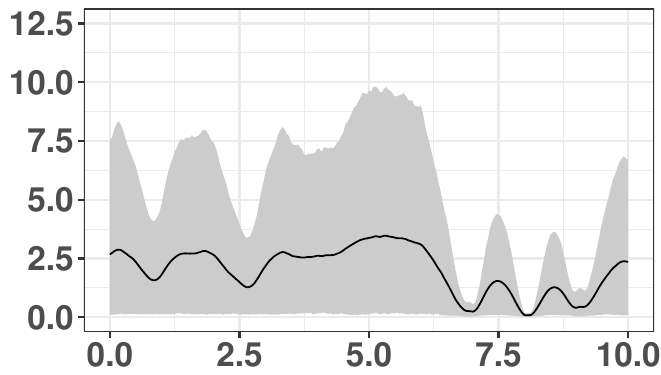}\label{subfig:MWGc}} \\
		\subfloat[{\tiny{$\mathbf{z}$, $n=85$}}]{\includegraphics[scale=.35]{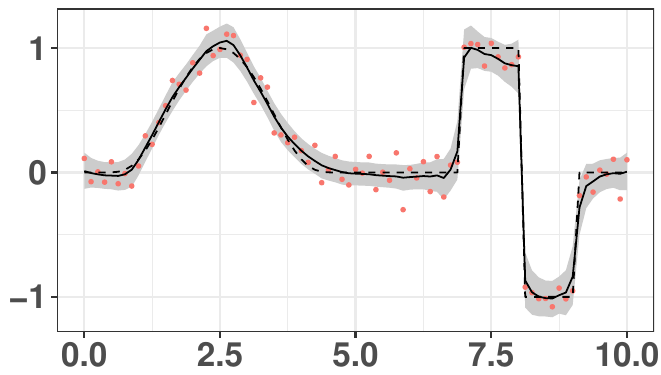} \label{subfig:MWGd}}
		\subfloat[{\tiny{$\mathbf{z}$, $n=169$}}]{\includegraphics[scale=.35]{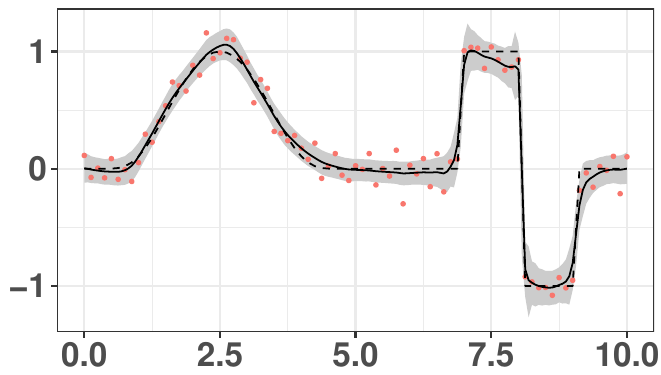} \label{subfig:MWGe}}
		\subfloat[{\tiny{$\mathbf{z}$, $n=253$}}]{\includegraphics[scale=.35]{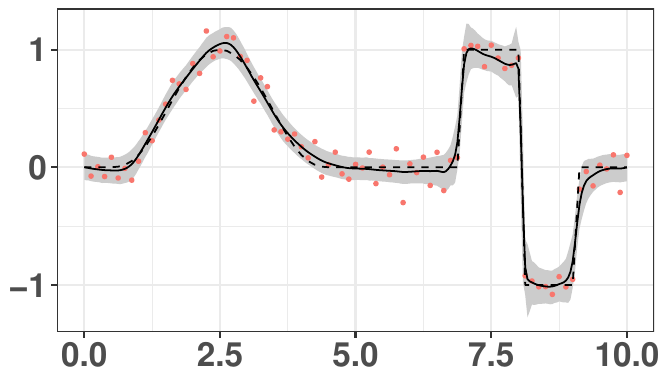} \label{subfig:MWGf}}\\
		\subfloat[{\tiny{$\boldsymbol{\ell}$, $n=85$}}]{\includegraphics[scale=.35]{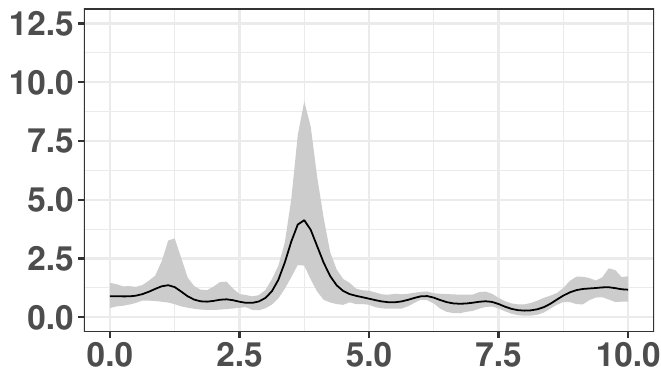}\label{subfig:MWGg}}
		\subfloat[{\tiny{$\boldsymbol{\ell}$, $n=169$}}]{\includegraphics[scale=.35]{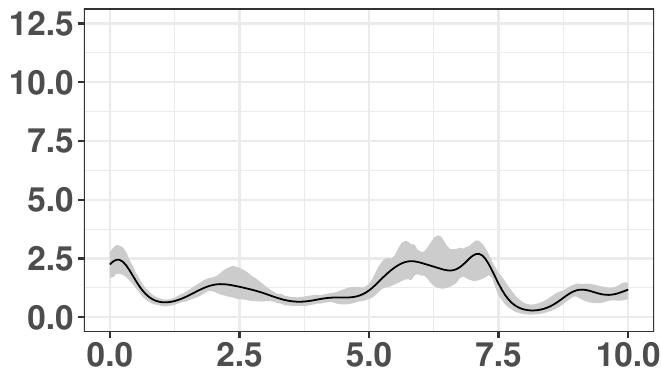}\label{subfig:MWGh}}
		\subfloat[{\tiny{$\boldsymbol{\ell}$, $n=253$}}]{\includegraphics[scale=.35]{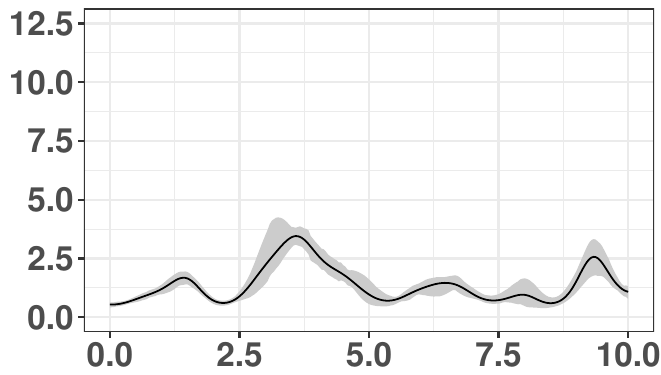}\label{subfig:MWGi}} \\
		\subfloat[{\tiny{$\mathbf{z}$, $n=85$}}]{\includegraphics[scale=.35]{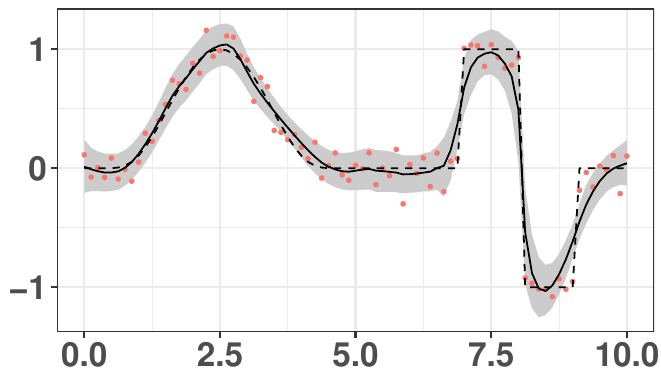}\label{subfig:MWGj}}
		\subfloat[{\tiny{$\mathbf{z}$, $n=169$}}]{\includegraphics[scale=.35]{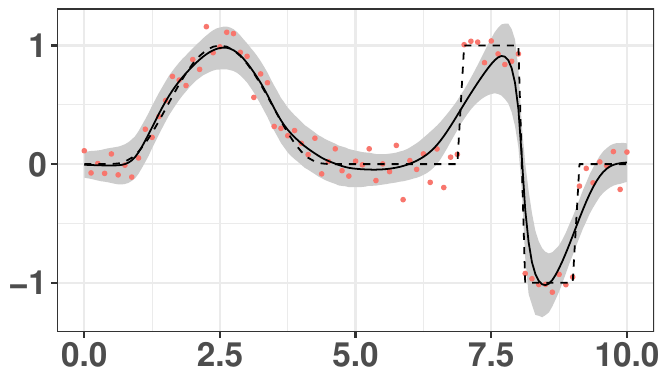}\label{subfig:MWGk}}
		\subfloat[{\tiny{$\mathbf{z}$, $n=253$}}]{\includegraphics[scale=.35]{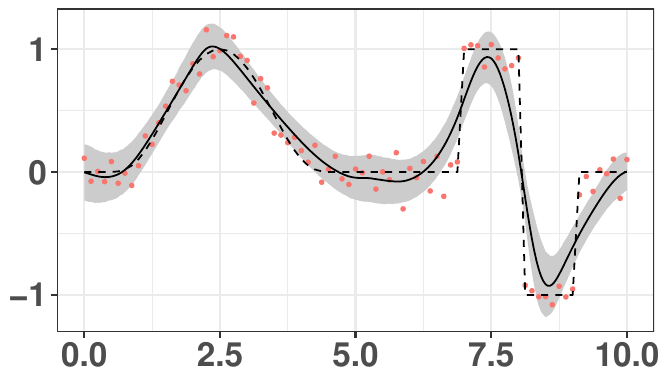} \label{subfig:MWGl}}
	\end{center}
	\caption{Results for Experiment 1 with MWG. (a)-(c): Estimated $\boldsymbol{\ell}$ process with $95\%$ credible intervals for AR(1) hyperprior on different grids. (d)-(f): Estimated $\mathbf{z}$ process with $95\%$ credible intervals for AR(1) hyperprior on different grids with observed data in red.
		(g)-(i): Estimated $\boldsymbol{\ell}$ process with $95\%$ credible intervals for SE hyperprior on different grids. (j)-(l): Estimated $\mathbf{z}$ process with $95\%$ credible intervals for SE hyperprior on different grids with observed data in red.} \label{fig:MWG}
\end{figure}

For all experiments, we use the same initialisation and run the chains for $T=200,000$ iterations. The burn-in period is algorithm specific, selected according to preliminary runs based on Raftery and Lewis's diagnostic \citep{raftery1992practical} for the second level length-scale. Numerical discretisation-invariance is studied by varying $n$ in the experiments, with $n=85,169,253$. The mean and variance of the prior length-scale process is set at zero and one, respectively. For the second level length-scale, we use a broad prior, $\log \lambda \sim \mathcal{N}(0, 3)$.

We start by presenting the results obtained with the MWG algorithm. Figure~\ref{fig:MWG} 
shows estimates of the spatially varying length-scales and the unknown function under both hyperpriors.  For the AR(1) hyperprior, an inspection of traceplots and cumulative averages of the estimates (not shown) suggest convergence of the chains for all discretisation schemes. In addition, the varying length-scale estimates exhibit the expected behaviour (i.e. decaying when the function has a sharp jump and increasing when the function is constant), and the interpolated estimates indicate a reasonable fit to the unknown function for all three discretisations schemes (Figure~\ref{fig:MWG}\subref{subfig:MWGa}-\subref{subfig:MWGf}). However, this is not the case for the SE hyperprior. Figure~\ref{fig:MWG}\subref{subfig:MWGg}-\subref{subfig:MWGl} illustrates the results obtained with this hyperprior for the same sampling algorithm. Under this setting, the effect of discretisation scheme is evident. As we increase $n$, the method fails to recover the unknown function. The strong correlation between the elements of $\mathbf{u}$ induced by the SE hyperprior makes the algorithm converge rather slowly to the target distribution. 

\begin{figure}[!t]   
	\begin{center}			
		\subfloat[{\tiny{$u_{199}$, MWG}}]{\includegraphics[width=0.3\textwidth,height=1.8cm]{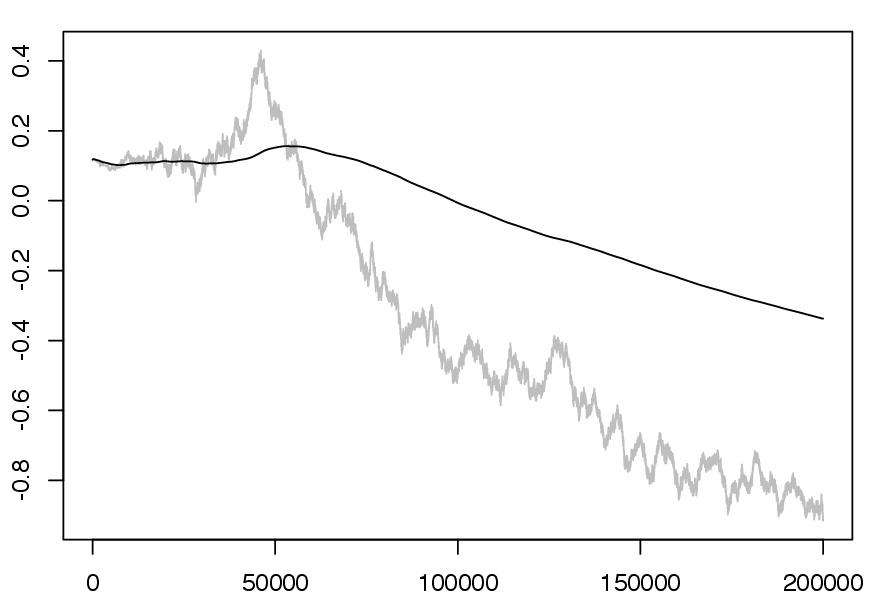}\label{subfig:a}}
\subfloat[{\tiny{$u_{185}$, w-ELL-SS}}]{\includegraphics[width=0.3\textwidth,height=1.8cm]{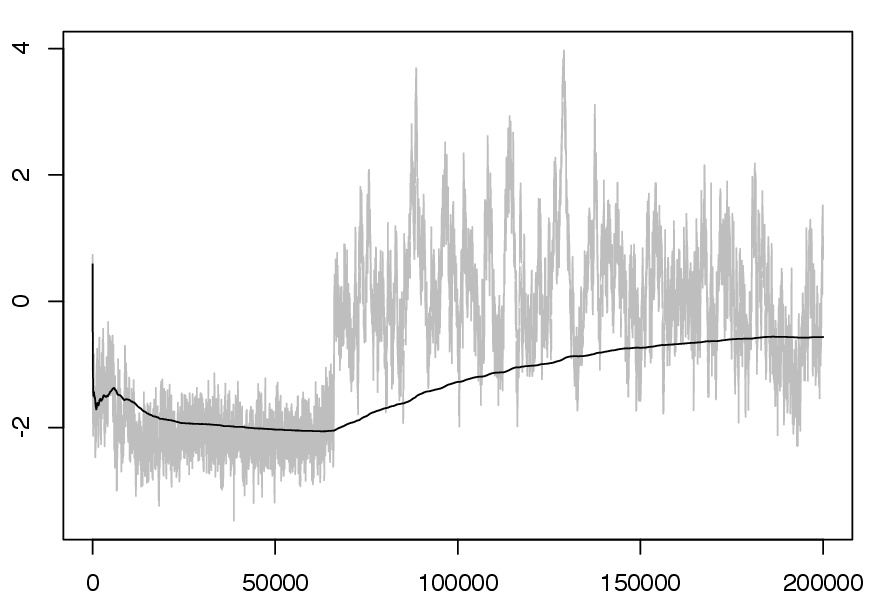}\label{subfig:b}}
\subfloat[{\tiny{$u_{190}$, m-ELL-SS}}]{\includegraphics[width=0.3\textwidth,height=1.8cm]{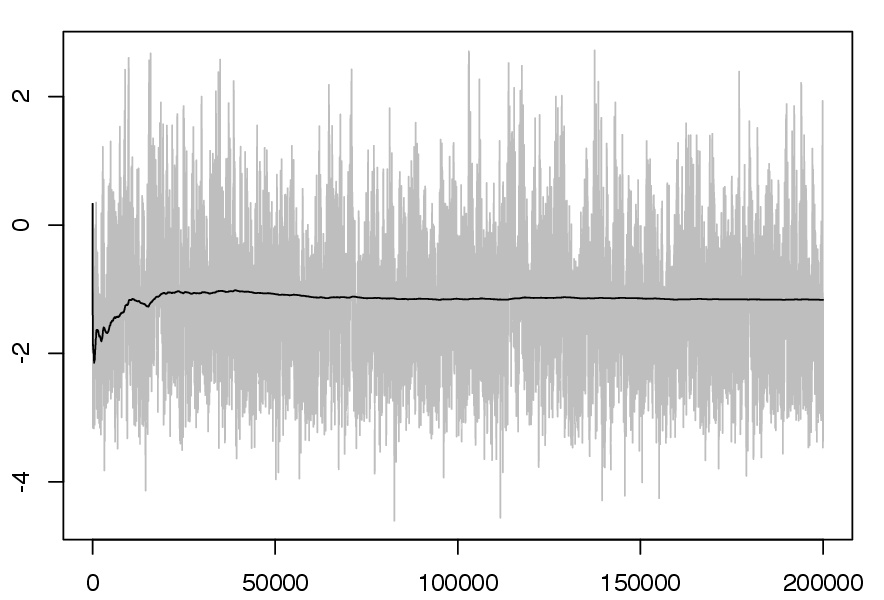}\label{subfig:c}} 
\\
\subfloat[{\tiny{$\lambda$, MWG}}]{\includegraphics[width=0.3\textwidth,height=1.8cm]{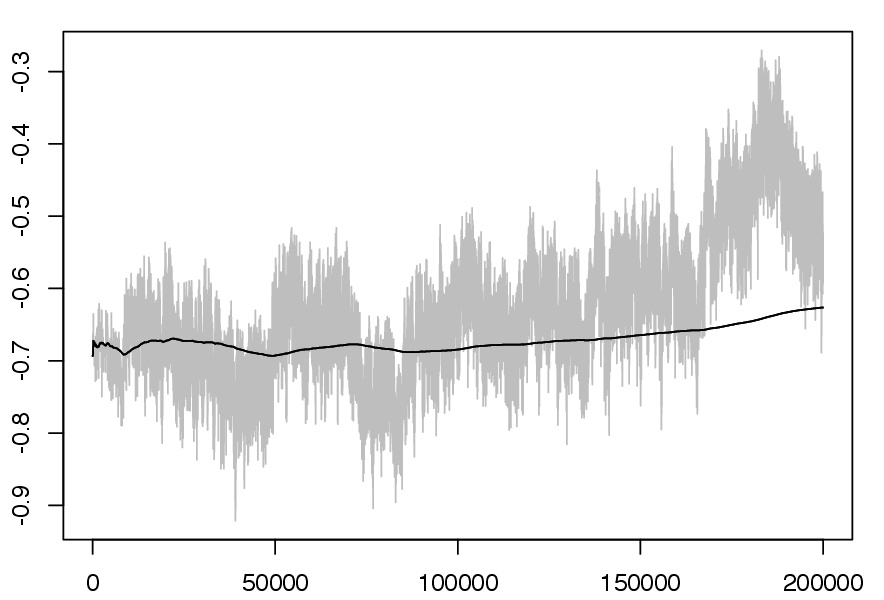} \label{subfig:d}}
\subfloat[{\tiny{$\lambda$, w-ELL-SS}}]{\includegraphics[width=0.3\textwidth,height=1.8cm]{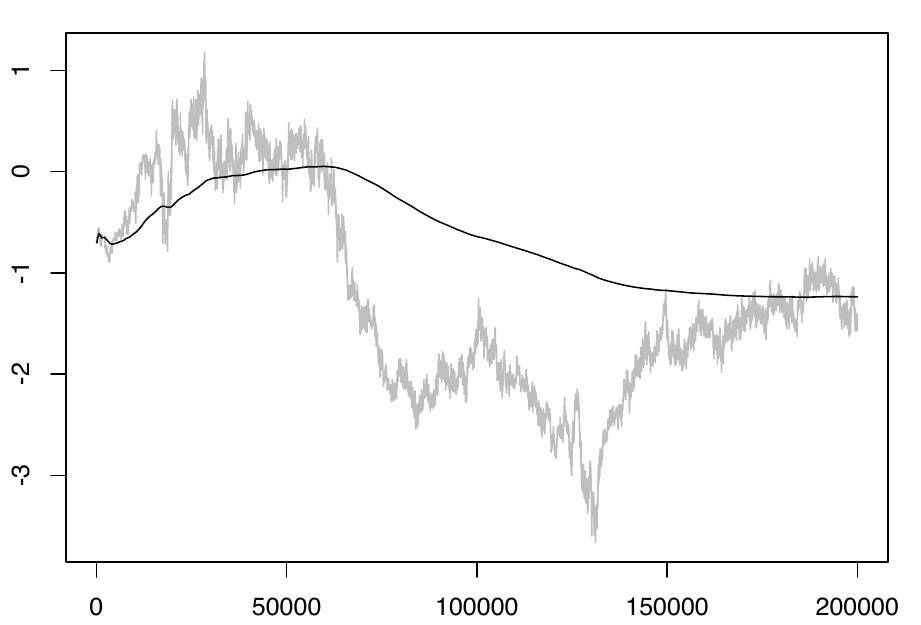}\label{subfig:e}}
\subfloat[{\tiny{$\lambda$, m-ELL-SS}}]{\includegraphics[width=0.3\textwidth,height=1.8cm]{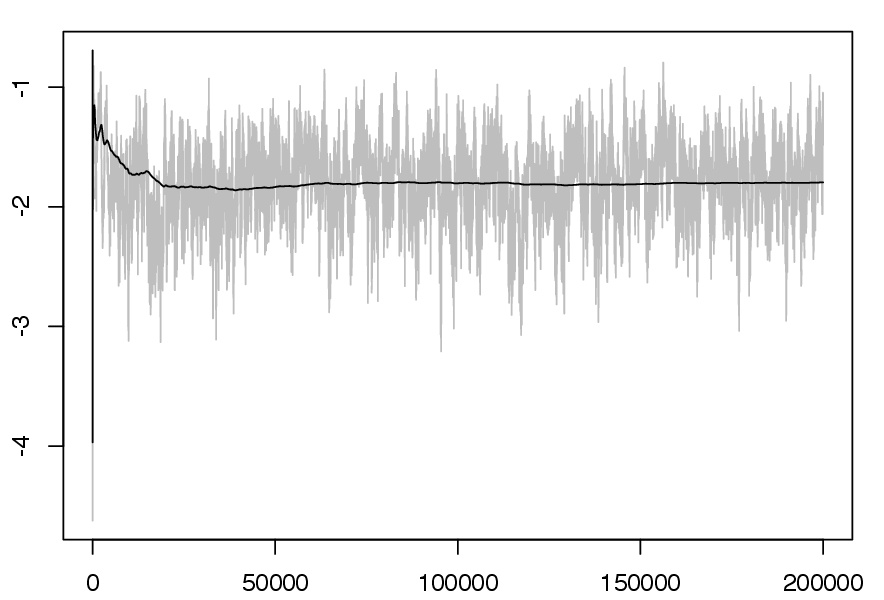}\label{subfig:f}}
	\end{center}
	\caption{Traceplots with cumulative averages of the chains for SE hyperprior with $n=253$. (Top row:) element of $\mathbf{u}$ with the lowest ESS. (Bottom row:) the hyperparameter.} \label{fig:traceplotsex1}
\end{figure}

In contrast to the results obtained with MWG, both w-ELL-SS and m-ELL-SS  demonstrate convergence for both hyperpriors and invariance to the discretisation 
(see Supplementary Figures~\ref{fig:RepESS} and \ref{fig:MarESS} for a complete analysis).  
Figure~\ref{fig:traceplotsex1} summarises succinctly important differences in mixing across the algorithms by showing traceplots with cumulative averages for a subset of parameters. The results are shown for the most challenging scenario, SE hyperprior at the highest resolution, $n=253$. Figure~\ref{fig:traceplotsex1}\subref{subfig:a}\subref{subfig:d} emphasises the lack of convergence for MWG.  Figure~\ref{fig:traceplotsex1}\subref{subfig:b}\subref{subfig:e} demonstrates the high autocorrelation of the chains and the slow convergence produced by w-ELL-SS. Finally, Figure \ref{fig:traceplotsex1}\subref{subfig:c}\subref{subfig:f} highlights the improvement offered by m-ELL-SS, fast convergence to the stationary distribution and low autocorrelation of the chains.

%
%

In order to evaluate the performance of the algorithms, we show in Table~\ref{tab:Efficiency} an overall efficiency score (OES) of the chains \citep{titsias2016auxiliary}. This measure considers both the CPU time (\stableref{tab:Comp_time1}) required to run the chains and the effective sample size (ESS) (\stableref{tab:EffSS_boxcar}). The score is computed as \(\OES=\ESS/\CPU\footnote{All experiments were run in an Intel Core i7-6700 CPU (3.40GHz, 16 GB of RAM).}
\).  For both multidimensional vectors, $\mathbf{z}$ and $\mathbf{u}$, we report the $\OES$ computed with the minimum $\ESS$ across all dimensions.
The results indicate that while MWG with the AR(1) hyperprior shows high efficiency for some parameters when $n=85$, its performance deteriorates as $n$ increases. This suggests that this sampling scheme will not perform efficiently for bigger datasets even when $m=n$ (this is explored in Experiment 2). Furthermore, despite the fact that MWG reports the lowest CPU time under the AR(1) hyperprior (\stableref{tab:Comp_time1}), its overall efficiency scores are outperformed by those obtained with m-ELL-SS; this is due to the low autocorrelation of the chains achieved by the marginal sampler (see \stableref{tab:EffSS_boxcar}). In contrast, chains of the parameters for w-ELL-SS result in the worse OES.
Notice also that the scores reported for MWG with the SE hyperprior are not informative as the chains show convergence problems. 
Table~\ref{tab:Efficiency} also reports mean absolute error (MAE) to evaluate the fit to the unknown function and the empirical coverage of the $95\%$ credible intervals (EC) to evaluate accuracy in uncertainty quantification. For the SE hyperprior, w-ELL-SS and m-ELL-SS report equivalent errors and EC, while MWG yields worse values.

\begin{table}[!t]
	\centering
	\scriptsize	
	\begin{threeparttable}
		\renewcommand{\tabcolsep}{3pt}
		\begin{tabular}{llrrrrrrrrr}	
			\toprule
			\multirow{2}[1]{*}{}& {\multirow{2}[1]{*}{{}}} & \multicolumn{3}{c}{{MWG}}  &\multicolumn{3}{c}{{w-ELL-SS}}  &\multicolumn{3}{c}{{m-ELL-SS}} \\	
			\cmidrule(lr){3-5}   \cmidrule(lr){6-8}  \cmidrule(lr){9-11}
			&	\multicolumn{1}{c}{} & \multicolumn{1}{c}{$n=85$}     & \multicolumn{1}{c}{$n=169$}   & \multicolumn{1}{c}{$n=253$}  &  \multicolumn{1}{c}{$n=85$}  & \multicolumn{1}{c}{$n=169$}  & \multicolumn{1}{c}{$n=253$} &  \multicolumn{1}{c}{$n=85$}   & \multicolumn{1}{c}{$n=169$}  & \multicolumn{1}{c}{$n=253$} \\
			\midrule	
			\multirow{6}[2]{*}{AR(1)} & $\sigma_\varepsilon^2$   & 622.76 &173.12 & 65.99  & 380.89 & 102.38& 38.91 & \bf{661.20}& \bf{257.81} & \bf{116.35} \\	
			&  $\ell_{min}$  & \bf{635.36} & 114.02 & 41.05 & 30.90  & 8.99  & 2.94  & 287.16 & \bf{114.36 }& \bf{59.71}\\	
			&  $z_{min}$ & \bf{203.80}& 42.10 & 13.91  & 9.12 & 2.34  & 0.86 & 129.75 & \bf{52.16} & \bf{22.30} \\
			&  $\lambda$ & 89.84 & 15.66  & 6.00  & 22.77 & 5.26  & 2.36  & \bf{111.80} & \bf{45.54 }& \bf{21.53}\\	
			\cmidrule{2-11}
			&MAE& 0.041 & 0.051 & 0.054 & 0.041 & 0.051 & 0.054 & 0.041 & 0.051 & 0.053 \\
			&EC& 0.988 & 0.975 & 0.971 & 0.988 & 0.975 & 0.975 & 0.988& 0.975& 0.975\\
			\midrule
			\multirow{6}[2]{*}{SE} & $\sigma_\varepsilon^2$ &  11.19&  4.88 & 7.49  & 246.24& 77.72 & 8.89 & \bf{856.15}& \bf{253.91} & \bf{125.97} \\	
			&   $\ell_{min}$  & 1.22& 0.73  & 0.64  & 21.69  & 10.22 & 2.79 & \bf{244.91} & \bf{122.57}& \bf{55.82} \\	
			& $z$  & 0.06 & 0.01  & 0.01 & 4.71 & 1.37  & 0.24 & \bf{76.80} & \bf{24.11} & \bf{9.87} \\	
			&  $\lambda$ &0.59  & 0.75 & 0.31 & 2.31  & 0.29  & 0.01 & \bf{16.59} &\bf{4.15}  &\bf{2.21}  \\	
			\cmidrule{2-11}
			&MAE& 0.078 & 0.100 &0.133 & 0.040 & 0.050 & 0.054 & 0.039 & 0.049& 0.052\\
			&EC& 0.889 & 0.826 &0.763 &0.988 &0.975 & 0.971 & 0.988 & 0.975 & 0.979 \\
			\bottomrule
		\end{tabular}
		\caption{Experiment 1: OES with both hyperpriors under various discretisation schemes ($n=86,169,253$) and  three different algorithms. $\ell_{min}$ and $z_{min}$ report OES for the minimum ESS across all dimensions. Highest values in boldface.}
		\label{tab:Efficiency}
	\end{threeparttable}%
\end{table}

\subsubsection*{Experiment 2: Damped sine wave}

This example explores the effect of increasing the sample size and measurement noise. Due to robustness of the estimates with respect to the discretisation in the first example, we only present experiments for the discretisation scheme when $m=n$. 
The chains are run for $T=100,000$ iterations with a burn-in period that is algorithm and prior specific.  In addition, we extend the domain with $40$ points on each side of the interval, such that $n=430$ and $m=350$. The prior distributions for $\mathbf{u}$ and $\log \lambda$ are as in Experiment 1.

\begin{figure}[!t]    
	\begin{center}
		\subfloat[{\tiny{$\mathbf{\ell}$, MWG with SE}}]{\includegraphics[scale=.35]{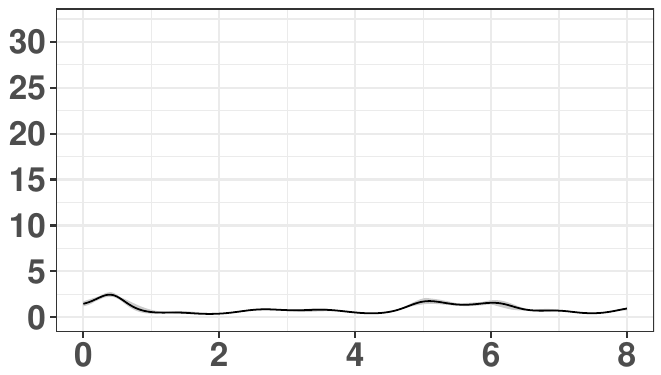}} 	
\subfloat[{\tiny{$\mathbf{\ell}$, w-ELL-SS with SE}}]{\includegraphics[scale=.35]{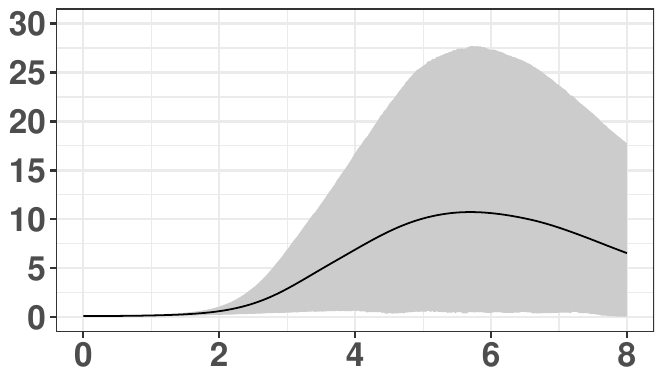}}
\subfloat[{\tiny{$\mathbf{\ell}$, m-ELL-SS with SE}}]{\includegraphics[scale=.35]{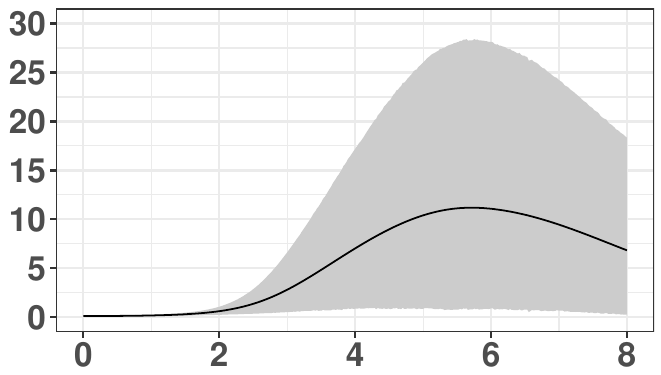}}
\\
\subfloat[{\tiny{$\mathbf{z}$, MWG with SE}}]{\includegraphics[scale=.35]{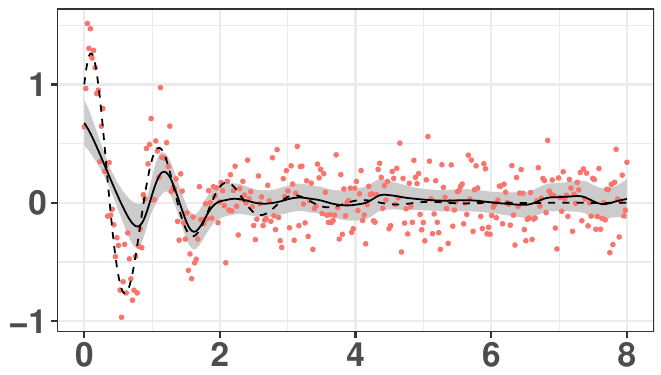}}
\subfloat[{\tiny{$\mathbf{z}$, w-ELL-SS with SE}}]{\includegraphics[scale=.35]{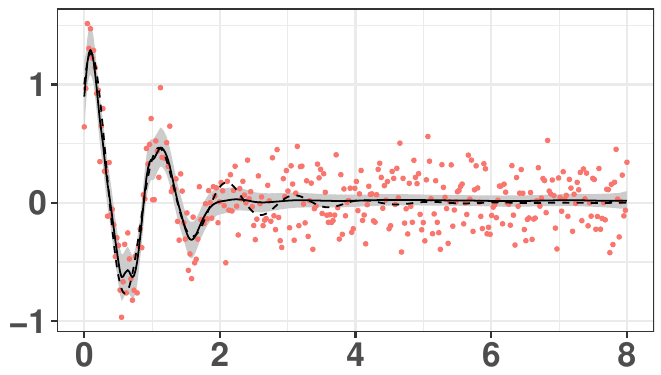}}
\subfloat[{\tiny{$\mathbf{z}$, m-ELL-SS with SE}}]{\includegraphics[scale=.35]{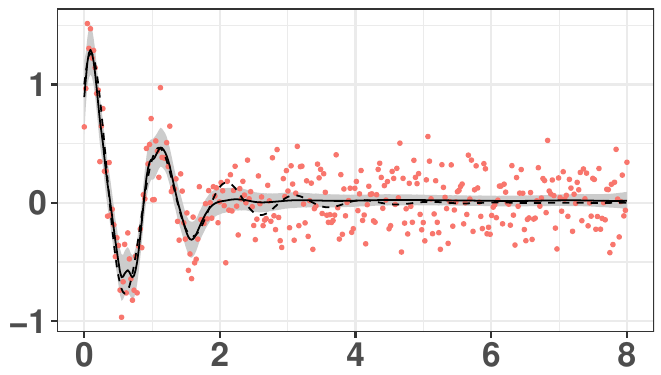}} 	
	\end{center}
	\caption{Results for Experiment 2. Top row: estimated $\boldsymbol{\ell}$ process with $95\% $ credible interval for SE hyperprior with (a) MWG, (b) w-ELL-SS and (c) m-ELL-SS. Second row: estimated $\mathbf{z}$ process with $95\% $ credible interval for SE hyperprior with (d) MWG, (e) w-ELL-SS and (f) m-ELL-SS.}
	\label{fig:SIN_SE}
\end{figure}

\begin{table}[!t]
	\centering
	\scriptsize	
	\begin{tabular}{lrrr rrr}
		\toprule
		\textcolor[rgb]{ 1,  0,  0}{}& \multicolumn{3}{c}{AR(1)} & \multicolumn{3}{c}{SE} \\
		\cmidrule(lr){2-4}
		\cmidrule(lr){5-7}
		& MWG   & w-ELL-SS & m-ELL-SS 	&   MWG   & w-ELL-SS & m-ELL-SS\\
		\midrule
		$\sigma_\varepsilon^2$ & 12.73 & \bf{27.54} & 14.21&  0.27  & \bf{32.29} & 15.27 \\
		$\ell_{min}$ & 0.06  & 0.14  & \bf{0.65} & 0.00  & 0.40  & \bf{1.04}  \\
		$z_{min}$ & 0.13  & 0.13  & \bf{0.75}  & 0.01  & 0.55  &  \bf{1.41}\\\
		$\lambda$  & 0.19  & 0.36  & \bf{0.95}  & 0.02  & 0.05  & \bf{0.25}\\
		\cmidrule{1-7}
		MAE & 0.038& 0.039& 0.039& 0.089& 0.038& 0.038\\
		EC & 0.920 & 0.934&  0.934 & 0.863&0.940 & 0.934\\
		\bottomrule
	\end{tabular}%
	\caption{Experiment 2: OES with AR(1) and SE hyperprior employing three different algorithms. $\ell_{min}$ and $z_{min}$ report OES for the minimum ESS across all dimensions. Highest values in boldface. }\label{tab:Sin_OES}%
\end{table}%

%

While the results with the AR(1) hyperprior appear satisfactory under the three sampling schemes (\sfref{fig:SIN_AR}), once again, SE hyperprior (Figure~\ref{fig:SIN_SE}) with MWG is not able to explore the posterior of $\mathbf{u}$, resulting in poor estimates and hence, the highest MAE and poor EC (see Table~\ref{tab:Sin_OES}). 
Analysing the efficiency of the samplers, first, for the AR hyperprior, we observe that while MWG is faster (Table \ref{tab:Comp_time2}), its ESS is consistently smaller (\stableref{tab:ESS_sin}), hence reducing its OES (Table~\ref{tab:Sin_OES}). 
In contrast to the findings in Experiment 1, w-ELL-SS reports better OES compared to MWG due to better mixing in the chains. We believe this is due to the noise level, which favours a whitened parametrisation. Finally, despite the fact that the marginal sampler reports larger CPU times, the low correlation of its chains (\stableref{tab:ESS_sin}) favours its OES.  Second, when using the SE hyperprior, the marginal sampler appears to be significantly faster and consistently reports the best OES.  
This, together with the negligible differences in MAE and EC, suggests that m-ELL-SS offers a good compromise between computational cost and efficiency, with the benefit of working well under highly correlated priors.

\subsubsection*{Experiment 3: Bumps}

\begin{figure}[!t]    
	\begin{center}
	\subfloat[{\tiny{$\boldsymbol{\ell}$, MWG with AR}}]{\includegraphics[scale=.35]{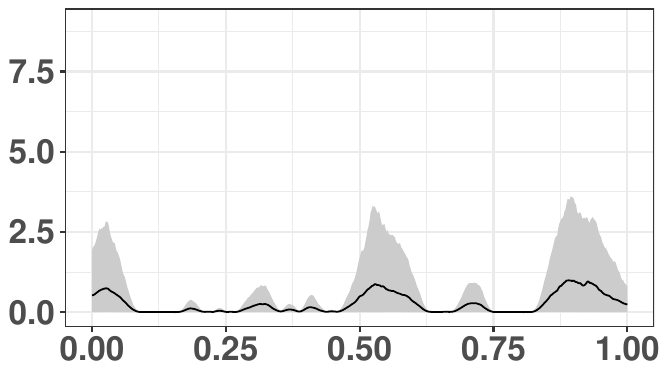}  \label{subfig:EX3a}} 	
	\subfloat[{\tiny{$\boldsymbol{\ell}$, w-ELL-SS with AR}}]{\includegraphics[scale=.35]{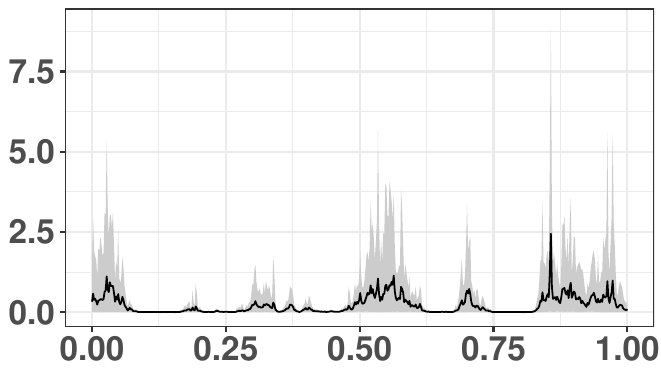} \label{subfig:EX3b}}
	\subfloat[{\tiny{$\boldsymbol{\ell}$, m-ELL-SS with AR}}]{\includegraphics[scale=.35]{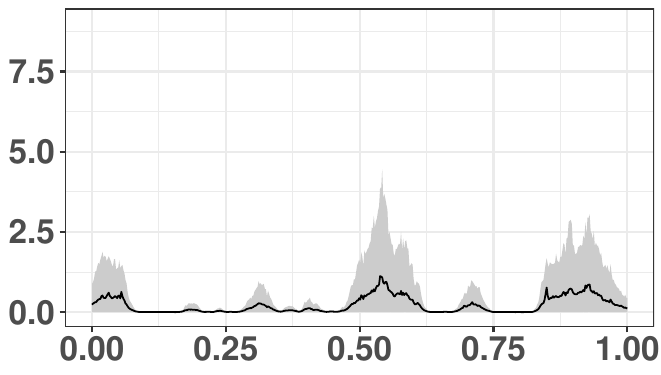} \label{subfig:EX3c}}
	\\
	\subfloat[{\tiny{$\mathbf{z}$, MWG with AR}}]{\includegraphics[scale=.35]{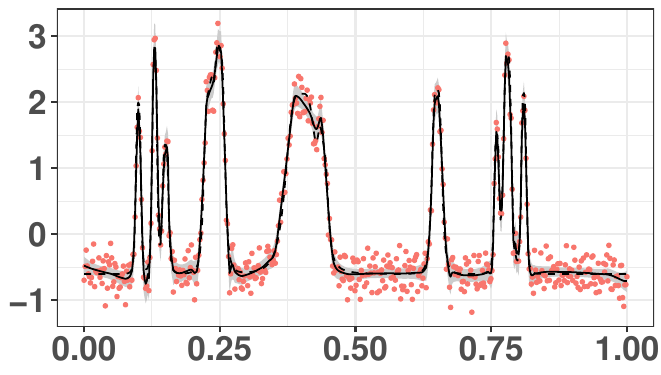} \label{subfig:EX3d}}
	\subfloat[{\tiny{$\mathbf{z}$, w-ELL-SS with AR}}]{\includegraphics[scale=.35]{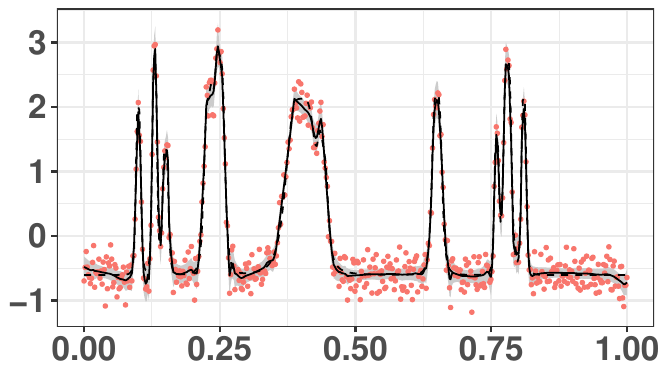}\label{subfig:EX3e}}
	\subfloat[{\tiny{$\mathbf{z}$, m-ELL-SS with  AR}}]{\includegraphics[scale=.35]{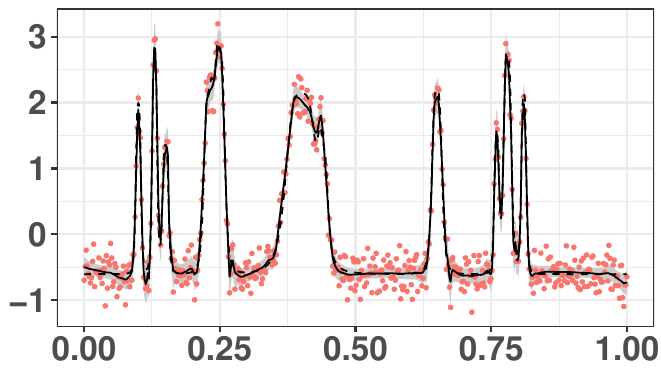} \label{subfig:EX3f}}\\
	\subfloat[{\tiny{$\boldsymbol{\ell}$, MWG SE}}]{\includegraphics[scale=.35]{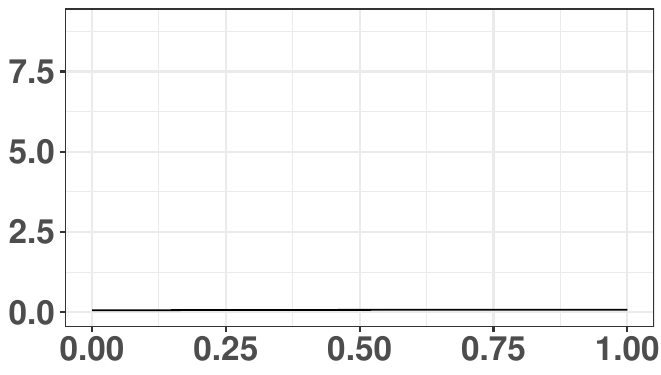}  \label{subfig:EX3g}} 	
	\subfloat[{\tiny{$\boldsymbol{\ell}$, w-ELL-SS with SE}}]{\includegraphics[scale=.35]{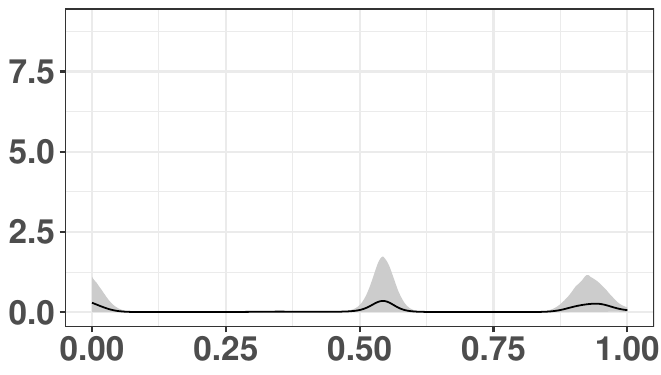} \label{subfig:EX3h}}
	\subfloat[{\tiny{$\boldsymbol{\ell}$, m-ELL-SS with SE} }]{\includegraphics[scale=.35]{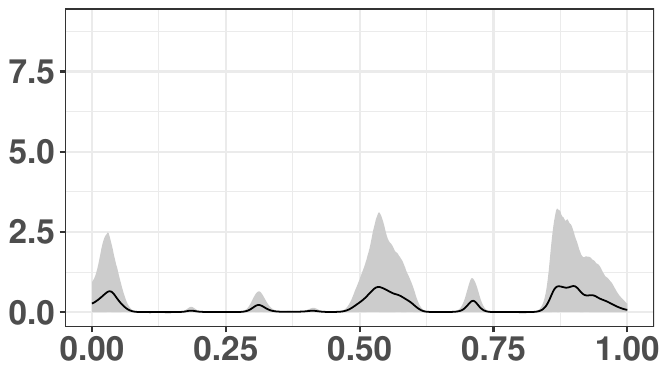} \label{subfig:EX3i}}
	\\
	\subfloat[{\tiny{$\mathbf{z}$, MWG with SE}}]{\includegraphics[scale= .35]{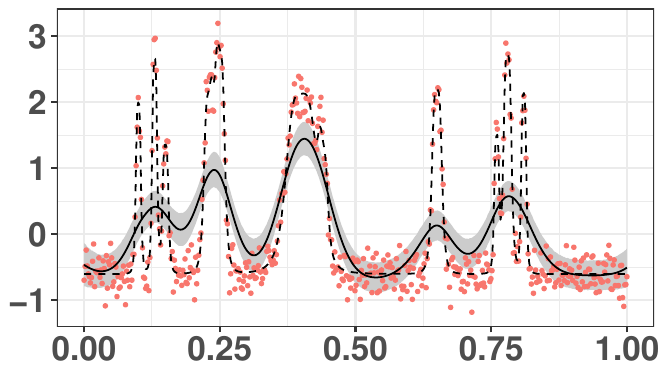} \label{subfig:EX3j}}
	\subfloat[{\tiny{$\mathbf{z}$, w-ELL-SS with SE}}]{\includegraphics[scale=.35]{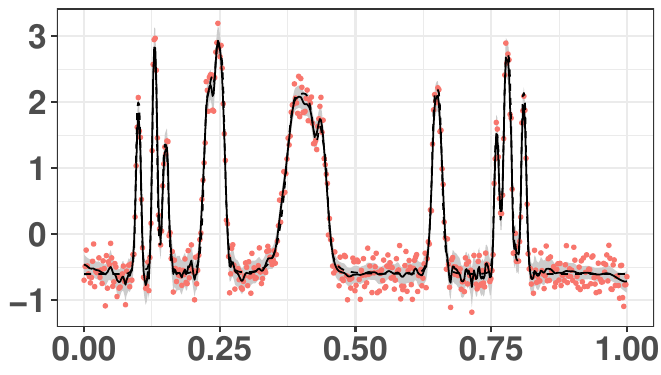}\label{subfig:EX3k}}
	\subfloat[{\tiny{$\mathbf{z}$, m-ELL-SS with SE}}]{\includegraphics[scale=.35]{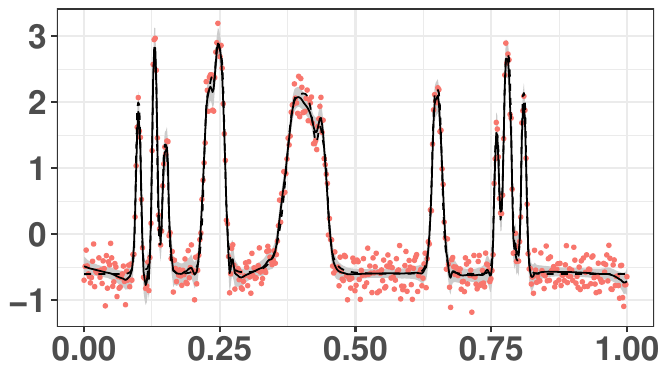} \label{subfig:EX3l}}
\end{center}
	\caption{Results for Experiment 3. Top row: estimated $\boldsymbol{\ell}$ process with $95\% $ credible interval for AR(1) hyperprior with (a) MWG, (b) w-ELL-SS and (c) m-ELL-SS. Second row: estimated $\mathbf{z}$ process with $95\% $ credible interval for AR(1) hyperprior with (d) MWG, (e) w-ELL-SS and (f) m-ELL-SS. Third row: estimated $\boldsymbol{\ell}$ process with $95\% $ credible interval for SE hyperprior with (g) MWG, (h) w-ELL-SS and (i) m-ELL-SS. Bottom row: estimated $\mathbf{z}$ process with $95\% $ credible interval for SE hyperprior with (j) MWG, (k) w-ELL-SS and (l) m-ELL-SS.}
	\label{fig:Bumps_AR}
\end{figure}

The data is generated employing the {\textit{Bumps}} function in \citet{donoho1995adapting} and scaled to have zero mean and unit variance. Following \citet{vannucci1999covariance}, we generate $m=512$ points in the interval [0,1] and  use a signal-to-noise ratio equal to 5, such that 
$\sigma_\varepsilon^2= .04$. To avoid a boundary problem, we extend the domain with $30$ points on each side of the interval, such that $n=572$.  Chains are run for $T=100,000$ iterations with algorithm and prior specific burn-in periods. We use empirical priors for the $\log$ length-scale process and $\log$ length-scale hyperparameter; namely, $\mu_\ell=-3.06$, $\tau_{\ell}^2=2.62$, and $ \log \lambda \sim \mathcal{N}\left( -3.06 , 2.62 \right)$ (see Supplementary Section \ref{Prior_elicitation_EX3} for more details on prior elicitation). 

This example highlights important differences between the two hyperpriors and the proposed MCMC algorithms. First, under the AR(1) hyperprior, the three sampling schemes show differences in the posterior length-scale process (Figure~\ref{fig:Bumps_AR}\subref{subfig:a}-\subref{subfig:c}). While MWG results in a smooth process, m-ELL-SS and w-ELL-SS appear to be more sensitive to the prior, with rougher estimates. 
Second, for the SE hyperprior, once more, MWG did not reach convergence. Also, the performance of w-ELL-SS has become impaired; the posterior length-scale process does not reflect the changes in the correlation structure, and the length-scale hyperparameter did not reach the stationary distribution. The posterior length-scale process obtained with m-ELL-SS appears more appropriate, although, still shows a prior effect.

\begin{table}[!t]
	\centering
	\scriptsize	
	\begin{tabular}{lrrr rrr}
		\toprule
		\textcolor[rgb]{ 1,  0,  0}{}& \multicolumn{3}{c}{AR(1)} & \multicolumn{3}{c}{SE} \\
		\cmidrule(lr){2-4}
		\cmidrule(lr){5-7}
		& MWG   & w-ELL-SS & m-ELL-SS 	&   MWG   & w-ELL-SS & m-ELL-SS\\
		\midrule
		$\sigma_\varepsilon^2$ &\bf{23.42}&	5.73&	5.70& 2.06&	5.48	&\bf{15.36} \\
		$\ell_{min}$ & 0.01	&	0.01	&\bf{0.13} & 0.00&	0.01&	\bf{0.15} \\
		$z_{min}$ &  \bf{2.43}	&	0.10	&0.24 & 0.56&	0.07&	\bf{0.85}\\\
		$\lambda$  & \bf{0.65	}&0.03	&0.13  & \bf{0.07}&	0.00 &	0.03\\
		\cmidrule{1-7}
		MAE &  0.060& 0.061& 0.062&0.461	&0.069	&0.060\\
		EC &0.955&	0.950	&0.959 &0.385&	0.961	&0.967\\
		\bottomrule
	\end{tabular}%
	\caption{Experiment 3: OES with AR(1) and SE hyperprior employing three different algorithms. $\ell_{min}$ and $z_{min}$ report OES for the minimum ESS across all dimensions. Highest values in boldface. }\label{tab:Bumps_OES}%
\end{table}%


The findings discussed above are also evidenced in the OES shown in Table~\ref{tab:Bumps_OES}, where MWG exhibits the highest scores and the lowest MAE under AR(1). In contrast, the m-ELL-SS scheme outperforms MWG and w-ELL-SS for a SE hyperprior.
We believe the differences illustrated in this experiment are a result of a key challenge of elliptical slice sampling. When the likelihood is strong, the sampler can result in poor mixing and, in extreme cases, can get stuck \citep{fagan2016elliptical}. In addition, when sampling kernel parameters in strong likelihood settings, one can expect a non-centred parametrisation (avoiding whitening) to be more efficient (see Section 3 in \citet{murray2010slice}).

The computational time required for this experiment is reported in \stableref{tab:Comp_time3}. Given the same initial values, the marginal sampler  converges to the stationary distribution faster; indeed, m-ELL-SS reports, across experiments, the smallest time spent in burn-in period. Finally, to highlight how the model can benefit from using a more powerful computer, we ran this experiment in an Intel Xeon E5-260V3 2.4GHz (Haswell), 8-core processors with 4GB per core, and we found that the inference procedure is sped up by a factor of $\approx 2.1$ for m-ELL-SS and w-ELL-SS (see \stableref{tab:Comp_time3_HPC}). However, for MWG, the speed up factor was only $\approx 1.2$.

\subsection{Two-dimensional synthetic data} \label{subsec:simulated_2d}

We study the performance of our approach on a $2$-$d$ synthetic dataset, by generating $m=20,449$ noisy observations in an expanded grid of $n_1=n_2=143$ equally spaced points in $\big[0, 10\big]$, employing $z(x_1,x_2)= z(x_1)+z(x_2)$, where both $z(x_1)$ and $z(x_2)$ correspond to the function used in Experiment 1. The noise variance is set to $\sigma^2_\varepsilon=.06$ and the sampler is run for $T=50,000$ iterations, with a burn-in of $10,000$. We use the same prior distributions of Experiment 1 for each of the length-scale processes and corresponding hyperparameters.

\begin{figure}[!t]
	\begin{minipage}[t]{.35\textwidth}
	\subfloat[True surface]{\adjincludegraphics[scale=.17,trim={2cm 1cm 2cm 2cm},clip]{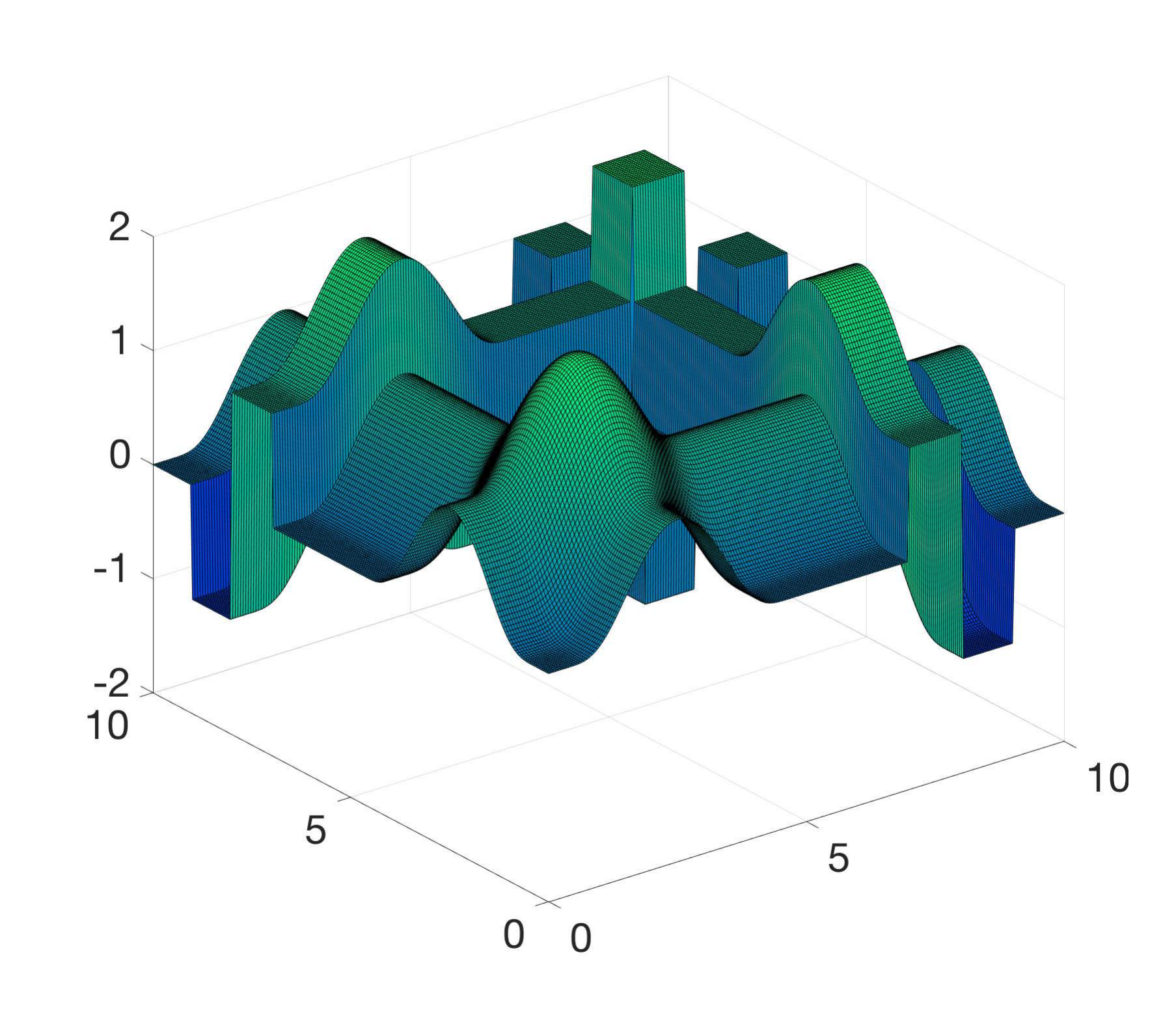}  \label{subfig:2dsimulated_data}}
\end{minipage} \hfill
\begin{minipage}[t]{.61\textwidth}
	\subfloat[Fitted surface and length-scale processes]	{\adjincludegraphics[scale=.17,trim={2cm 1cm 2cm 2.2cm},clip]{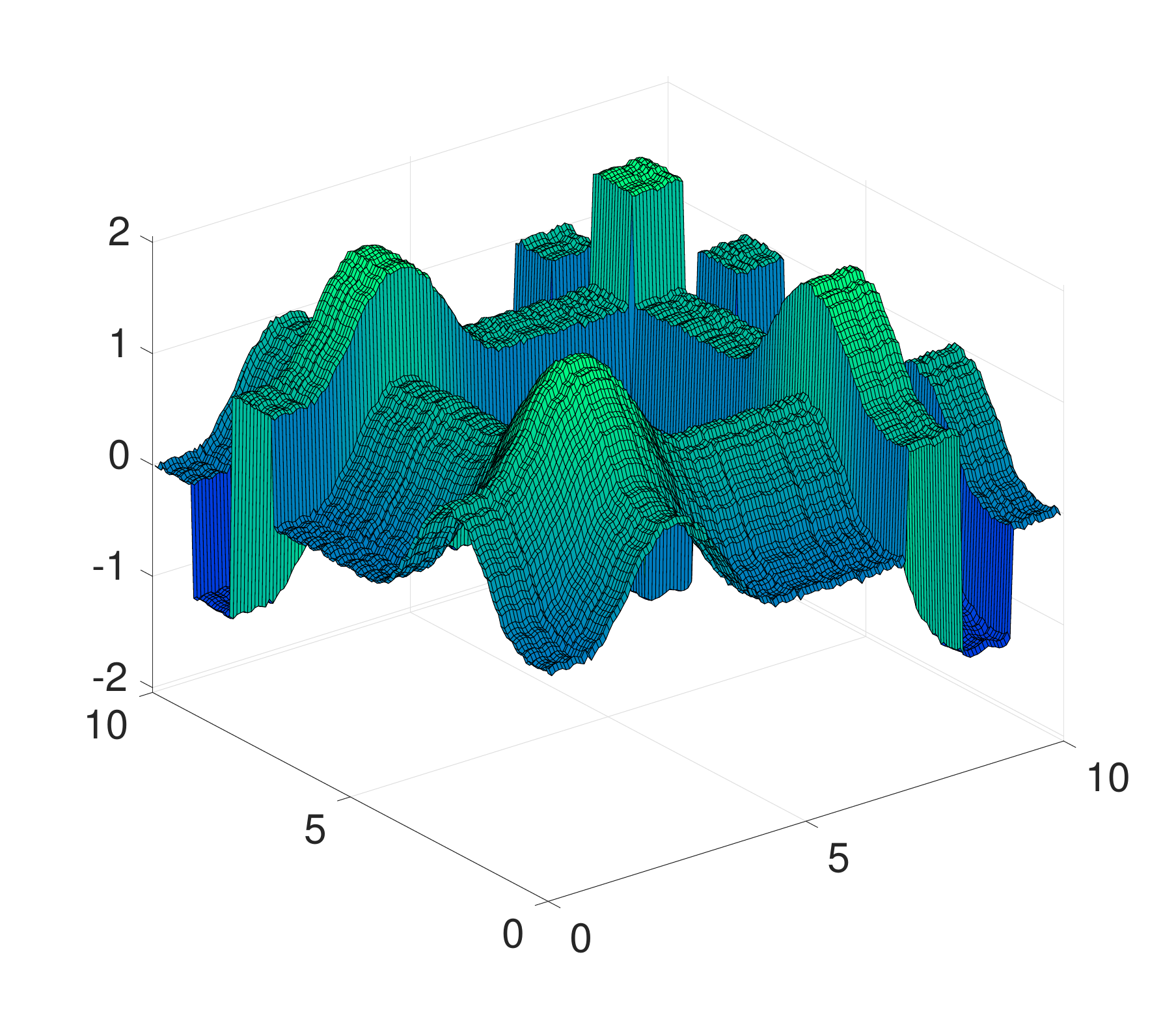} { \includegraphics[height=3.2cm, width=3.1cm]{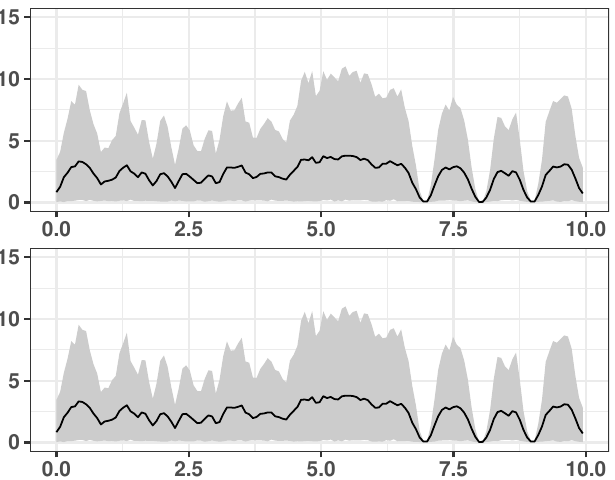}  \label{subfig:2dsimulated_fit}}}
\end{minipage}
	\caption{Results for $2$-dimensional synthetic data. (a): True surface. (b): Posterior mean surface and one-dimensional length-scale processes with 95\% credible intervals.} \label{fig:2d}
\end{figure}

Figure \ref{fig:2d} depicts the true surface versus the posterior mean obtained from a 2-level AGP model (without interaction term), employing the Block-m-ELL-SS algorithm. Our model is able to capture the smooth areas and edges of the surface. In addition, it provides information about the correlation structure along each axis (Figure~\ref{fig:2d}\subref{subfig:2dsimulated_fit}).
The 2-level AGP correctly learns the varying correlation along the surface; for instance, the true function in the region $[ 5,6 ]\times [5,6]$ is constant, and in the same region, the 1-$d$ length-scale processes depict strong correlation. 
The required total computational time for this experiment was 99.26 minutes (19.67 in burn-in and 79.59 in non-burned).

\subsection{Comparative evaluation} \label{Comparative_eval}
We offer a comparative evaluation of our model for the synthetic examples from Section \ref{Simulated_examples} and \ref{subsec:simulated_2d}, against: 1) stationary M\'atern Gaussian process (STAT) with $\nu=1.5$ and 2) Bayesian treed Gaussian process (TGP). For the stationary model, the  length scale and noise variance are inferred via MCMC, employing a marginal sampler with adaptive random walks. The GP prior mean and magnitude are fixed at $0$ and $1$, respectively, as in the 2-level GP model. For the TGP, we consider a stationary Matern kernel with $\nu=1.5$ and a constant mean function. 
The magnitude is also inferred, in contrast to the stationary and the 2-level model. In order to make use of the default prior distributions, we rescale the response and inputs, as recommended by the authors. 

In all the experiments, the chains are run for the same number of iterations ($100,000$), with the same burnin period ($20,000$), and initialised with the same values for STAT and 2-level GP. For our two-dimensional simulated dataset (Experiment 4), we were unable to run the TGP model\footnote{A single iteration of TGP took more than 24 hours on an Intel Core i7-6700 CPU (3.40GHz, 16 GB of RAM). Also, we used TGP in an iMac Pro (2.3GHz 18-core Intel Xeon W processor, Turbo Boost up to 4.3GHz, 128GB 2666MHz DDR4 ECC memory) and after 2 weeks, the code was still running.}, due to the size of the dataset. To offer a comparison, we consider a subset of the original data, reducing the data size from $20,449$ to $441$ observations.

\begin{figure}[!h]   
	\begin{center}
	\subfloat[{\tiny{STAT}}]{\includegraphics[scale=.35]{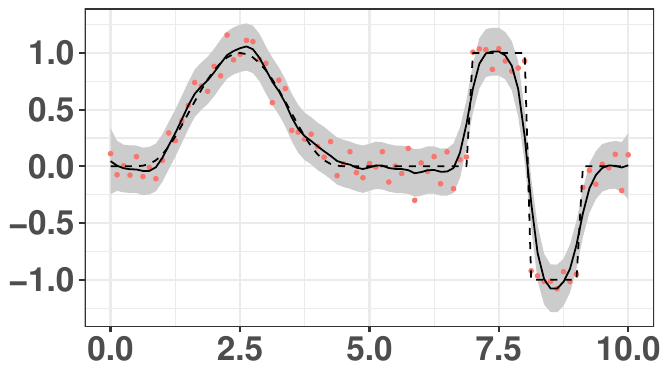} \label{subfig:Comparison_a}}
	\subfloat[{\tiny{TGP}}]{\includegraphics[scale=.35]{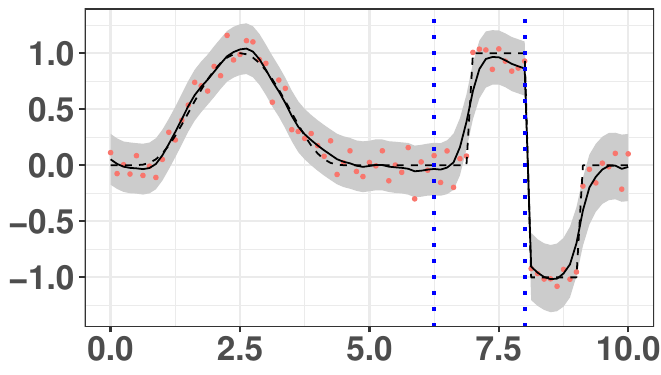} \label{subfig:Comparison_b}}
	\subfloat[{\tiny{m-ELL-SS with SE}}]{\includegraphics[scale=.35]{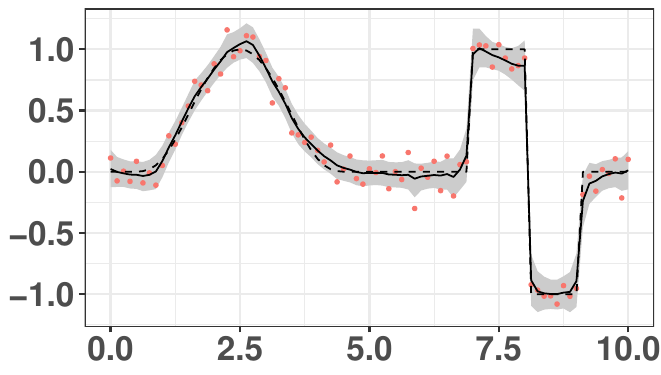}}\\
	\subfloat[{\tiny{STAT}}]{\includegraphics[scale=.35]{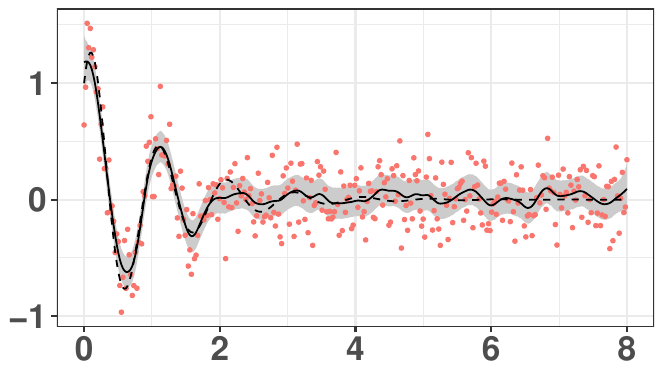}\label{subfig:Comparison_d}}
	\subfloat[{\tiny{TGP}}]{\includegraphics[scale=.35]{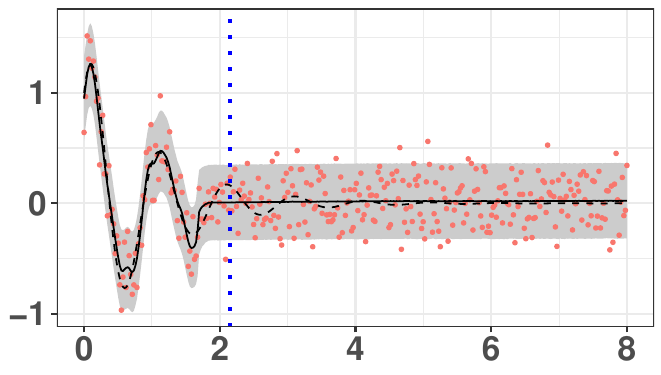}\label{subfig:Comparison_e}}
	\subfloat[{\tiny{m-ELL-SS with SE}}]{\includegraphics[scale=.35]{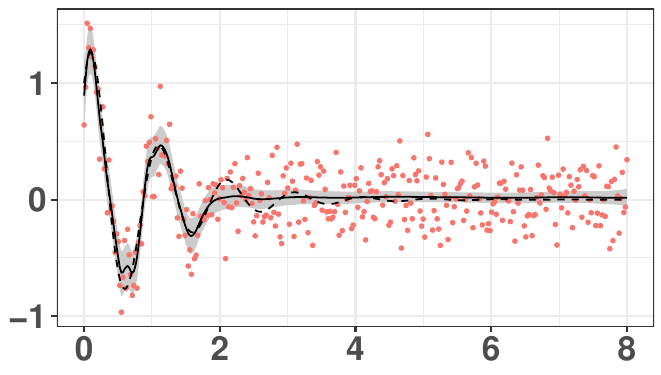}}\\
	\subfloat[{\tiny{STAT}}]{\includegraphics[scale=.35]{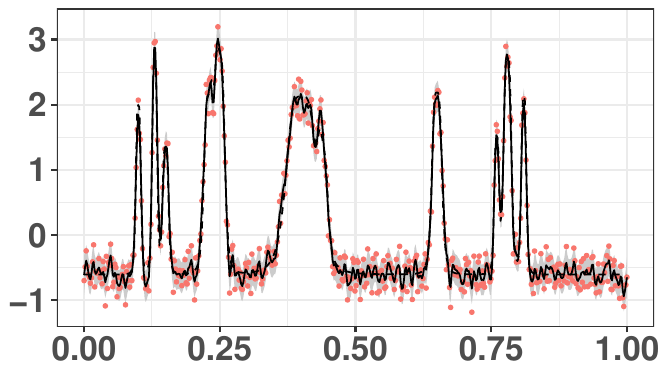}\label{subfig:Comparison_g}}
	\subfloat[{\tiny{TGP}}]{\includegraphics[scale=.35]{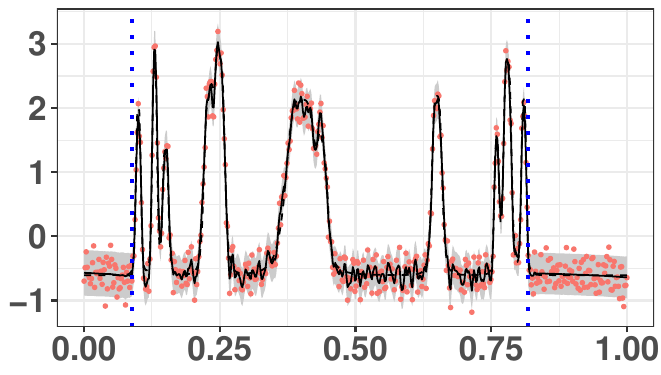} \label{subfig:Comparison_h}}
	\subfloat[{\tiny{m-ELL-SS with SE}}]{\includegraphics[scale=.35]{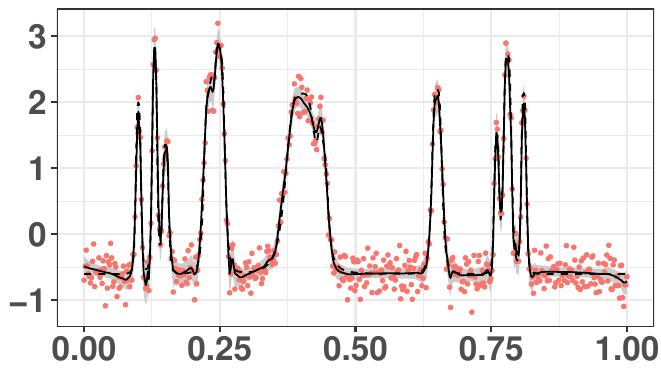}}
\end{center}
	\caption{Comparative evaluation for 1-$d$ experiments. Each row shows one of the simulated experiments. Red dots depict observed data, dotted lines show the true signal, solid lines show the posterior mean, and grey areas depict 95\% credible intervals. (a)(d)(g)(j): Stationary GP (b)(e)(h)(k): TGP, with blue dotted lines depicting MAP cut-off points. (c)(f)(i)(l): 2-level GP with m-ELL-SS algorithm and the hyperprior with lowest MAE.} 
	\label{fig:Comparison}
\end{figure}

\begin{figure}[!h]    
	\centering
	\subfloat[{\tiny{STAT}}]{\includegraphics[scale=.21]{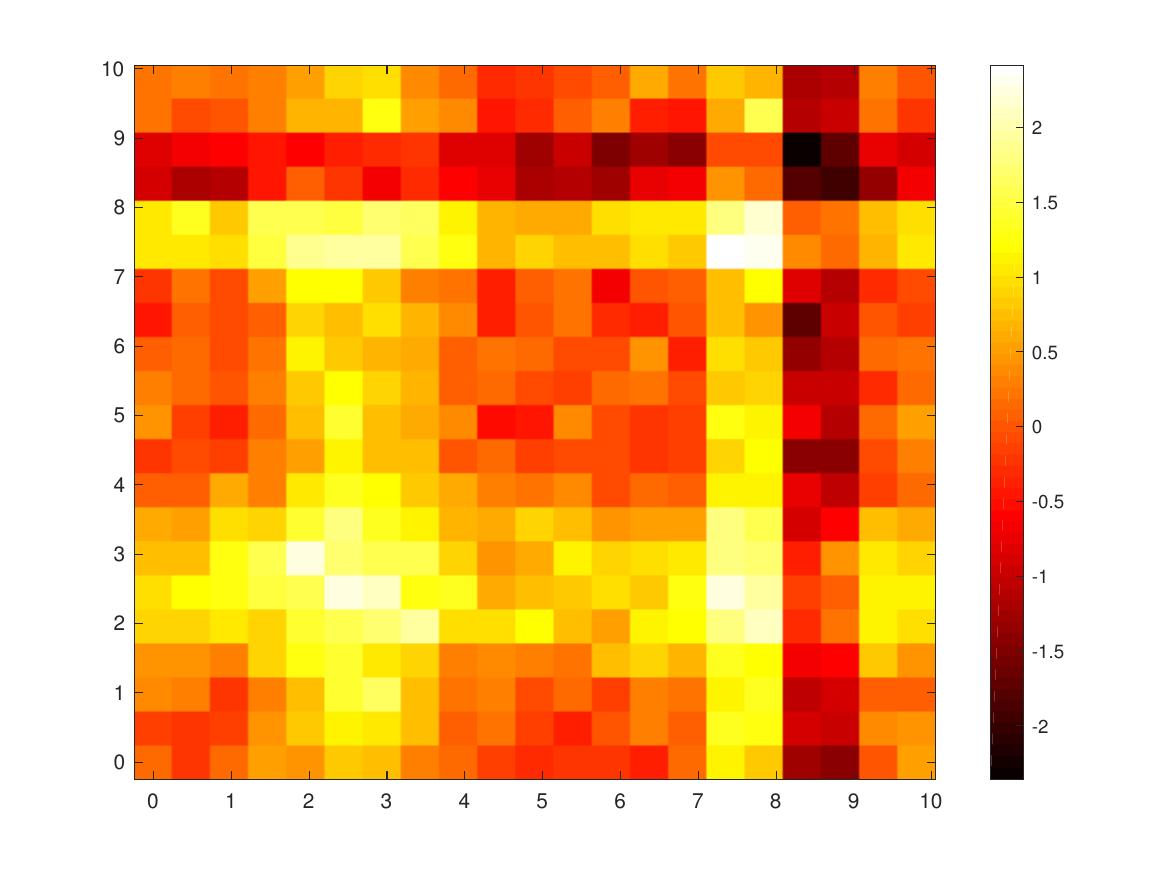}\label{subfig:Comparison2d_a} } 
	\subfloat[{\tiny{TGP}}]{\includegraphics[scale=.21]{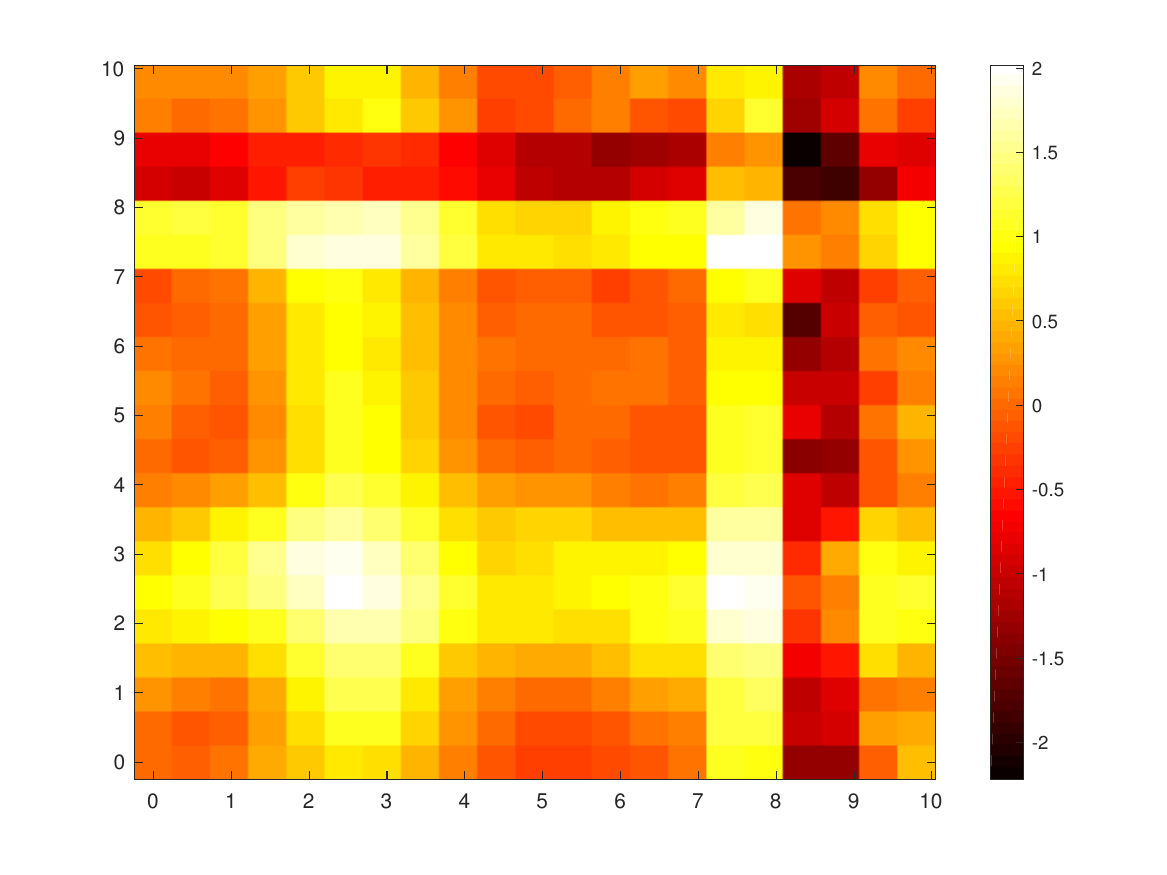} \label{subfig:Comparison2d_b}}
	\subfloat[{\tiny{2-level GP}}]{\includegraphics[scale=.21]{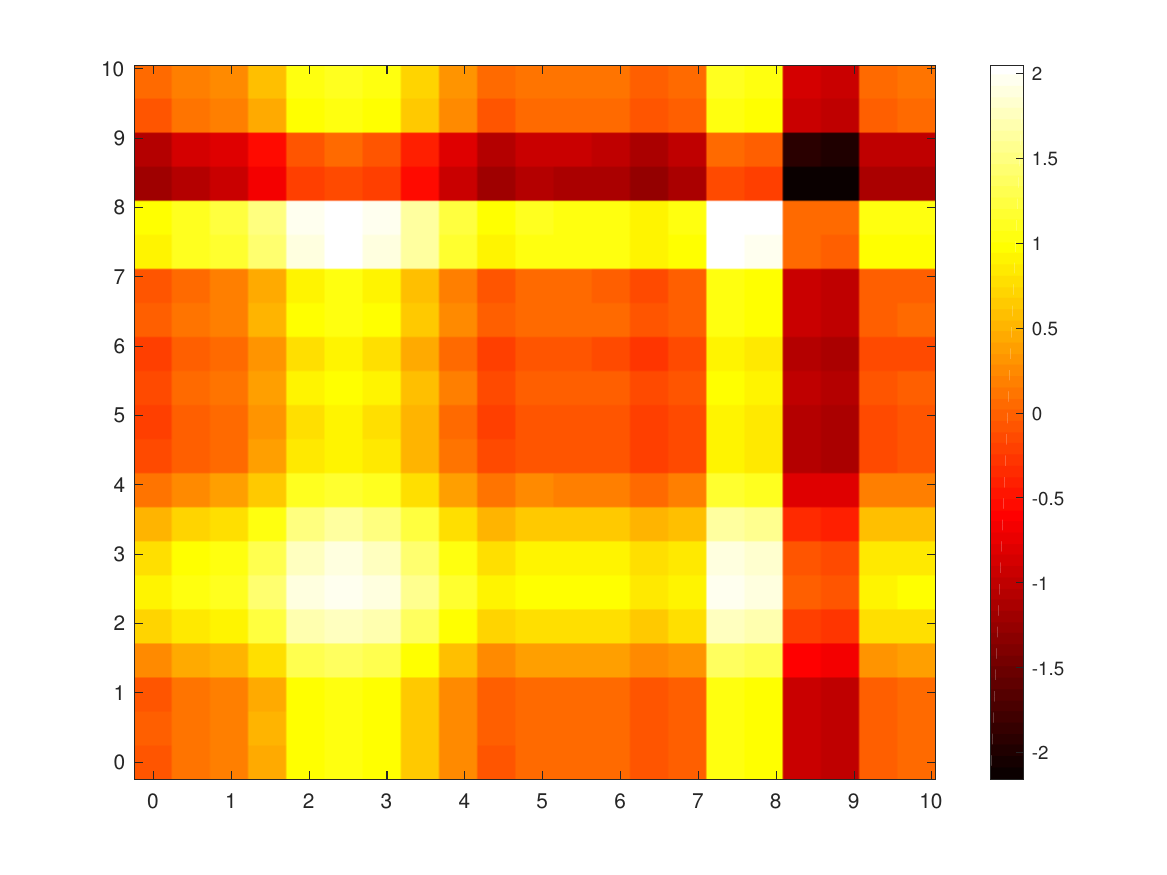} \label{subfig:Comparison2d_c}} 
	\caption{Comparative evaluation for 2-$d$ experiment. Posterior mean surface for (a): anisotropic stationary model, (b): TGP, (c): 2-level AGP with first order terms. } 	\label{fig:Comparison2d}
\end{figure}

Figure~\ref{fig:Comparison} shows the posterior mean estimates of the unknown under the three models for the three different 1-$d$ synthetic datasets, and Figure~\ref{fig:Comparison2d} illustrates the posterior mean surface for the subset of data in Experiment 4. In addition, Table~\ref{tab:Comparison} reports MAE and EC of the experiments. Note that the grey areas depict the $95\%$ credible intervals of the unknown function for STAT and 2-level GP but, instead, depict the $95\%$ credible intervals of the noisy observations for TGP. This is because storing region-specific traces is memory intensive, and the storage is not supported in the tgp package without doing predictions. Similarly, we report EC of the noisy process for TGP in Table \ref{tab:Comparison}.

\begin{table}[!h]
	\scriptsize
	\centering
	\begin{tabular}{lrrrrrrr}
		\toprule
		&    & \multicolumn{2}{c}{STAT} & \multicolumn{2}{c}{TGP} & \multicolumn{2}{c}{2-level GP (AR/SE)} \\
		\cmidrule(lr){3-4}   \cmidrule(lr){5-6}   \cmidrule(lr){7-8}
		& \multicolumn{1}{c}{m} & \multicolumn{1}{c}{MAE} & \multicolumn{1}{c}{EC} & \multicolumn{1}{c}{MAE} & \multicolumn{1}{c}{EC$^\star$} & \multicolumn{1}{c}{MAE} & \multicolumn{1}{c}{EC} \\
		\midrule
		Experiment 1 & 81    &    0.076   &  0.914     &0.056    &   0.963   &  0.041/\bf{0.039}     & 0.988/0.988 \\
		Experiment 2 & 350   &   0.047    & 0.946    &    0.043     &   0.934    &  0.039/\bf{0.038}    & 0.934/0.940  \\
		Experiment 3 & 512   &    0.094   &    0.947   & 0.079      &   0.963    & 0.062/\bf{0.060}  &   0.959/0.967 \\
		Experiment 4 (subset) & 441  &   0.195   &  0.501     &   0.122   &   0.980   & \bf{0.072} & 0.963 \\
		\bottomrule
	\end{tabular}%
	\caption{Comparative evaluation. For Experiments 1-3 with 2-level GP model, we employ m-ELL-SS algorithm for both hyperpriors. Experiment 4 uses Block-m-ELL-SS with AR hyperprior. EC$^\star$ for TGP is reported for the noisy process. Best values in boldface. }
	\label{tab:Comparison}%
	
\end{table}

First, the results make clear the downside of applying a stationary model to non-stationary data in all four experiments. In Experiment 1, STAT is oversmoothing and unable to capture the edges in the function (see Figure~\ref{fig:Comparison}\subref{subfig:Comparison_a}). Example 2 and 3 (Figures~\ref{fig:Comparison}\subref{subfig:Comparison_d}\subref{subfig:Comparison_g}) illustrate how a stationary model tends to overfit when the function is constant, as a result of the different characteristics of the unknown. The same behaviour is repeated in the two-dimensional synthetic example (Figure~\ref{fig:Comparison2d}\subref{subfig:Comparison2d_a}).

Second, while TGP offers an improvement, compared with a stationary setting, the model still oversmooths where the function possesses an edge. For instance, in Figure~\ref{fig:Comparison}\subref{subfig:Comparison_b}, the partition found around $6.2$ is misplaced, and a third partition should be included around $9$ to capture correctly the edges.  In Experiment 2 (Figure~\ref{fig:Comparison}\subref{subfig:Comparison_e}), the partition is also misplaced; this is however more reasonable (compared to Experiment 1) due to the smooth change in the behaviour. In Experiment 3, despite the fact that TGP fit is good when the function is constant (Figure~\ref{fig:Comparison}\subref{subfig:Comparison_h}), the main limitation appears to be in finding some of the partitions that are required to ameliorate the issues resulting from fitting piecewise stationary models. Note that we ran TGP with a different number of iterations ($100,000$; $200,000$ and $500,000$) to verify the results shown in Figure~\ref{fig:Comparison} and~\ref{fig:Comparison2d}  (see Supplementary Section~\ref{Sup_Comparative_eval}
for the results). In Experiment 3, while increasing the number of iterations has a positive effect on the partitions found (and therefore on MAE), it was not enough to outperform the 2-level GP model. Also, this was not the case for the other experiments, where increasing the number of iterations either did not affect the fit or worsened it. Moreover, without knowing the ground truth, it would be hard to know beforehand if the algorithm has been run for long enough to find the appropriate partitions. 

In summary, the 2-level GP is an alternative model for non-stationary data that resolves the issues discussed above. It does not overfit or oversmooth and appears to be more efficient in dealing with different types of non-stationarities, such as, edges, smooth changes, and sharp peaks. Moreover, the 2-level GP clearly benefits from the additive structure, making the model scalable, while retaining flexibility. Notice that evaluating the methods solely on running time can be misleading, as STAT and 2-level GP are implemented in R using standard libraries, while TGP uses R as front end to call \code{C} and \code{C++} optimised code.

\subsection{Real data: NASA rocket booster vehicle} \label{NASA_data}
The analysed dataset in this experiment comes from a computer simulator of  a NASA rocket booster vehicle, the Langley Glide-Back Booster \citep{gramacy2012bayesian}. NASA scientists are interested in understanding the behaviour of the rocket when it re-enters the atmosphere. To do so, the computer experiment considers six different variables; lift, drag, pitch, side force, yaw, and roll; all forces that keep the rocket up.  
Here, we focus on how the lift force is affected as a function of the speed (mach)  and the angle of attack (alpha) for a particular value of the slide-slip angle (beta=0).  
The data is, by nature, non-stationary, with different levels of smoothness along the surface and with a ridge showing the change from subsonic to supersonic flow at mach$=$1 and large alpha. 

\begin{figure}[!t]
	\centering
	\begin{minipage}{.4\textwidth}
		\includegraphics[scale=.3]{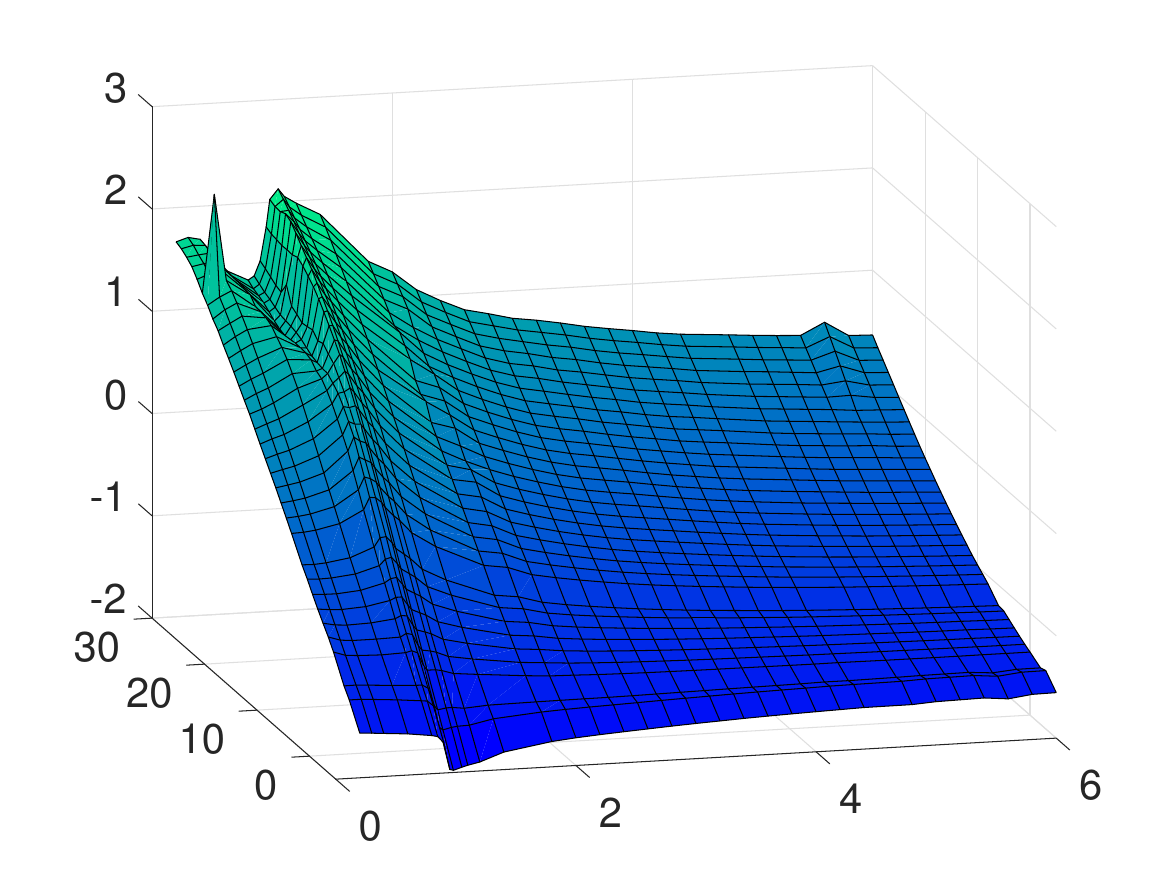}
	\end{minipage}
	\hspace{10mm}
	\begin{minipage}{.35\textwidth}
		{\includegraphics[scale=.3]{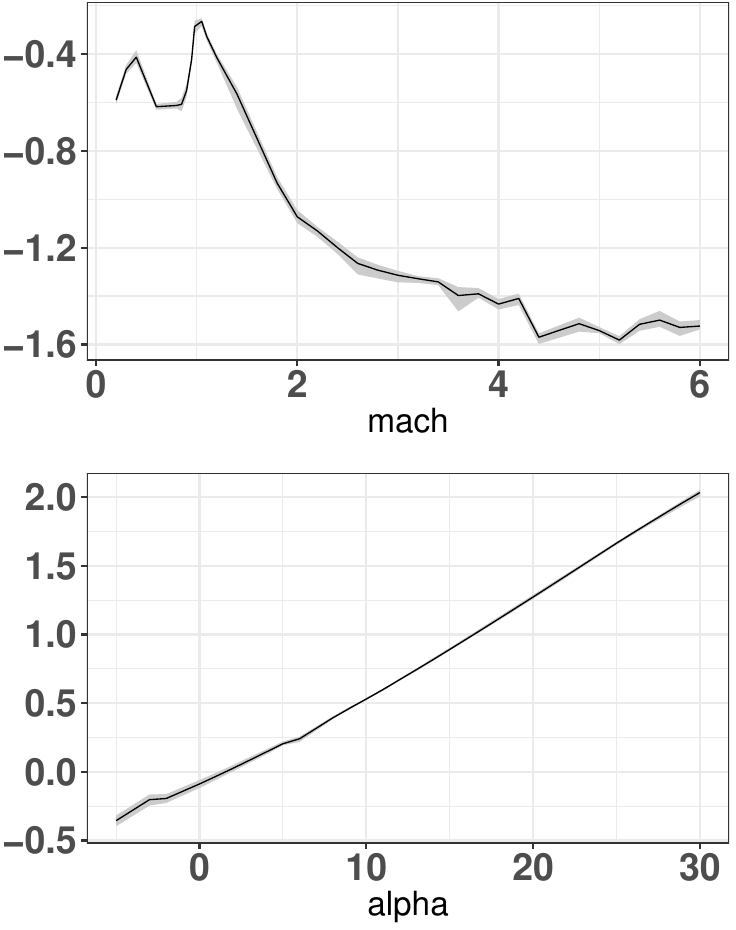}} 
	\end{minipage}
	\caption{Results for NASA rocket booster vehicle. (Left:) Posterior mean. (Right:) Posterior mean of the two one-dimensional processes with 95\% credible intervals.}  \label{fig:Nasa_fit}
\end{figure} 

The data consists on $861$ observations on a $34 \times 33$ grid where the speed ranges from  $[.2, 6]$ and the angle of attack from $[-5,30]$. The data is more dense for mach values around one. Thus, the data  is available on 
an incomplete, non-equally spaced, rectangular grid. We consider the 2-level AGP model with interaction term, employing the Block-m-ELL-SS algorithm for inference.  In order to deal with missing values, we use the model to impute them at each iteration of the MCMC. The chain is run for $50,000$ iterations with a burn-in period of $10,000$.

Figure~\ref{fig:Nasa_fit} shows the posterior mean obtained. The model is able to capture the expected ridge around mach$=$1 and a sharp peak in the boundary around alpha$=$25, where the latter seems to be an error in the convergence of the simulator \citep{gramacy2012bayesian}. Furthermore, the figure illustrates the posterior mean of each of the one-dimensional processes. The results suggest that fitting a stationary process for the angle of attack (alpha) may be enough. 
Depictions of the posterior mean of the second-order interaction term and all length scale processes are provided in the Supplementary Material. 
The required computational time for this experiment was $5.78$ hours in a high performance cluster.

\section{Discussion} \label{Discussion}

We constructed non-stationary hierarchical models based on stochastic parameters and Gaussian Markov random fields, 
ameliorating the computational constraints of doing exact inference in 2-level GP models through sparsity in the finite-dimensional approximation of the inverse covariance matrix of the non-stationary field. 
Different hyperpriors were also explored for the spatially varying length-scale, from  strong prior smoothness assumptions through a squared exponential covariance to rough hyperpriors of an autoregressive AR(1) model, with the latter benefiting from further computational gains.
Strong dependence between the model layers makes efficient inference challenging, and to address this, we introduced and investigated the performance of three different MCMC algorithms. First, we found that the Metropolis-within-Gibbs scheme performs poorly for highly correlated hyperpriors and exhibits deteriorating efficiency as the number of observations or discretisation size increase. Second, the whitened
elliptical slice sampler performs well for weak likelihoods, regardless the hyperprior employed, at the price of highly correlated chains. Finally, the marginal elliptical slice sampler appears to be an efficient strategy to break the correlation between latent process and hyperparameters and offers a good compromise between computational complexity and efficiency of the chains.

We also proposed a novel extension to $D$-dimensional settings by combining additive Gaussian process models with 2-level GPs. 
The additive structure and use of Kronecker algebra for the interaction term result in an inference procedure that is tractable and scalable. Our experiments show that the additive structure retains the flexibility of the 2-level GP and favours its interpretability. Moreover, while we focus on the two-dimensional setting, the additive 2-level model and inference scheme naturally extend to higher dimensions. 
Overall, the comparative evaluation highlights the benefits of our approach, over stationary and popular non-stationary GP models, to recover edges, peaks and smooth variations in the data in both one-dimensional and two-dimensional settings. In addition, the methodology may benefit greatly from using powerful computational resources.

The experiments presented here suggest that the algorithms based on elliptical slice sampling do not deteriorate as the resolution becomes finer or the sample size increases, similar to the schemes discussed by \citet{chen2018robust}. However, it is important to emphasise that elliptical slice sampling is known to perform well for weak data likelihoods; therefore, care must be taken in the small noise limit. 
Furthermore, it would be interesting to explore the performance of the auxiliary gradient-based sampling scheme recently proposed by \citet{titsias2016auxiliary}; 
however, notice that this scheme requires derivatives, which for our model are expensive and not straightforward to compute.  
We also highlight the recent work of \citet{Durrande}, implementing banded matrix operators in TensorFlow, which, combined with GPflow \cite{GPflow}, could provide a promising direction for automatic differentiation for our model.

A natural extension of this work is to the 3-level GP model or, more generally, the deep GP models studied in \citet{dunlop2017deep}. Other interesting directions for future research include exploring higher-order autoregressive hyperpriors; more general kernels;
and alternative likelihoods for problems beyond regression, such as the classification and inverse problems discussed in \citet{chen2018robust}.

%

\section*{Acknowledgements}

\noindent The work reported in this paper was funded by the Mexican National Council of Science and Technology (CONACYT) grant no. CVU609843; the Engineering and Physical Sciences Research Council, grant no. EP/K034154/1; and the Academy of Finland, grant nos. 326240 and 326341, and with support from the Alan Turing Institute - Lloyd’s Register Foundation programme on data-centric engineering.

\bibliographystyle{plainnat}
\bibliography{sample.bib}
\appendix

\pagebreak

\beginsupplement
\setcounter{page}{1}
\setcounter{equation}{0}
\setcounter{section}{0}
\renewcommand{\theequation}{S\arabic{equation}}

\begin{center}
	\textbf{\Large {Supplementary material:
			Posterior Inference for Sparse Hierarchical Non-stationary Models}}
\end{center}
\begin{changemargin}{-1cm}{-1cm}

\section{Fixing the hyperparameters} \label{FixingHyperparameters}
The non-identifiability of covariance hyperparameters in Gaussian process models is a known issue in the literature \citep{zhang2004inconsistent}. A common approach is to set the magnitude parameter to one and only infer the corresponding length-scale, or to employ a re-parametrisations of the hyperparameters. Here, we use the observed data to constrain the prior information of $\mathbf{z}$, $\mathbf{u}$ and $\lambda$. 
First, for the non-stationary process $z(\cdot)$, one can simply re-scale the data to have zero mean and unit variance; such that $\mathbf{z}\sim \mathcal{N}(0,  Q_{\mathbf{u}}^{-1})$. 
Second, for the spatially varying log length-scale prior, $\mathbf{u}\sim \mathcal{N}(\boldsymbol{\mu}_{\ell}, C_{\lambda})$, we empirically fix its mean and magnitude, and only infer the length-scale $\lambda$.  We start by computing the minimum covariate distance, $\alpha$ and the maximum covariante distance, $\beta$. Because identifiability issues arise for length scales outside of $[\alpha,\beta]$, we want to place most of the prior mass within this range for each $\ell_j$. To accomplish this, we can use the quantile function of a Gaussian random variable and solve the following following system of equations,
\begin{align}
\mu_{\ell}-1.96 \tau_{\ell} &= \log \alpha \\
\mu_{\ell}+1.96 \tau_{\ell}&= \log \beta,
\end{align}
to find $\mu_{\ell}$ and $\tau^2_{\ell}$.
Finally, the same approach can be used to set a Gaussian prior for the $\log \lambda$ parameter. 

\section{Additive 2-level GPs} \label{Additive2GP}

\begin{figure}[H]   
	\begin{center}
		\subfloat[1st-order $C^{\text{NS}}_{\mathbf{u}_1}$ ]{\includegraphics[width=4cm, height=4cm]{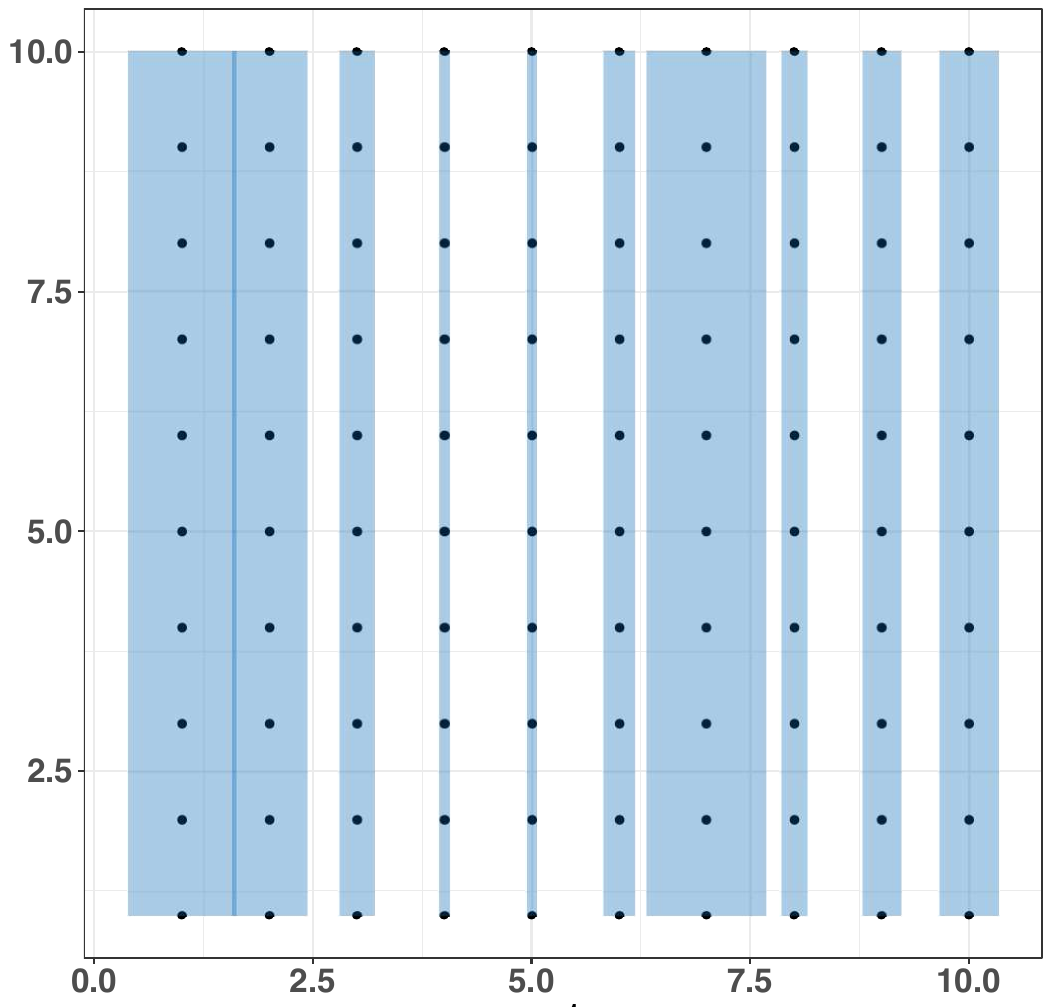}} \hfill
\subfloat[1st-order  $C^{\text{NS}}_{\mathbf{u}_2}$ ]{\includegraphics[width=4cm, height=4cm]{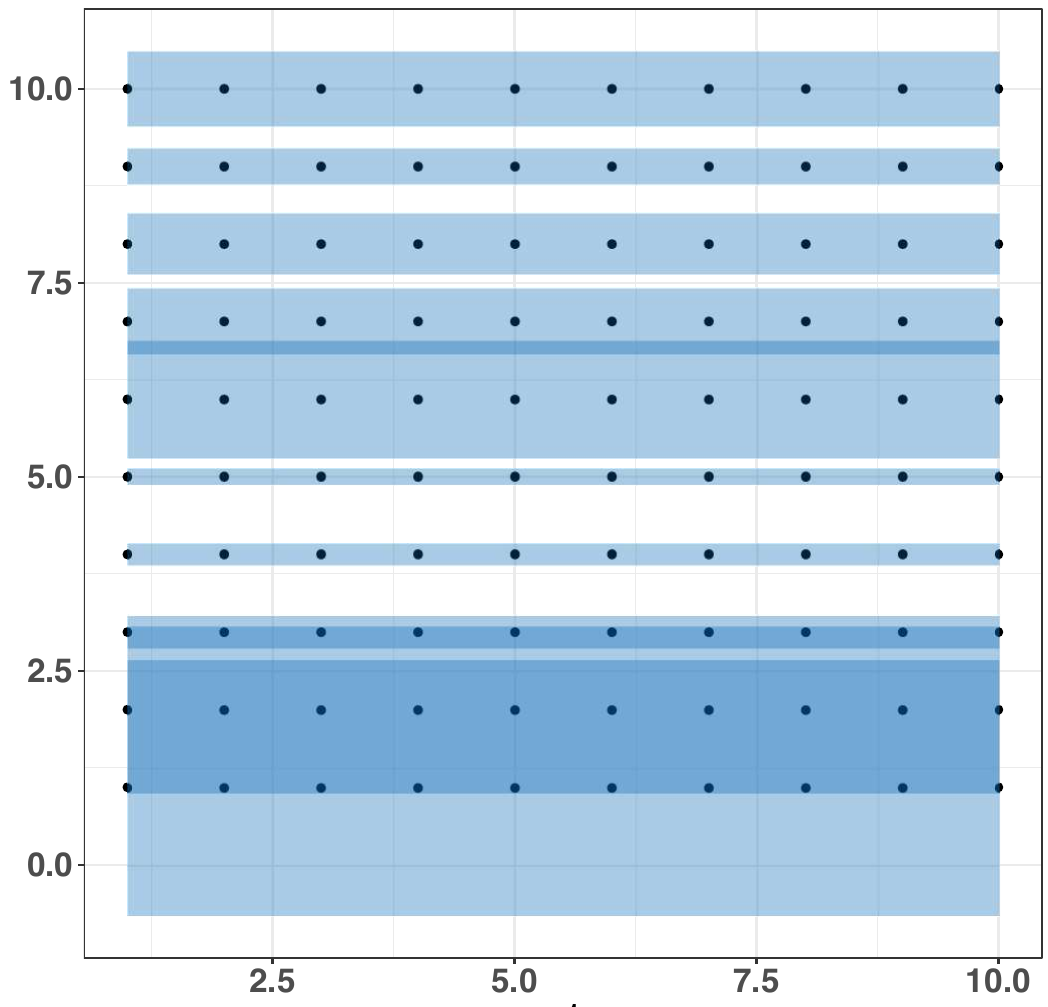}}\hfill
\subfloat[2nd-order $C^{\text{NS}}_{\mathbf{u}_3,\mathbf{u}_4}$ ]{\includegraphics[width=4cm, height=4cm]{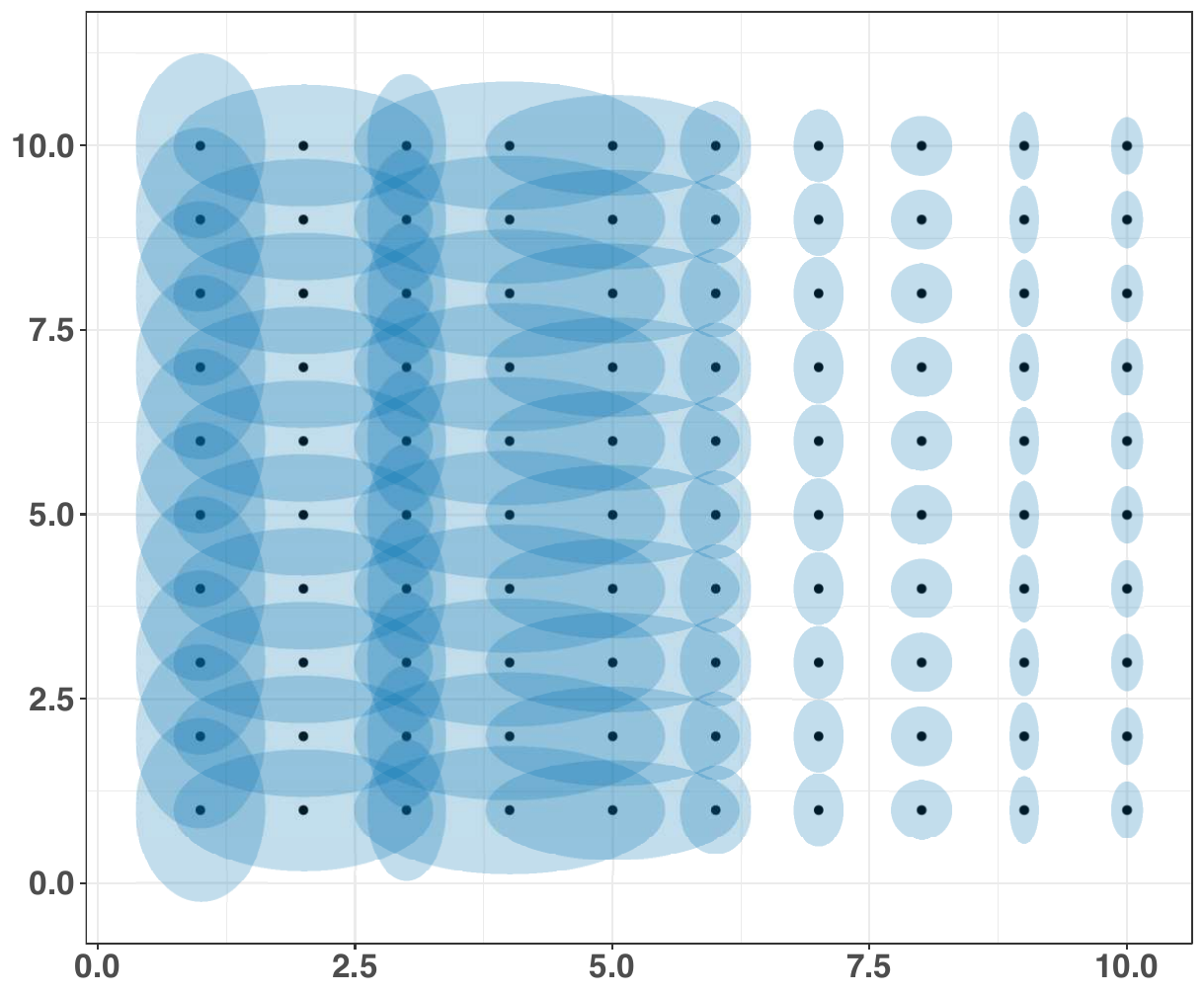}}
	\end{center}
	\caption{The non-stationary additive covariance function in $2$-$d$ with main effects and an interaction is the sum of the three terms: $C^{\text{NS}}$=$C^{\text{NS}}_{\mathbf{u}_1}$ +$C^{\text{NS}}_{\mathbf{u}_2}$ +$C^{\text{NS}}_{\mathbf{u}_3,\mathbf{u}_4}$. At each location the covariance function will make use the data contained within the shaded region in each of the plots. The 1st-order terms can pool together data across dimensions for long-range correlations, while the 2nd-order terms can capture local behavior in both dimensions.} \label{fig:kernels}
\end{figure}

\section{Inference for one-dimensional problems} \label{Algorithmsapp}
\begin{algorithm}[H]
	\small
	\caption{Metropolis-within-Gibbs (MWG)}\label{alg:MWG}
	\begin{algorithmic}[1]	
		\REQUIRE \(A\), \( \sigma{_\varepsilon^2}^{(0)}\), \(\mathbf{u}^{(0)} \), \(\mathbf{z}^{(0)} \) and \( \lambda^{(0)} \)
		\FOR{$t=1$ to $T$}	\vspace{1mm}
		\STATE Draw: \(\log\sigma^{2}_\varepsilon \mid {\log\sigma_\varepsilon^{2}}^{(t-1)} \sim\mathcal{N} ( {\log\sigma_\varepsilon^{2}}^{(t-1)},s_1)\)			\vspace{1mm}
		\STATE Compute: \(\alpha_{\sigma^{2}_\varepsilon}={\text{min}} \left\{ 1, \frac{  \prod_i \mathcal{N} \left(y_i \mid   Az_i^{(t-1)}, {\sigma_\varepsilon^2}  \right) \pi \left(\log{\sigma^2_\varepsilon} \right)}{  \prod_i \mathcal{N} \left(y_i \mid      Az_i^{(t-1)}, {\sigma_\varepsilon^{2}}^{(t-1)}\right)\pi \left(  {\log\sigma_\varepsilon^{2}}^{(t-1)} \right)}\right\} \)		\vspace{1mm}
		\STATE With probability  \(\alpha_{\sigma^{2}_\varepsilon}\) set \({\log\sigma_\varepsilon^{2}}^{(t)} =\log\sigma^2_\varepsilon\),
		otherwise set \({\log\sigma_\varepsilon^{2}}^{(t)} ={\log\sigma_\varepsilon^{2}}^{(t-1)} \)	\vspace{1mm}
		\STATE Run Adaptation for $s_1$  \vspace{1mm}
		\STATE Draw: \(\boldsymbol{\eta} \sim \mathcal{N}(0,{I}_{m+n})
		\)  \vspace{1mm}
		\STATE Set: \( \mathbf{z}^{(t)}=
		{\begin{pmatrix}
			{\sigma_\varepsilon^{-1}}^{(t)} A\\  
			{L} (\boldsymbol{\mathbf{u}}^{(t-1)})				\end{pmatrix}}^{\dagger} \begin{pmatrix}   \begin{pmatrix}
		{\sigma_\varepsilon^{-1} }^{(t)}\mathbf{y}\\
		{	0}
		\end{pmatrix}+  \boldsymbol{\eta}
		\end{pmatrix}\)	
		\COMMENT{ ${^\dagger}$ denotes matrix pseudoinverse. Use QR decomposition} \vspace{1mm}
		\STATE Draw:  \( \mathbf{u}\sim \mathcal{N}( \mathbf{u}^{(t-1)}, P)\)	 \COMMENT{$P=\text{diag}(\sigma^2_{u_{1}}, \hdots,\sigma^2_{u_{n}} )$} 	\vspace{1mm}
		\STATE Set: \(\mathbf{u}^{\prime}=  \mathbf{u}^{(t-1)}\)  and \(\mathbf{u}^{(t)}=  \mathbf{u}^{(t-1)}\)  \vspace{1mm} 
		\FOR{$k=1$ to $n$}  \vspace{1mm} 
		\STATE  Set: \(\mathbf{u}_{j \neq k} = ( u_1^{(t)},  \hdots,u_{k-1}^{(t)} ,u_k,u_{k+1}^{(t-1)}, \hdots, u_{n}^{(t-1)} ) ^{T}\)  \vspace{1mm}   
		\STATE Compute:  \(\alpha_{u_k}={\text{min}} \left\{ 1, \frac{ \mathcal{N} \left( \mathbf{z}^{(t)} \mid 0,  C_{\mathbf{u}_{j \neq k}} \right) \mathcal{N}\left(\mathbf{u}_{j \neq k} \mid \boldsymbol{\mu}_{\ell}, C_{\lambda}^{(t-1)}\right)} { \mathcal{N} \left( \mathbf{z}^{(t)} \mid 0,  C_{\mathbf{u}^{\prime}} \right) \mathcal{N}\left(\mathbf{u}^{\prime} \mid \boldsymbol{\mu}_{\ell}, C_{\lambda}^{(t-1)}\right)}\right\}\)  \vspace{1mm}
		\STATE With probability  $\alpha_{u_k}$ set $u_k^{{(t)}}=u_k$ and $u^{\prime}_k=u_k$;
		otherwise set $u_k^{(t)}=u^{{(t-1)}}_k $\vspace{1mm}
		\ENDFOR\vspace{1mm}
		\STATE Run Adaptation for $P$ \vspace{1mm}
		\STATE  Draw: \(\log\lambda|\log\lambda^{(t-1)} \sim\mathcal{N} (\log\lambda^{(t-1)},s_2 )\) \vspace{1mm}
		\STATE Compute: \( \alpha_\lambda={\text{min}} \left\{ 1, \frac{ \mathcal{N} \left( \mathbf{u}^{(t)}|  \boldsymbol{\mu}_{\ell}, C_{\lambda} \right) \pi \left(\log{\lambda} \right)}{  \mathcal{N} \left( \mathbf{u}^{(t)}|  \boldsymbol{\mu}_{\ell}, C_{\lambda^{(t-1)}} \right) \pi \left(\log{\lambda^{(t-1)}} \right)}\right\}\) \vspace{1mm}
		\STATE With probability  \(\alpha_\lambda\) set \(\log\lambda^{{(t)}}=\log\lambda\),
		otherwise set \(\log\lambda^{{(t)}}=\log\lambda^{{(t-1)}}\)\vspace{1mm}
		\STATE Run Adaptation for $s_2$\vspace{1mm}
		\ENDFOR
	\end{algorithmic}
\end{algorithm}

\begin{algorithm}[htbp]
	\footnotesize
	\caption{Whitened Elliptical Slice Sampling (w-ELL-SS)}\label{alg:W_ESS}
	\begin{algorithmic}[1]
		\REQUIRE \(A\), \( \sigma{_\varepsilon^2}^{(0)}\), \( \boldsymbol{\zeta}^{(0)}\), \( \boldsymbol{\xi}^{(0)}\), \( \lambda^{(0)} \) ,  \( \mathbf{u} =R_{\lambda^{(0)}}\boldsymbol{\zeta}^{(0) } + \boldsymbol{\mu}_{\ell}\) and \(  \mathbf{z}=  {L(\mathbf{u})}^{-1}\boldsymbol{\xi}^{(0)} \) \vspace{1mm} 
		\FOR{$t=1$ to $T$}	\vspace{1mm}
		\STATE Draw: \(\log\sigma^{2}_\varepsilon \mid {\log\sigma_\varepsilon^{2}}^{(t-1)} \sim\mathcal{N} ( {\log\sigma_\varepsilon^{2}}^{(t-1)},s_1)\)			\vspace{1mm}
		\STATE Compute: \(\alpha_{\sigma^{2}_\varepsilon}={\text{min}} \left\{ 1, \frac{  \prod_i \mathcal{N} \left(y_i \mid   Az_i, {\sigma_\varepsilon^2}  \right) \pi \left(\log{\sigma^2_\varepsilon} \right)}{   \prod_i \mathcal{N} \left(y_i \mid   Az_i, {\sigma_\varepsilon^{2}}^{(t-1)}  \right)\pi \left(  {\log\sigma_\varepsilon^{2}}^{(t-1)} \right)}\right\} \) 	\vspace{1mm}
		\STATE With probability  \(\alpha_{\sigma^{2}_\varepsilon}\) set \({\log\sigma_\varepsilon^{2}}^{(t)} =\log\sigma^2_\varepsilon\),
		otherwise set \({\log\sigma_\varepsilon^{2}}^{(t)} ={\log\sigma_\varepsilon^{2}}^{(t-1)} \)	\vspace{1mm}
		\STATE Run Adaptation for $s_1$  \vspace{1mm}
		\STATE Draw: \(\boldsymbol{\nu}\sim \mathcal{N}(0, {I_{n}})\)  \vspace{1mm}
		\STATE Draw:  \(\beta \sim \mathcal{U}[0,1]\)\vspace{1mm}
		\STATE Compute: \(\kappa= \log   \prod_i  \mathcal{N} (  y_i \mid Az_i,  {\sigma^2_{\varepsilon}}^{(t)} ) + \log \beta\)  \vspace{1mm}
		\STATE Draw: \( \theta \sim \mathcal{U}[0,2\pi] \)\vspace{1mm}
		\STATE Define: \([\theta_{\text{min}},\theta_{\text{max}}]= [\theta-2\pi,\theta]\) \vspace{1mm}
		\STATE Propose: \(\boldsymbol{\zeta}^{\prime}=\boldsymbol{\zeta}^{(t-1)} \cos\theta+\boldsymbol{\nu}\sin \theta\) \vspace{1mm}
		\STATE Update: \(\mathbf{u}=R_{\lambda^{(t-1)}}\boldsymbol{\zeta}^{\prime}+ \boldsymbol{\mu}_{\ell} \) \vspace{1mm}
		\STATE Solve: \({L(\mathbf{u})}\mathbf{z}=\boldsymbol{\xi}^{(t-1)}\) \vspace{1mm}
		\IF{ \(\log   \prod_i\mathcal{N} ( y_i \mid  A z_i , {\sigma^2_{\varepsilon}}^{(t)} )  > \kappa\)  }   \vspace{1mm}
		\STATE  Set: \(\boldsymbol{\zeta}^{(t)}=\boldsymbol{\zeta}^{\prime} \) \vspace{1mm}
		\ELSE   \vspace{1mm}
		\IF{\(\theta<0\)} \vspace{1mm} \STATE {$\theta_{\text{min}}=\theta$ } \ELSE \vspace{1mm}  \STATE {$\theta_{\text{max}}=\theta$} \vspace{1mm}
		\ENDIF  \vspace{1mm}
		\STATE Draw: \(\theta \sim \mathcal{U} [\theta_{\text{min}},\theta_{\text{max}}]\)  \vspace{1mm}
		\STATE Go back to step 11.  \vspace{1mm}
		\ENDIF \vspace{1mm}
		\STATE  Draw: \(\log\lambda|\log\lambda^{(t-1)} \sim\mathcal{N} (\log\lambda^{(t-1)},s_2)\)	 \vspace{1mm}
		\STATE Compute: \( \mathbf{u}^{\prime}=R_{\lambda}\boldsymbol{\zeta}^{(t)}+ \boldsymbol{\mu}_{\ell}\) \vspace{1mm}
		\STATE Solve: \({L}(\mathbf{u^{\prime}})\mathbf{z}^{\prime}=\boldsymbol{\xi}^{(t-1)}\) \vspace{1mm}
		\STATE Compute: \( \alpha_\lambda={\text{min}} \left\{ 1, \frac{ \prod_i \mathcal{N} \left( y_i|  Az_i^{\prime} , {\sigma^2_{\varepsilon}}^{(t)} \right) \pi \left(\log{\lambda} \right)}{   \prod_i \mathcal{N} \left(y_i| Az_i,  {\sigma^2_{\varepsilon}}^{(t)} \right) \pi \left(\log{\lambda^{(t-1)}} \right)}\right\}\)  \vspace{1mm}
		\STATE With probability  \(\alpha_\lambda\) set \( \log\lambda^{{(t)}}= \{  \log \tau^2_\ell,  \log\lambda \}\), and \( \mathbf{u}=\mathbf{u}^{\prime}\);
		otherwise, set \( \log\lambda^{{(t)}}= \{ \log \tau^2_\ell,  \log\lambda^{(t-1)} \}\), \vspace{1mm}
		\STATE Run Adaptation for $s_2$  \vspace{1mm}
		\STATE Draw: \(\boldsymbol{\eta} \sim \mathcal{N}(0,{I_{m+n}})\)  \vspace{1mm}
		\STATE Set: \( \mathbf{z}=
		{\begin{pmatrix}
			{\sigma_\varepsilon^{-1}}^{(t)} A\\  
			{L} (\boldsymbol{\mathbf{u}})				\end{pmatrix}}^{\dagger} \begin{pmatrix}   \begin{pmatrix}
		{\sigma_\varepsilon^{-1} }^{(t)}\mathbf{y}\\
		{	0}
		\end{pmatrix}+  \boldsymbol{\eta}
		\end{pmatrix}\)	
		\COMMENT{ ${^\dagger}$ denotes the matrix pseudoinverse. Use QR decomposition} \vspace{1mm}
		\STATE 	Solve: \( {L{(\mathbf{u})}} \boldsymbol{\xi}^{(t)}=\mathbf{z}\)	 \vspace{1mm}
		\ENDFOR
	\end{algorithmic}
\end{algorithm}

\begin{algorithm}[H]
	\footnotesize
	\caption{Marginal Elliptical Slice Sampling (m-ELL-SS)}\label{alg:M_ESS}
	\begin{algorithmic}[1]
		\REQUIRE \(A\), \( \sigma{_\varepsilon^2}^{(0)}\), \(\boldsymbol{\zeta}^{(0)} \), \( \lambda^{(0)} \), and  \( \mathbf{u}=R_{\lambda^{(0)}}\boldsymbol{\zeta}^{(0)} +\boldsymbol{\mu}_{\ell} \) \vspace{1mm} 
		\FOR{$t=1$ to $T$}	\vspace{1mm}
		\STATE Draw: \(\log\sigma^{2}_\varepsilon \mid {\log\sigma_\varepsilon^{2}}^{(t-1)} \sim\mathcal{N} ( {\log\sigma_\varepsilon^{2}}^{(t-1)},s_1)\)			\vspace{1mm}
		\STATE Compute: \(\alpha_{\sigma^{2}_\varepsilon}={\text{min}} \left\{ 1, \frac{ \mathcal{N} \left(
			\mathbf{y} \mid 0, AQ^{-1}_{\mathbf{u}}A^{\text{\tiny{T}}}+\sigma^2_{\varepsilon}I_m\right)  \pi \left(\log{\sigma^2_\varepsilon} \right)}{\mathcal{N} \left(\mathbf{y} \mid 0,    AQ^{-1}_{\mathbf{u}}A^{\text{\tiny{T}}}+  {\sigma_\varepsilon^{2}}^{(t-1)} {I_m} \right)\pi \left(  {\log\sigma_\varepsilon^{2}}^{(t-1)} \right)}\right\} \)		\vspace{1mm}
		\STATE With probability  \(\alpha_{\sigma^{2}_\varepsilon}\) set \({\log\sigma_\varepsilon^{2}}^{(t)} =\log\sigma^2_\varepsilon\),
		otherwise set \({\log\sigma_\varepsilon^{2}}^{(t)} ={\log\sigma_\varepsilon^{2}}^{(t-1)} \)	\vspace{1mm}
		\STATE Run Adaptation for $s_1$  \vspace{1mm}
		\STATE Draw: \(\boldsymbol{\nu}\sim \mathcal{N}(0, {I_n})\)  \vspace{1mm}
		\STATE Draw:  \(\beta \sim \mathcal{U}[0,1]\)\vspace{1mm}
		\STATE Compute: \(\kappa= \log \mathcal{N} ( \mathbf{y} \mid 0, AQ_{\mathbf{u}}^{-1}A^{\text{\tiny{T}}}+ {\sigma^2_{\varepsilon}}^{(t)}{I_m} ) + \log \beta\) \vspace{1mm}
		\STATE Draw: \( \theta \sim \mathcal{U}[0,2\pi] \)\vspace{1mm}
		\STATE Define: \([\theta_{\text{min}},\theta_{\text{max}}]= [\theta-2\pi,\theta]\) \vspace{1mm}
		\STATE Propose: \(\boldsymbol{\zeta}^{\prime}=\boldsymbol{\zeta}^{(t-1)} \cos\theta+\boldsymbol{\nu}\sin \theta\) \vspace{1mm}
		\STATE Compute: \(\mathbf{u}=R_{\lambda^{(t-1)}}\boldsymbol{\zeta}^{\prime} + \boldsymbol{\mu}_{\ell}\) \vspace{1mm}
		\IF{ \(\log \mathcal{N} ( \mathbf{y} \mid 0, AQ_{\mathbf{u}}^{-1}A^{\text{\tiny{T}}}+ {\sigma^2_{\varepsilon}}^{(t)}{I_{m}} )  > \kappa\)}  \vspace{1mm}
		\STATE  Set: \(\boldsymbol{\zeta}^{(t)}=\boldsymbol{\zeta}^{\prime} \)  \vspace{1mm}
		\ELSE   \vspace{1mm}
		\IF{\(\theta<0\)} \vspace{1mm} \STATE {$\theta_{\text{min}}=\theta$ } \ELSE \vspace{1mm}  \STATE {$\theta_{\text{max}}=\theta$} \vspace{1mm}
		\ENDIF  \vspace{1mm}
		\STATE Draw: \(\theta \sim \mathcal{U} [\theta_{\text{min}},\theta_{\text{max}}]\)  \vspace{1mm}
		\STATE Go back to step 11.  \vspace{1mm}
		\ENDIF \vspace{1mm}
		\STATE  Draw: \(\log\lambda|\log\lambda^{(t-1)} \sim\mathcal{N} (\log\lambda^{(t-1)},s_2)\)	 \vspace{1mm}
		\STATE Compute: \( \mathbf{u}^{\prime}=R_{\lambda}\boldsymbol{\zeta}^{(t)} +\boldsymbol{\mu}_{\ell}\) 
		\COMMENT{${\lambda}=\{ \tau_{\ell}^2, \lambda \} $}
		\vspace{1mm}
		\STATE Compute: \( \alpha_\lambda={\text{min}} \left\{ 1, \frac{ \mathcal{N} \left( \mathbf{y} \mid  0, AQ^{-1}_{\mathbf{u}^{\prime}}A^{\text{\tiny{T}}}+  {\sigma_\varepsilon^{2}}^{(t)}I_m \right)  \pi \left(\log{\lambda} \right) }{   \left( \mathbf{y} \mid  0, AQ^{-1}_{\mathbf{u}}A^{\text{\tiny{T}}}+  {\sigma_\varepsilon^{2}}^{(t)}I_m \right) \pi \left(\log{\lambda^{(t-1)}} \right)}\right\}\)  \vspace{1mm}
		\STATE With probability  \(\alpha_\lambda\) set \( \log\lambda^{{(t)}}= \{  \log \tau^2_\ell,  \log\lambda \}\), and \( \mathbf{u}=\mathbf{u}^{\prime}\);
		otherwise, set \( \log\lambda^{{(t)}}= \{ \log  \tau^2_\ell,  \log\lambda^{(t-1)} \}\), \vspace{1mm}
		\STATE Run Adaptation for $s_2$  \vspace{1mm}
		\ENDFOR
	\end{algorithmic}
\end{algorithm}
\section{Inference for two-dimensional problems}
\begin{algorithm}[H]
	\footnotesize
	\caption{Block Marginal Elliptical Slice Sampling (Block-m-ELL-SS)}\label{alg:M_ESS_2d}
	\begin{algorithmic}[1]
		\REQUIRE \(A_1\), \(A_2\), \(A_3\) \( \sigma{_\varepsilon^2}^{(0)}\), \( \mathbf{z}_1^{(0)}\), \( \mathbf{z}_2^{(0)}\), \( \mathbf{z}_3^{(0)}\), \( \boldsymbol{\xi}_1^{(0)}\) , \( \boldsymbol{\xi}_2^{(0)}\) , \( \boldsymbol{\xi}_3^{(0)}\),  \( \boldsymbol{\xi}_4^{(0)}\), \( \lambda_1^{(0)} \), \( \lambda_2^{(0)} \), \( \lambda_3^{(0)} \) and \( \lambda_4^{(0)} \) 	\vspace{1mm}
		\FOR{$t=1$ to $T$}	\vspace{1mm}
		\STATE Draw: \(\log\sigma^{2}_\varepsilon \mid {\log\sigma_\varepsilon^{2}}^{(t-1)} \sim\mathcal{N} ( {\log\sigma_\varepsilon^{2}}^{(t-1)},s_1)\)			\vspace{1mm}
		\STATE Compute: \(\alpha_{\sigma^{2}_\varepsilon}={\text{min}} \left\{ 1, \frac{ \mathcal{N} \left(\mathbf{y} \mid   A_1 {\mathbf{z}_1}+  A_2 {\mathbf{z}_2}+ A_3 {\mathbf{z}_3}, {\sigma_\varepsilon^2} {I_{m}} \right) \pi \left(\log{\sigma^2_\varepsilon} \right)}{\mathcal{N} \left(\mathbf{y} \mid      A_1{\mathbf{z}_1}+ A_2 {\mathbf{z}_2}+ A_3{\mathbf{z}_3}, {\sigma_\varepsilon^{2}}^{(t-1)}  {I_{m}} \right)\pi \left(  {\log\sigma_\varepsilon^{2}}^{(t-1)} \right)}\right\} \) 	\vspace{1mm}
		\STATE With probability  \(\alpha_{\sigma^{2}_\varepsilon}\) set \({\log\sigma_\varepsilon^{2}}^{(t)} =\log\sigma^2_\varepsilon\),
		otherwise set \({\log\sigma_\varepsilon^{2}}^{(t)} ={\log\sigma_\varepsilon^{2}}^{(t-1)} \)	\vspace{1mm}
		\STATE Run Adaptation for $s_1$  \vspace{1mm}
		\STATE Draw: \(\boldsymbol{\nu}\sim \mathcal{N}(0, {I_{n_1}})\)  \vspace{1mm}
		\STATE Draw:  \(\beta \sim \mathcal{U}[0,1]\)\vspace{1mm}
		\STATE Compute: \(\kappa= \log \mathcal{N} ( \mathbf{y}-A_2\mathbf{z}_2^{(t-1)}-A_3\mathbf{z}_3^{(t-1)} \mid 0, A_1Q_{\mathbf{u}_1}^{-1}A_1^{\text{\tiny{T}}}+ {\sigma^2_{\varepsilon}}^{(t)}{I_m} ) + \log \beta\) \vspace{1mm}
		\STATE Draw: \( \theta \sim \mathcal{U}[0,2\pi] \)\vspace{1mm}
		\STATE Define: \([\theta_{\text{min}},\theta_{\text{max}}]= [\theta-2\pi,\theta]\) \vspace{1mm}
		\STATE Propose: \(\boldsymbol{\zeta}_1^{\prime}=\boldsymbol{\zeta}_1^{(t-1)} \cos\theta+\boldsymbol{\nu}\sin \theta\) \vspace{1mm}
		\STATE Compute: \(\mathbf{u}_1=R_{{\lambda_1}^{(t-1)}}\boldsymbol{\zeta}_1^{\prime} + \boldsymbol{\mu}_{\ell_1}\) 
		\COMMENT { ${{\lambda_1}^{(t-1)}}= \left\lbrace  \tau_{\ell_1}^2, \lambda_1^{(t-1)} \right\rbrace$}\vspace{1mm}
		\IF{ \(\log \mathcal{N} ( \mathbf{y} -A_2\mathbf{z}_2^{(t-1)}-A_3\mathbf{z}_3^{(t-1)}\mid 0, A_1Q_{\mathbf{u}_1}^{-1}A_1^{\text{\tiny{T}}}+ {\sigma^2_{\varepsilon}}^{(t)}{I_{m}} )  > \kappa\)}  \vspace{1mm}
		\STATE  Set: \(\boldsymbol{\zeta}_1^{(t)}=\boldsymbol{\zeta}_1^{\prime} \)  \vspace{1mm}
		\ELSE   \vspace{1mm}
		\IF{\(\theta<0\)} \vspace{1mm} \STATE {$\theta_{\text{min}}=\theta$ } \ELSE \vspace{1mm}  \STATE {$\theta_{\text{max}}=\theta$} \vspace{1mm}
		\ENDIF  \vspace{1mm}
		\STATE Draw: \(\theta \sim \mathcal{U} [\theta_{\text{min}},\theta_{\text{max}}]\)  \vspace{1mm}
		\STATE Go back to step 13.  \vspace{1mm}
		\ENDIF \vspace{1mm}
		\STATE  Draw: \(\log\lambda_1|\log\lambda_1^{(t-1)} \sim\mathcal{N} (\log\lambda_1^{(t-1)},s_2)\)	 \vspace{1mm}
		\STATE Compute: \( \mathbf{u}=R_{{\lambda}_1}\boldsymbol{\zeta}_1^{(t)} +\boldsymbol{\mu}_{\ell_1} \)
		\COMMENT { ${{\lambda_1}}= \left\lbrace  \tau_{\ell_1}^2, \lambda_1 \right\rbrace$}\vspace{1mm} 
		\STATE Compute: \( \alpha_{\lambda_1}={\text{min}} \left\{ 1, \frac{ \mathcal{N} \left( \mathbf{y} -A_2\mathbf{z}_2^{(t-1)}-A_3\mathbf{z}_3^{(t-1)}\mid  0, A_1Q^{-1}_{\mathbf{u}^{\prime}}A_1^{\text{\tiny{T}}}+  {\sigma_\varepsilon^{2}}^{(t)}I_m \right)  \pi \left(\log{\lambda}_1 \right) }{   \left( \mathbf{y}-A_2\mathbf{z}_2^{(t-1)}-A_3\mathbf{z}_3^{(t-1)} \mid  0, A_1Q^{-1}_{\mathbf{u}_1}A_1^{\text{\tiny{T}}}+  {\sigma_\varepsilon^{2}}^{(t)}I_m \right) \pi \left(\log{\lambda_1^{(t-1)}} \right)}\right\}\)  \vspace{1mm}
		\STATE With probability  \(\alpha_{\lambda_1}\) set \( \log\lambda_1^{{(t)}}=\log\lambda_1\) and \( \mathbf{u}_1=\mathbf{u}^{\prime}\),otherwise set \(\log\lambda_1^{{(t)}}=\log\lambda_1^{{(t-1)}}\)  \vspace{1mm}
		\STATE Run Adaptation for $s_2$  \vspace{1mm}
		\STATE Draw \( \mathbf{z}_1^{(t)}= \mathcal{N} \left( {\sigma_{\varepsilon}^{-2}}^{(t)} \Sigma_{z_1} A_1^{\text{\tiny{T}}} ( \mathbf{y}- A_2\mathbf{z}_2^{(t-1)}-A_3\mathbf{z}_3^{(t-1)}) , \Sigma_{z_1}  \right)\)
		\COMMENT{ \(  \Sigma_{z_1} =   \left(Q_{{\mathbf{u}}_{1}^{(t)}} + {\sigma^{-2}_{\varepsilon}}^{(t)}  A_1^{\text{\tiny{T}}} A_1 \right)^{-1}  \)}	\vspace{1mm}
		\STATE Repeat steps 6-29 for $\mathbf{z}_2, \boldsymbol{\zeta}_2, \lambda_2$ \vspace{1mm}
		\STATE Draw: \(\boldsymbol{\nu}_{3,4}\sim \mathcal{N}(0, {I_{n_1n_2}})\)  \vspace{1mm}
		\STATE Draw:  \(\beta \sim \mathcal{U}[0,1]\)\vspace{1mm}
		\STATE Compute: \(\kappa= \log \mathcal{N} ( \mathbf{y}-A_1\mathbf{z}_1^{(t)}-A_2\mathbf{z}_2^{(t)} \mid 0, A_3(Q_{\mathbf{u}_3}^{-1}\otimes Q_{\mathbf{u}_4}^{-1})A_3^{\text{\tiny{T}}}+ {\sigma^2_{\varepsilon}}^{(t)}{I_m} ) + \log \beta\) \vspace{1mm}
		\STATE Draw: \( \theta \sim \mathcal{U}[0,2\pi] \)\vspace{1mm}
		\STATE Define: \([\theta_{\text{min}},\theta_{\text{max}}]= [\theta-2\pi,\theta]\) \vspace{1mm}
		\STATE Propose: \( \boldsymbol{\zeta}_{3,4}^{\prime}= \boldsymbol{\zeta}_{3,4}^{(t-1)} \cos\theta_1+\boldsymbol{\nu}_{3,4}\sin \theta_1\) \vspace{1mm}
		\COMMENT{ $\boldsymbol{\zeta}_{3,4}$ is formed by stacking $\boldsymbol{\zeta}_3$ and $\boldsymbol{\zeta}_4$ } \vspace{1mm}	
		\STATE Update: \(\mathbf{u}_3=R_{{\lambda}_3^{(t-1)}}\boldsymbol{\zeta}_3^{\prime} + \boldsymbol{\mu}_{\ell_3}\) and   \(\mathbf{u}_4=R_{\lambda_4^{(t-1)}}\boldsymbol{\zeta}_4^{\prime} + \boldsymbol{\mu}_{\ell_4}\) \vspace{1mm}
		\algstore{2dm}
	\end{algorithmic}
\end{algorithm}

\begin{algorithm}
	\footnotesize
	\begin{algorithmic}[1]
		\algrestore{2dm}	
		\IF{ \(\log \mathcal{N} ( \mathbf{y}-A_1\mathbf{z}_1^{(t)}-A_2\mathbf{z}_2^{(t)}  \mid  0, A_3(Q_{\mathbf{u}_3}^{-1}\otimes Q_{\mathbf{u}_4}^{-1})A_3^{\text{\tiny{T}}}+ {\sigma^2_{\varepsilon}}^{(t)}{I_m} )  > \kappa\)  }   \vspace{1mm}
		\STATE  Set: \(\boldsymbol{\zeta}_3^{(t)}=\boldsymbol{\zeta}_3^{\prime} \) and   \(\boldsymbol{\zeta}_4^{(t)}=\boldsymbol{\zeta}_4^{\prime} \)\vspace{1mm}
		\ELSE   \vspace{1mm}
		\IF{\(\theta_<0\)} \vspace{1mm} \STATE {$\theta_{\text{min}}=\theta$ } \ELSE \vspace{1mm}  \STATE {$\theta_{\text{max}}=\theta$} \vspace{1mm} 		
		\ENDIF  \vspace{1mm}
		\STATE Draw: \(\theta_\sim \mathcal{U} [\theta_{\text{min}},\theta_{\text{max}}]\)  \vspace{1mm}
		\STATE Go back to step 36.  \vspace{1mm}
		\ENDIF \vspace{1mm}
		\STATE  Draw: \(\log {\lambda}_{3}|\log {\lambda}_{3}^{(t-1)} \sim\mathcal{N} (\log {\lambda}_{3}^{(t-1)},s_3)\)	 \vspace{1mm}
		\STATE Compute: \( \mathbf{u}_3^{\prime}=R_{{\lambda}_3}\boldsymbol{\zeta}_3^{(t)} +\boldsymbol{\mu}_3 \) 
		\COMMENT { ${{\lambda_3}}= \left\lbrace  \tau_{\ell_3}^2, \lambda_3 \right\rbrace$}\vspace{1mm}
		\STATE Compute: \( \alpha_{\lambda_{3}}={\text{min}} \left\{ 1, \frac{ \mathcal{N} \left( \mathbf{y} -A_1\mathbf{z}_1^{(t)}-A_2\mathbf{z}_2^{(t)}| 0,  A_3(Q_{\mathbf{u}_3^{\prime}}^{-1}\otimes Q_{\mathbf{u}_4^{\prime}}^{-1})A_3^{\text{\tiny{T}}}+{\sigma^2_{\varepsilon}}^{(t)}{I_{m}} \right) \pi \left(\log{\lambda}_3 \right)}{  \mathcal{N} \left( \mathbf{y}-A_1\mathbf{z}_1^{(t)}-A_2\mathbf{z}_2^{(t)}| 0,A_3(Q_{\mathbf{u}_3}^{-1}\otimes Q_{\mathbf{u}_4}^{-1})A_3^{\text{\tiny{T}}}+ {\sigma^2_{\varepsilon}}^{(t)}{I_{m}} \right) \pi \left(\log{\lambda_3^{(t-1)}} \right)}\right\}\)  \vspace{1mm}
		\STATE With probability  \(\alpha_{\lambda_{3}}\) set \( \log {\lambda}_3^{{(t)}}=\log {\lambda}_3\), and \( \mathbf{u}_3=\mathbf{u}_3^{\prime}\);
		otherwise, set \(\log {\lambda}_3^{{(t)}}=\log {\lambda}_3^{{(t-1)}}\). \vspace{1mm}
		\STATE Run Adaptation for $s_3$ \vspace{1mm}
		\STATE  Repeat 49-53 for $\lambda_4$\vspace{1mm}
		\STATE Draw \( \mathbf{z}_3^{(t)}=
		\mathcal{N} \left( {\sigma_{\varepsilon}^{-2}}^{(t)} \Sigma_{z_3} A_3^{\text{\tiny{T}}} ( \mathbf{y}- A_1\mathbf{z}_1^{(t-1)}-A_2\mathbf{z}_2^{(t-1)}) , \Sigma_{z_3}  \right)\)
		\COMMENT{ \(  \Sigma_{z_1} =   \left(Q_{\mathbf{u}_{3}^{(t)}} \otimes Q_{\mathbf{u}_{4}^{(t)}} + {\sigma^{-2}_{\varepsilon}}^{(t)}  A_3^{\text{\tiny{T}}} A_3 \right)^{-1}  \)}	\vspace{1mm}
		\ENDFOR
	\end{algorithmic}
\end{algorithm}

\section{Experiments}

\begin{figure}[!htb]   
	\begin{center}
	\subfloat[Experiment 1]{\includegraphics[scale=.42
	]{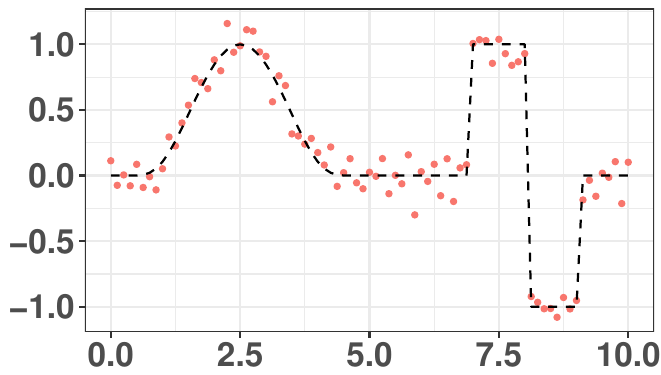}} \hfil
\subfloat[Experiment 2]{\includegraphics[scale=.42]{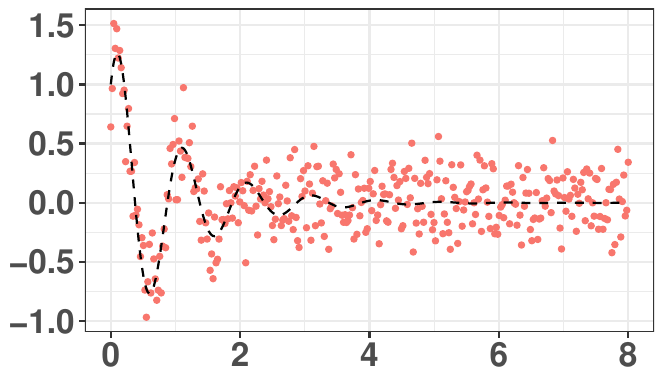}} \hfil
\subfloat[Experiment 3]{\includegraphics[scale=.42]{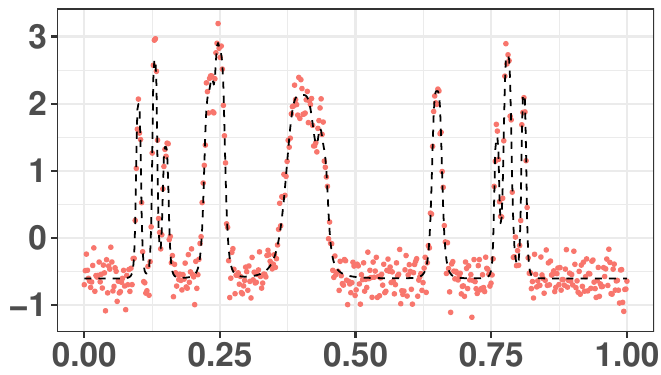}}
	\end{center}
	\caption{One-dimensional simulated dataset. (a): 81 observations with domain $\big[0, 10 \big]$ and noise variance $\sigma^2_{\varepsilon}=0.01$. (b): 350 observations with domain $\big[0, 8\big]$ and noise variance $\sigma^2_{\varepsilon}=0.04$. (c): 512 observations with domain$\big[0, 1\big]$ and noise variance $\sigma^2_{\varepsilon}=0.04$} \label{fig:simulated_data}
\end{figure}

We consider three simulated datasets with different characteristics. The first example is a function which has smooth parts and edges, and it is also piecewise constant,
\begin{equation*}
z(x)= \begin{cases}         
\exp \left( 4-\frac{25}{x(5-x)}\right) &x \in (0,5)\\
1 &x \in [7,8]\\
-1 &x \in (8,9]\\
0 &\text{otherwise}
\end{cases}.
\end{equation*}
The second corresponds to a damped sine wave function, \[
z(x)= \exp{(-x)}\cos(2\pi x).
\]
The data was generated employing the {\textit{Bumps}} function in \cite{donoho1995adapting} and scaled to have zero mean and unit variance. Following \cite{vannucci1999covariance}, we generate $m=512$ points in the interval [0,1] and  use a signal-to-noise ratio equal to 5, such that the noise variance $\sigma_\varepsilon^2= 0.04$.

\subsection{Experiment 1}

\begin{table}[H]  	
	\centering
	\scriptsize
	\begin{threeparttable}
		\renewcommand{\tabcolsep}{4pt}
		\begin{tabular}{llrrrrrrrrr}	
			\toprule
			\multirow{2}[1]{*}{}&  \multicolumn{1}{c}{\multirow{2}[1]{*}{{}}} & \multicolumn{3}{c}{{MWG}}  &\multicolumn{3}{c}{{w-ELL-SS}}  &\multicolumn{3}{c}{{m-ELL-SS}} \\	
			\cmidrule(lr){3-5}   \cmidrule(lr){6-8}  \cmidrule(lr){9-11}
			&	\multicolumn{1}{c}{} & \multicolumn{1}{c}{\(n=85\)}     & \multicolumn{1}{c}{\(n=169\)}   & \multicolumn{1}{c}{\(n=253\)}  &  \multicolumn{1}{c}{\(n=85\)}  & \multicolumn{1}{c}{\(n=169\)}  & \multicolumn{1}{c}{\(n=253\)} &  \multicolumn{1}{c}{\(n=85\)}   & \multicolumn{1}{c}{\(n=169\)}  & \multicolumn{1}{c}{\(n=253\)} \\
			\midrule
			\multirow{4}[2]{*}{{AR(1)}} & \(\sigma_\varepsilon^2\) & 0.014 & 0.015 & 0.015 & 0.014 & 0.014 & 0.014 & 0.014 & 0.014 & 0.014 \\	
			\multicolumn{1}{c}{} &  \(\ell_{j}\) & 2.416 & 2.653 & 2.785 & 2.350 & 1.912 & 2.015 & 2.118 & 2.163 & 1.968  \\	
			\multicolumn{1}{c}{} & $z_{j}$  & 0.687 & 0.686 & 0.685 & 0.690 & 0.693 & 0.693 & 0.692 & 0.692 & 0.692  \\	
			\multicolumn{1}{c}{} & $\lambda$& 0.435 & 0.405 & 0.385 & 0.312 & 0.408 & 0.379 & 0.381 & 0.358 & 0.338 \\
			\midrule	
			\multirow{4}[2]{*}{SE} & $\sigma_\varepsilon^2$ & 0.031 &0.043 &0.055  & 0.013 & 0.013 & 0.015 & 0.013 & 0.013 & 0.013 \\	
			\multicolumn{1}{c}{} & $\ell_{j}$& 0.678 & 1.183& 1.165 & 1.888 & 2.147 & 1.709 & 2.119 & 2.142 & 2.145 \\	
			\multicolumn{1}{c}{} & $z_{j}$&0.690  & 0.698& 0.674 & 0.692 & 0.692 & 0.691 & 0.692 & 0.693 & 0.693 \\	
			\multicolumn{1}{c}{} & $\lambda$  & 0.543 &0.545  &0.539 & 0.188 & 0.191 &0.476 &0.186  &0.181 & 0.174 \\
			\bottomrule
		\end{tabular}	
		\caption{Experiment 1: Posterior mean estimates 
			with both hyperpriors under various discretisation schemes ($n=85,169,253$) and three different algorithms.} 
		\label{tab:Post_est_boxcar}
	\end{threeparttable}
\end{table}

\begin{table}[H] 
	\centering
	\scriptsize
	\begin{tabular}{llrrrrrr}
		\toprule
		& \textcolor[rgb]{ 1,  0,  0}{} & \multicolumn{3}{c}{AR(1)} & \multicolumn{3}{c}{SE} \\
		\cmidrule(lr){3-5}
		\cmidrule(lr){6-8}
		&       & \multicolumn{1}{l}{Burned} & \multicolumn{1}{l}{Non-burned} & \multicolumn{1}{l}{Total time} & \multicolumn{1}{l}{Burned} & \multicolumn{1}{l}{Non-burned} & \multicolumn{1}{l}{Total time} \\
		
		\cmidrule{1-8}
		\multirow{3}[2]{*}{MWG} & \(n=85\)  & \textbf{0.01} & 16.78 & 16.80 &    28.02   &   NA    & 28.02 \\
		& \(n=169\)& 0.04  & \textbf{40.66} & \textbf{40.69} &  103.55     &    NA   & 103.55 \\
		& \(n=253\) & 0.10  & \textbf{76.84} & \textbf{76.94} &  265.16     &   NA    & 265.16 \\
		\cmidrule{1-8}
		\multirow{3}[2]{*}{w-ELL-SS} &  \(n=85\)   & 0.04  & \textbf{14.55} & \textbf{14.58} & 0.18  & 24.84 & 25.02 \\
		&  \(n=169\) & 0.30  & 51.90 & 52.20 & 0.82  & 103.05 & 103.86 \\
		&  \(n=253\) & 0.70  & 127.67 & 128.37 & 3.22  & 249.15 & 252.37 \\
		\cmidrule{1-8}
		\multicolumn{1}{c}{\multirow{3}[2]{*}{m-ELL-SS}} &  \(n=85\)  & \textbf{0.01} & 18.50 & 18.52 & \bf{0.03} & \bf{22.17} & \bf{22.20} \\
		&  \(n=169\) & \textbf{0.03} & 46.54 & 46.57 & \bf{0.18}  & \bf{59.42} & \bf{59.60} \\
		& \(n=253\) & \textbf{0.06} & 104.20 & 104.26 & \bf{0.37} & \bf{132.97} & \bf{133.35} \\
		\bottomrule
	\end{tabular}
	\caption{Experiment 1: CPU time (minutes) for $200,000$ iterations. NA denotes that MWG for the SE hyperprior did not converge. Best values in boldface.}
	\label{tab:Comp_time1}%
\end{table}

\begin{table}[!htbp] 
	\scriptsize
	\begin{threeparttable}
		\begin{tabular}{llrrrrrrrrr}	
			\toprule
			\multirow{2}[1]{*}{}&  {\multirow{2}[1]{*}{{}}} & \multicolumn{3}{c}{{MWG}}  &\multicolumn{3}{c}{{w-ELL-SS}}  &\multicolumn{3}{c}{{m-ELL-SS}} \\
			\cmidrule(lr){3-5}  \cmidrule(lr){6-8}  \cmidrule(lr){9-11}
			&	\multicolumn{1}{c}{} & \multicolumn{1}{c}{$n=85$}     & \multicolumn{1}{c}{$n=169$}   & \multicolumn{1}{c}{$n=253$}  &  \multicolumn{1}{c}{$n=85$}  & \multicolumn{1}{c}{$n=169$}  & \multicolumn{1}{c}{$n=253$} &  \multicolumn{1}{c}{$n=85$}   & \multicolumn{1}{c}{$n=169$}  & \multicolumn{1}{c}{$n=253$} \\
			\midrule	
			\multirow{6}[2]{*}{AR(1)} & $\sigma_\varepsilon^2$ & 10452.5 &  7038.0 &  5070.5  &  5541.0 &  5313.1  &  4967.9  & \bf{ 12234.4 } & \bf{11999.4} & \bf{12124.1} \\	
			\multicolumn{1}{c}{} &  $\ell_{15}$ & \bf{5424.4} & 2150.5 & 1317.3 & 181.1 & 192.0 & 201.6& 3146.7 & \bf{3391.2} & \bf{3282.9}\\
			\multicolumn{1}{c}{} & $\ell_{66}$ &\bf{22539.7} & \bf{11131.5} & \bf{6901.8} & 773.2 &467.1 & 268.0 & 9337.0 & 3736.3 & 3557.9 \\	
			\multicolumn{1}{c}{} & $z_{15}$  & 25449.8 & 11648.3 & 7878.2 &  4635.0& 5981.3 & 5264.1 & \bf{30601.6} & \bf{35096.7} & \bf{47895.6}\\
			\multicolumn{1}{c}{} & $z_{66}$& \bf{42146.1} & 27135.4 & 21528.7&8343.2 & 7485.8 & 8127.4& \bf{26530.5}& \bf{27856.4} &\bf{26881.2}  \\
			\multicolumn{1}{c}{} & $\lambda$  & 1507.9& 636.6& 460.8 & 331.2 & 272.8 &300.9 & \bf{2068.6} & \bf{2119.5} & \bf{2243.5} \\
			\midrule		
			\multirow{6}[2]{*}{SE} & $\sigma_\varepsilon^2$ & 313.4 &505.5    &1986.6   &6117.6 & 8008.2 & 2214.8 & \bf{18983.7} & \bf{15087.8} & \bf{16750.4}\\	
			\multicolumn{1}{c}{} & $\ell_{15}$ &  2.1& 7.5 & 6.7    & 214.0 &195.7 & 289.1 & \bf{3401.0} & \bf{3498.8} & \bf{3381.2}  \\
			\multicolumn{1}{c}{} & $\ell_{66}$ &  2.1 &2.1  &1.4 & 961.5 & 717.8 & 309.1 & \bf{8434.1} & \bf{7023.2} & \bf{7391.8} \\		
			\multicolumn{1}{c}{} & $z_{15}$ & \bf{91330.7} &22391.1  &117111.0  & 4992.2 & 5113.6 & 5989.6 & 28060.0 & \bf{30737.0} & \bf{28382.6}   \\
			\multicolumn{1}{c}{} &$z_{66}$ & 48.4 & 83.1 &8678.6  & 11139.8 & 12676.2 & 2561.6 & \bf{31456.3} & \bf{33268.2} & \bf{41623.7}  \\
			\multicolumn{1}{c}{} & $\lambda$ & 16.6& 77.4  & 82.3  & 57.5  & 29.5  & 3.6  & \bf{367.8} & \bf{246.9} & \bf{293.3} \\	
			\bottomrule
		\end{tabular}%
		\caption{Results Experiment 1: ESS after burn-in period for both hyperpriors under various discretisation schemes ($n=85,169,253$) and employing three different sampling algorithms. Highest values in boldface.}	
		\label{tab:EffSS_boxcar}
	\end{threeparttable}
\end{table}

\begin{figure}[!htbp]   
		\begin{center}
	\subfloat[{\tiny{$\boldsymbol{\ell}$, $n=85$}}]{\includegraphics[scale=.42]{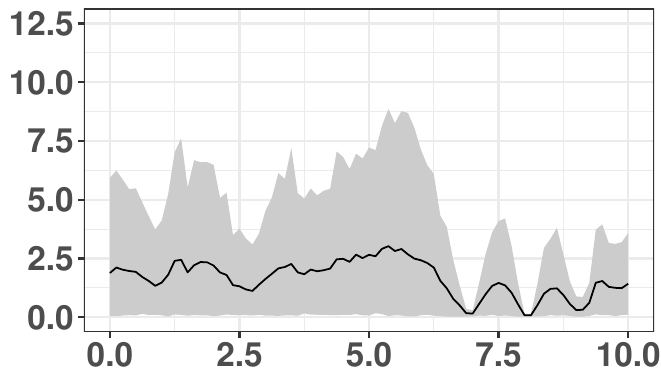}}
	\subfloat[{\tiny{$\boldsymbol{\ell}$, $n=169$}}]{\includegraphics[scale=.42]{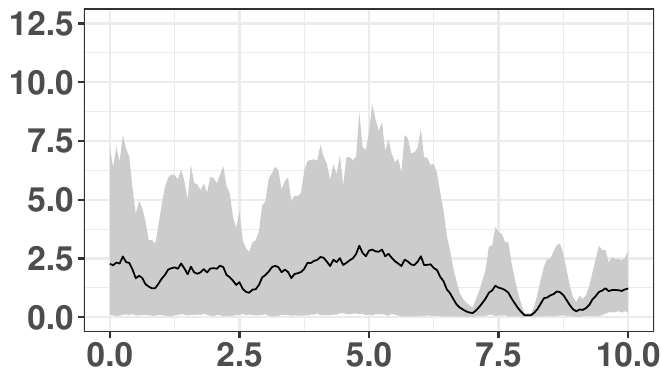}}
	\subfloat[{\tiny{$\boldsymbol{\ell}$, $n=253$}}]{\includegraphics[scale=.42]{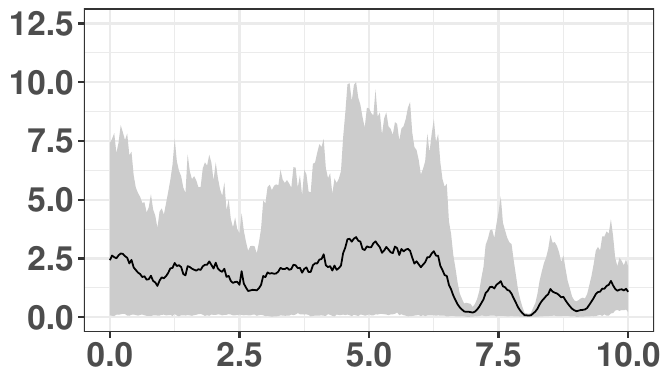}} 
	\\
	\subfloat[{\tiny{$\mathbf{z}$, $n=85$}}]{\includegraphics[scale=.42]{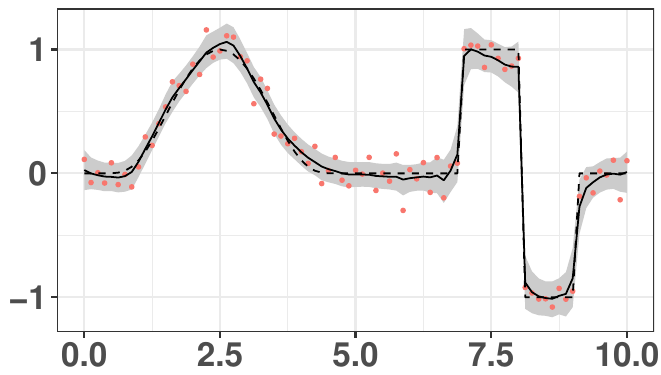}} 
	\subfloat[{\tiny{$\mathbf{z}$, $n=169$}}]{\includegraphics[scale=.42]{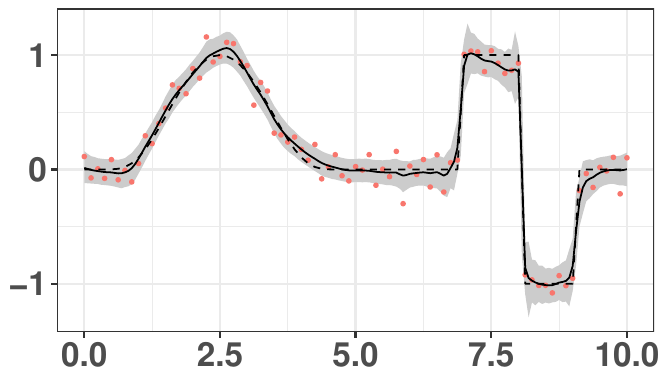}}
	\subfloat[{\tiny{$\mathbf{z}$, $n=253$}}]{\includegraphics[scale=.42]{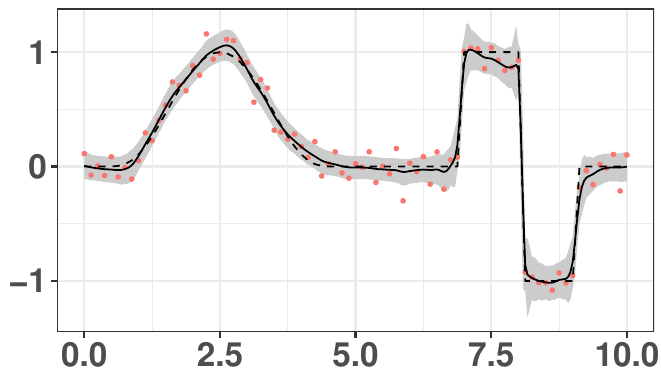}}\\
	\subfloat[{\tiny{$\boldsymbol{\ell}$, $n=85$}}]{\includegraphics[scale=.42]{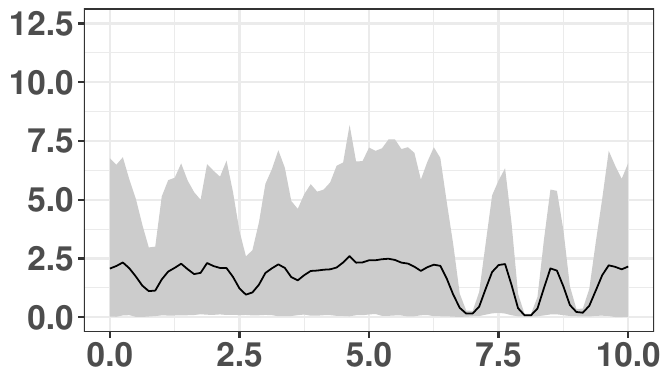}}
	\subfloat[{\tiny{$\boldsymbol{\ell}$, $n=169$}}]{\includegraphics[scale=.42]{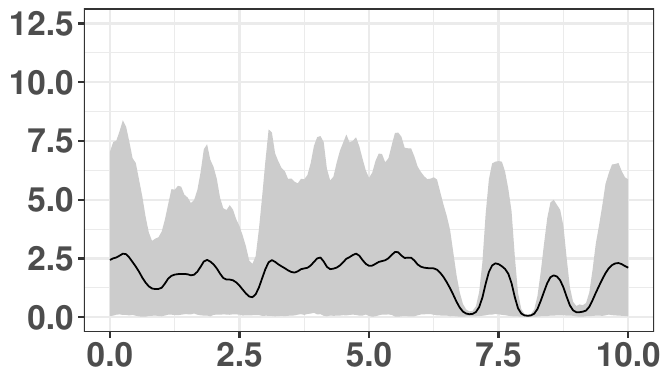}}
	\subfloat[{\tiny{$\boldsymbol{\ell}$, $n=253$}}]{\includegraphics[scale=.42]{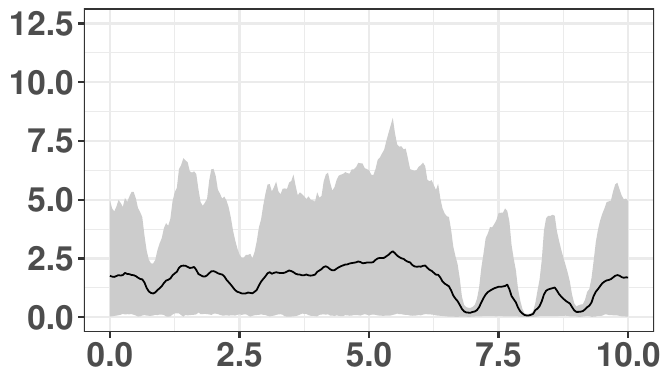}} 
	\\
	\subfloat[{\tiny{$\mathbf{z}$, $n=85$}}]{\includegraphics[scale=.42]{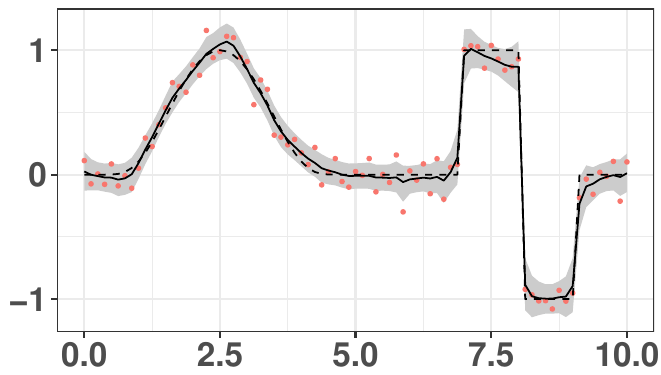}} 
	\subfloat[{\tiny{$\mathbf{z}$, $n=169$}}]{\includegraphics[scale=.42]{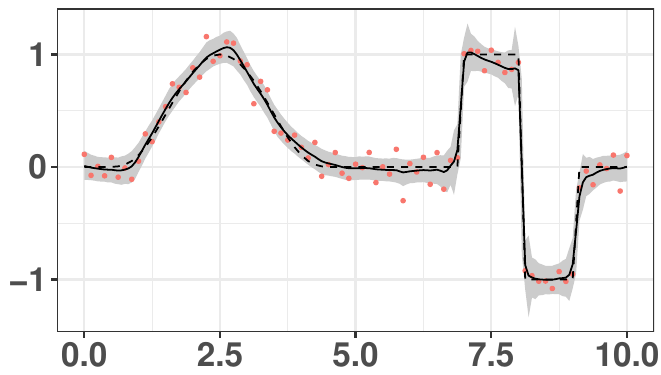}}
	\subfloat[{\tiny{$\mathbf{z}$, $n=253$}}]{\includegraphics[scale=.42]{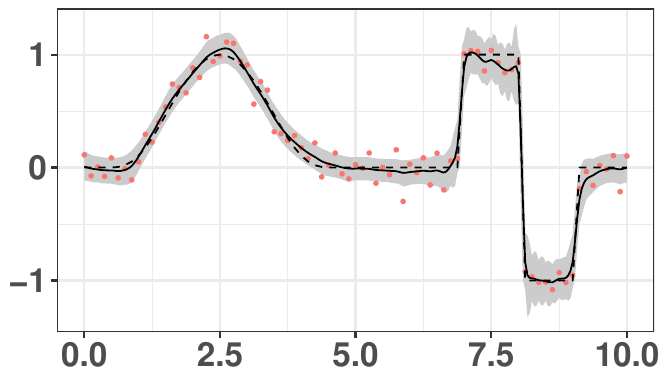}}	
\end{center}
	\caption{Experiment 1 with w-ELL-SS algorithm. (a)-(c): Estimated $\boldsymbol{\ell}$ process with $95\%$ credible intervals for AR(1) hyperprior on different grids. (d)-(f): Estimated $\mathbf{z}$ process with $95\%$ credible intervals for AR(1) hyperprior on different grids with observed data in red. (g)-(i): Estimated $\boldsymbol{\ell}$ process with $95\%$ credible intervals for SE hyperprior on different grids. (j)-(l): Estimated $\mathbf{z}$ process with $95\%$ credible intervals for SE hyperprior on different grids with observed data in red.} \label{fig:RepESS}
\end{figure}
\begin{figure}[!htbp]   
		\begin{center}
	\subfloat[{\tiny{$\boldsymbol{\ell}$, $N=85$}}]{\includegraphics[scale=.42]{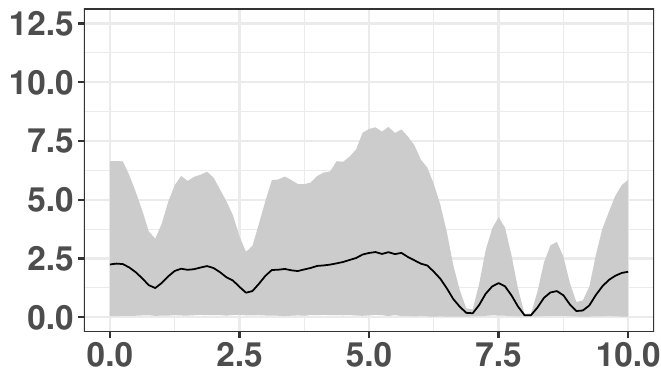}}
	\subfloat[{\tiny{$\boldsymbol{\ell}$, $N=169$}}]{\includegraphics[scale=.42]{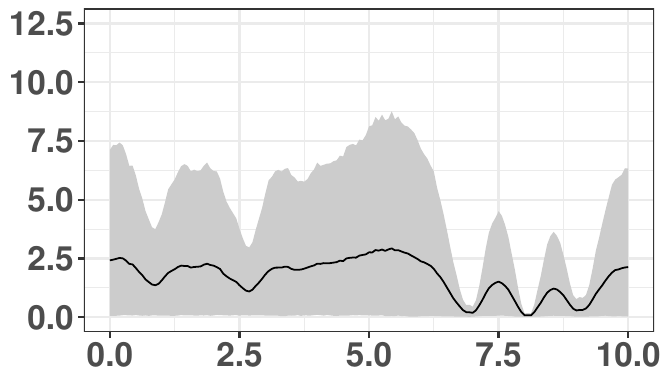}}
	\subfloat[{\tiny{$\boldsymbol{\ell}$, $N=253$}}]{\includegraphics[scale=.42]{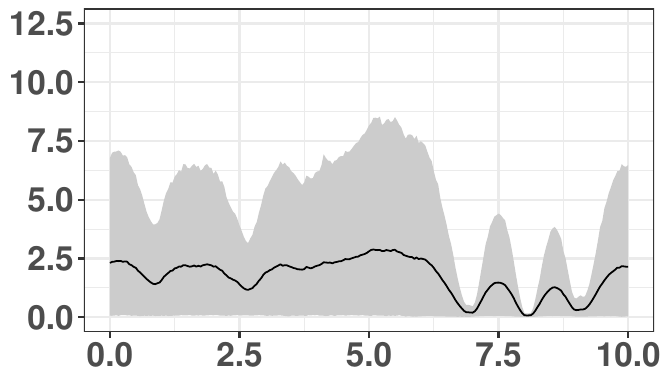}} 
	\\
	\subfloat[{\tiny{$\mathbf{z}$, $N=85$}}]{\includegraphics[scale=.42]{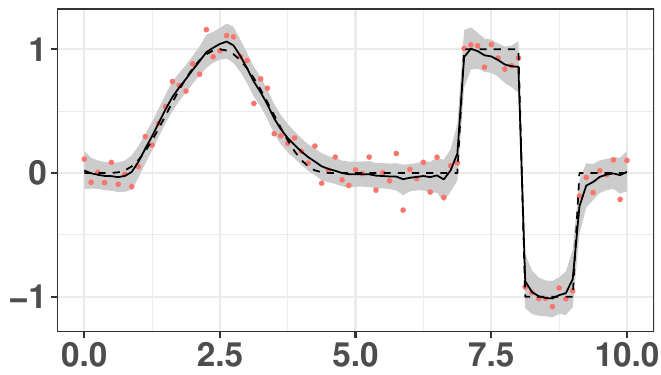}} 
	\subfloat[{\tiny{$\mathbf{z}$, $N=169$}}]{\includegraphics[scale=.42]{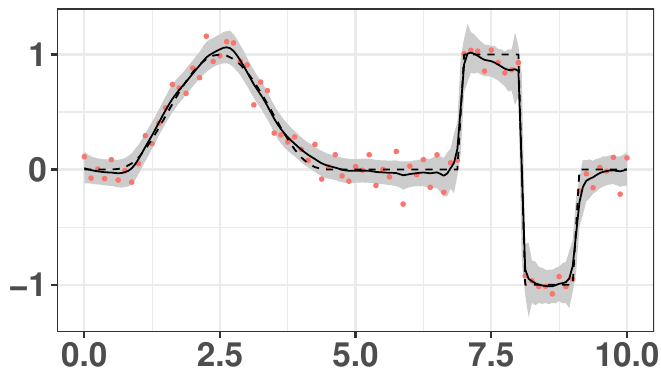}}
	\subfloat[{\tiny{$\mathbf{z}$, $N=253$}}]{\includegraphics[scale=.42]{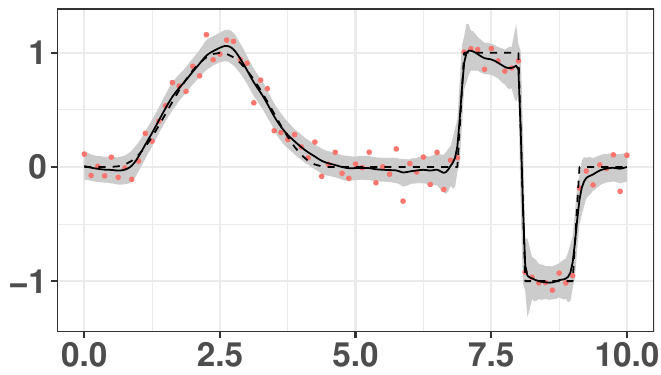}}\\
	\subfloat[{\tiny{$\boldsymbol{\ell}$, $n=85$}}]{\includegraphics[scale=.42]{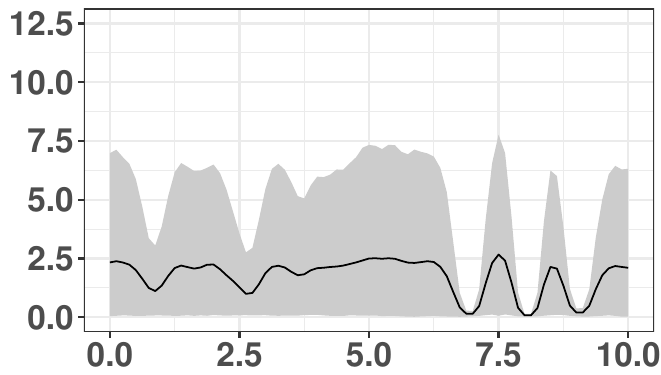}}
	\subfloat[{\tiny{$\boldsymbol{\ell}$, $n=169$}}]{\includegraphics[scale=.42]{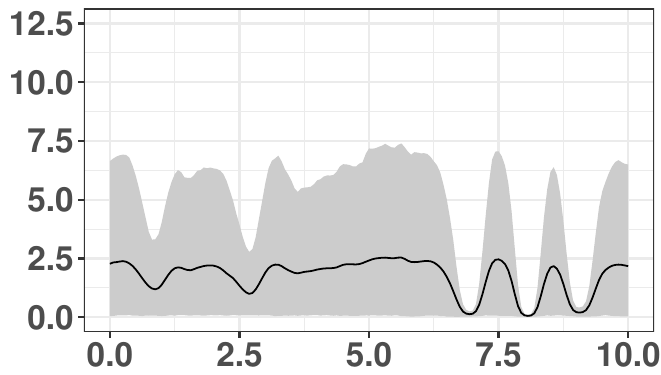}}
	\subfloat[{\tiny{$\boldsymbol{\ell}$, $n=253$}}]{\includegraphics[scale=.42]{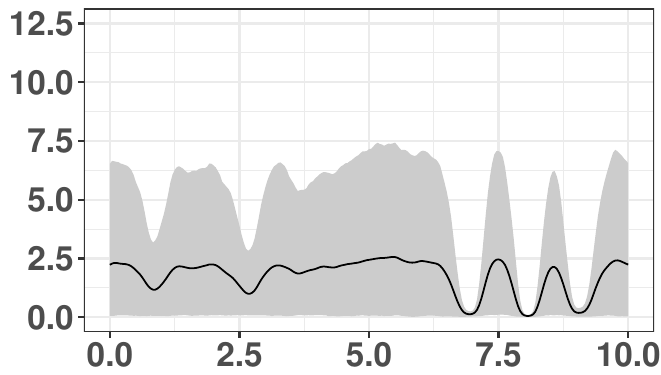} }
	\\
	\subfloat[{\tiny{$\mathbf{z}$, $n=85$}}]{\includegraphics[scale=.42]{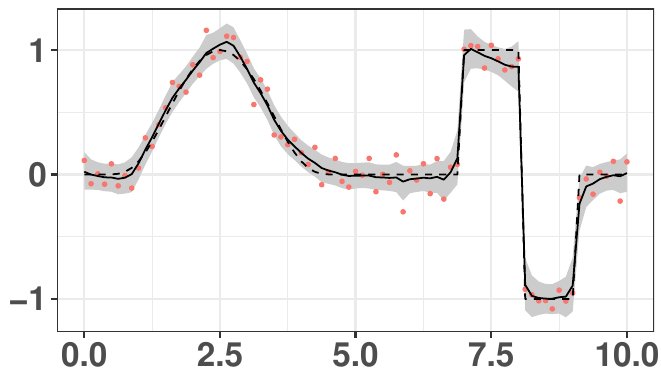}} 
	\subfloat[{\tiny{$\mathbf{z}$, $n=169$}}]{\includegraphics[scale=.42]{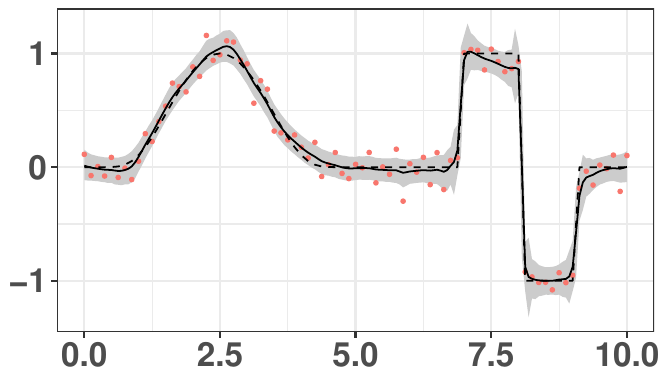}}
	\subfloat[{\tiny{$\mathbf{z}$, $n=253$}}]{\includegraphics[scale=.42]{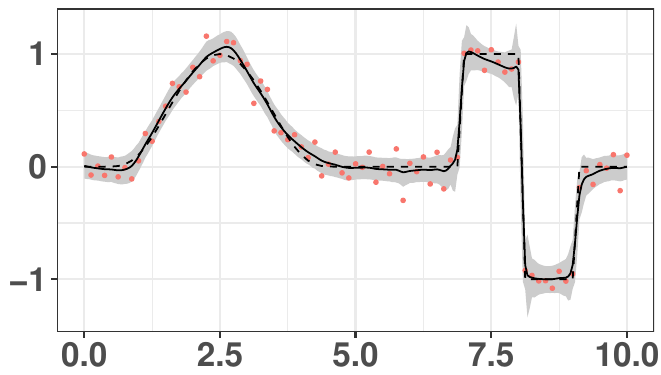}}
\end{center}
	\caption{Experiment 1 with m-ELL-SS algorithm. (a)-(c): Estimated $\boldsymbol{\ell}$ process with $95\%$ credible intervals for AR(1) hyperprior on different grids. (d)-(f): Estimated $\mathbf{z}$ process with $95\%$ credible intervals for AR(1) hyperprior on different grids with observed data in red.. 	(g)-(i): Estimated $\boldsymbol{\ell}$ process with $95\%$ credible intervals for SE hyperprior on different grids. (j)-(l): Estimated $\mathbf{z}$ process with $95\%$ credible intervals for SE hyperprior on different grids with observed data in red.} \label{fig:MarESS}
\end{figure}

\newpage
\subsection{Experiment 2}

\begin{figure}[!h]    
	\begin{center}

	\subfloat[{\tiny{$\boldsymbol{\ell}$, MWG with AR}}]{\includegraphics[scale=.37]{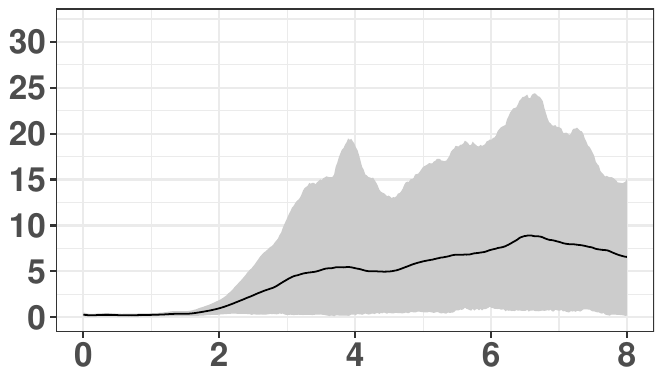}} 	
	\subfloat[{\tiny{$\boldsymbol{\ell}$, w-ELL-SS with AR}}]{\includegraphics[scale=.37]{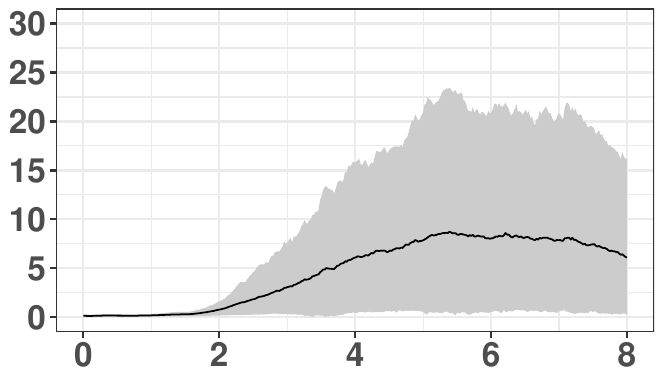}}
	\subfloat[{\tiny{$\boldsymbol{\ell}$, m-ELL-SS with AR}}]{\includegraphics[scale=.37]{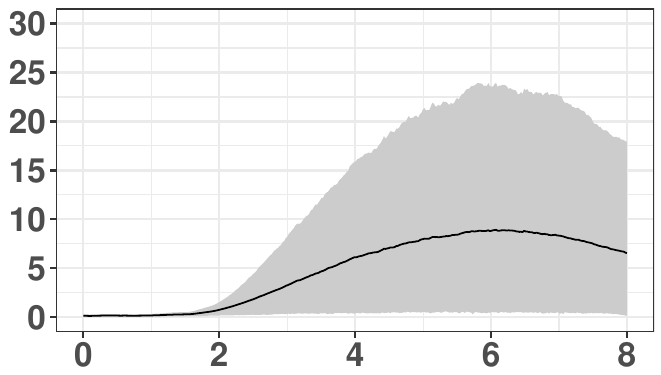}}
	\\
	\subfloat[{\tiny{$\mathbf{z}$, MWG with AR}}]{\includegraphics[scale=.37]{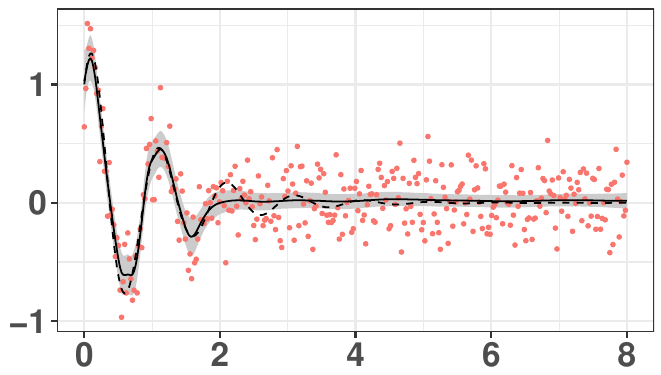}}
	\subfloat[{\tiny{$\mathbf{z}$, w-ELL-SS with AR}}]{\includegraphics[scale=.37]{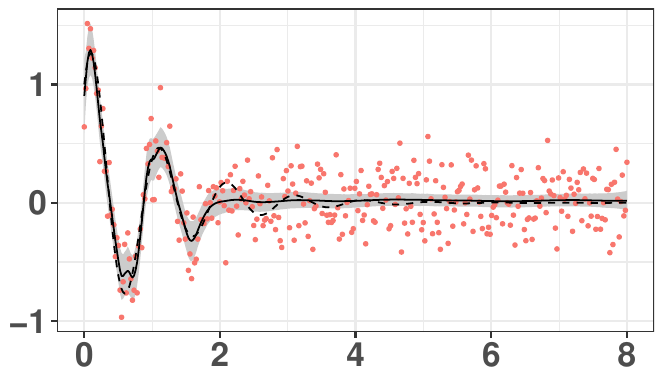}}
	\subfloat[{\tiny{$\mathbf{z}$, m-ELL-SS with AR}}]{\includegraphics[scale=.37]{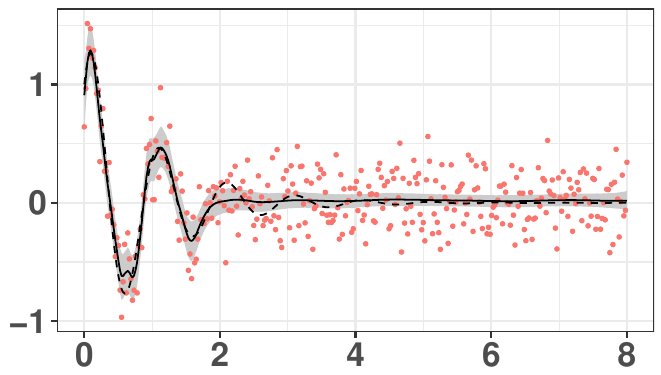}} 
\end{center}
	\caption{Experiment 2. Top row: estimated $\boldsymbol{\ell}$ process with $95\% $ credible interval for AR(1) hyperprior with (a) MWG, (b) w-ELL-SS and (c) m-ELL-SS. Second row: estimated $\mathbf{z}$ process with $95\% $ credible interval for AR(1) hyperprior with (d) MWG, (e) w-ELL-SS and (f) m-ELL-SS.}
	\label{fig:SIN_AR}
\end{figure}

\begin{table}[!b]
	\centering
	\scriptsize
	\begin{tabular}{lllrrrrrr}
		\toprule
		&   \textcolor[rgb]{ 1,  0,  0}{}& \multicolumn{3}{c}{AR(1)} & \multicolumn{3}{c}{SE} \\
		\cmidrule(lr){3-5}
		\cmidrule(lr){6-8}
		&         & \multicolumn{1}{l}{Burned} & \multicolumn{1}{l}{Non-burned} & \multicolumn{1}{l}{Total time} & \multicolumn{1}{l}{Burned} & \multicolumn{1}{l}{Non-burned} & \multicolumn{1}{l}{Total time} \\
		\cmidrule{1-9}
		\multirow{1}[1]{*}{MWG} & \(n=430\)  & 0.60& \bf{155.82} & \bf{156.43}  & 572.36    &  NA      & 572.36\\
		\multirow{1}[1]{*}{w-ELL-SS} &  \(n=430\)   & 1.42  & 306.60 & 308.02& 3.60 & 500.49 & 504.09 \\
		\multicolumn{1}{c}{\multirow{1}[1]{*}{m-ELL-SS}} &  \(n=430\)  & \textbf{0.25} & 308.67 & 308.92 & \bf{1.17}  & \bf{330.04} & \bf{331.22} \\
		\bottomrule
	\end{tabular}%
	\caption{Experiment 2: CPU time (minutes) for $100,000$ iterations. NA denotes that MWG for SE hyperprior did not converge. Best values in boldface.}
	\label{tab:Comp_time2}%
\end{table}%

\begin{table}[!htb]
	\centering
	\scriptsize
	\begin{tabular}{llrrr}
		\toprule
		& & MWG &  w-ELL-SS &  m-ELL-SS\\
		\midrule
		\multicolumn{1}{c}{\multirow{6}[1]{*}{AR(1)}} & $\sigma_\varepsilon^2$ & 0.045 & 0.044 & 0.044 \\	
		\multicolumn{1}{c}{} & $\ell_{100}$  & 1.694 & 1.379 & 1.287 \\
		\multicolumn{1}{c}{} & $\ell_{200}$  & 5.051 & 6.922 &7.131 \\
		\multicolumn{1}{c}{} & $z_{100}$& 0.021 & 0.025 & 0.027 \\
		\multicolumn{1}{c}{} & $z_{200}$  & 0.031 & 0.027 & 0.027  \\
		\multicolumn{1}{c}{} & $\lambda$ & 2.598 & 2.771 & 2.710  \\     
		\midrule
		\multicolumn{1}{c}{\multirow{6}[1]{*}{SE}}& $\sigma_\varepsilon^2$   & 0.072 & 0.044 & 0.044 \\
		\multicolumn{1}{c}{} & $\ell_{100}$ & 0.594 & 0.965 & .951\\
		\multicolumn{1}{c}{} & $\ell_{200}$& 0.677 & 8.967 & 9.187 \\
		\multicolumn{1}{c}{} & $z_{100}$ & 0.032& 0.029 &   0.029\\
		\multicolumn{1}{c}{} & $z_{200}$ & 0.060 & 0.025 &   0.024\\
		\multicolumn{1}{c}{} & $\lambda$ & 0.450 & 1.877& 1.970 \\
		\bottomrule
	\end{tabular}%
	\caption{Experiment 2: Posterior mean estimates obtained with both hyperpriors and employing three different sampling algorithms. Estimates are consistent across sampling algorithms, except for SE with MWG because the sampler did not reach convergence.}
	\label{tab:Post_est_sin}%
\end{table}
\begin{table}[!htb]
	\centering
	\scriptsize	
	\begin{tabular}{llrrr}
		\toprule
		&  & MWG   & w-ELL-SS & m-ELL-SS \\
		\midrule
		\multicolumn{1}{c}{\multirow{6}[1]{*}{AR(1)}} & $\sigma_\varepsilon^2$ & 14505.3 & 17446.5& \bf{20673.4} \\
		\multicolumn{1}{c}{} & $\ell_{100}$ &116.3& 282.6 & \bf{2485.3} \\
		\multicolumn{1}{c}{} & $\ell_{200}$ & 56.3 &385.5 & \bf{2421.7}  \\
		\multicolumn{1}{c}{} & $z_{100}$& 7002.5 & 13637.9 & \bf{37023.4}  \\
		\multicolumn{1}{c}{} & $z_{200}$&3424.5& 8179.6 & \bf{27585.5} \\
		\multicolumn{1}{c}{} & $\lambda$ &92.6 & 145.7 & \bf{1312.8} \\
		\midrule
		\multicolumn{1}{c}{\multirow{6}[1]{*}{SE}} & $\sigma_\varepsilon^2$  &444.5 & 18804.2 & \bf{21169.3} \\
		\multicolumn{1}{c}{} & $\ell_{100}$ & 5.0  & 1145.9 & \bf{5996.4} \\
		\multicolumn{1}{c}{} & $\ell_{200}$  & 7.4   & 919.4&  \bf{3563.6} \\
		\multicolumn{1}{c}{} & $z_{100}$ & \bf{100000.0} & 37550.5  & 76574. \\
		\multicolumn{1}{c}{} & $z_{200}$ & \bf{98891.7} & 14476.0&  49195.2\\
		\multicolumn{1}{c}{} & $\lambda$ & 44.8  &91.0  & \bf{668.4}  \\
		\bottomrule
	\end{tabular}
	\caption{Experiment 2: ESS after burnin period for both hyperprior and employing three different sampling algorithms. Highest values in boldface. m-ELL-SS results in the highest efficiency scores.}
	\label{tab:ESS_sin}%
\end{table}

\newpage
\subsection{Experiment 3}
\begin{table}[!htb]
	\centering
	\scriptsize
	\begin{tabular}{llrrr}
		\toprule
		& & MWG &  w-ELL-SS &  m-ELL-SS\\
		\midrule
		\multicolumn{1}{c}{\multirow{6}[1]{*}{AR(1)}} & $\sigma_\varepsilon^2$ & 0.041 & 0.040 & 0.040 \\	
		\multicolumn{1}{c}{} & $\ell_{100}$  &  1.821& 3.780 & 1.520 \\
		\multicolumn{1}{c}{} & $\ell_{200}$  &0.519& 0.375 & 0.510\\
		\multicolumn{1}{c}{} & $z_{100}$& -0.519 & -0.538& -0.535 \\
		\multicolumn{1}{c}{} & $z_{200}$  & 2.097& 2.110& 2.086  \\
		\multicolumn{1}{c}{} & $\lambda$ &  0.033& 0.029& 0.033 \\     
		\midrule
		\multicolumn{1}{c}{\multirow{6}[1]{*}{SE}}& $\sigma_\varepsilon^2$   &  0.504 & 0.039 &  0.039 \\
		\multicolumn{1}{c}{} & $\ell_{100}$ & 1.414 & 0.126 & 0.666\\
		\multicolumn{1}{c}{} & $\ell_{200}$&  1.523& 	0.310 & 0.381\\
		\multicolumn{1}{c}{} & $z_{100}$ &0.178& -0.499  & -0.523   \\
		\multicolumn{1}{c}{} & $z_{200}$ & 1.303& 2.046&2.053 \\
		\multicolumn{1}{c}{} & $\lambda$ & 1.058 &0.106 & 0.024 \\
		\bottomrule
	\end{tabular}%
	\caption{Experiment 3: Posterior mean estimates obtained with both hyperpriors and employing three different sampling algorithms. }
	\label{tab:Post_est_bumps}%
\end{table}
\begin{table}[!htb]
	\centering
	\scriptsize	
	\begin{tabular}{llrrr}
		\toprule
		&  & MWG   & w-ELL-SS & m-ELL-SS \\
		\midrule
		\multicolumn{1}{c}{\multirow{6}[1]{*}{AR(1)}} & $\sigma_\varepsilon^2$ & \bf{6975.6} & 3398.3	& 4638.5 \\
		\multicolumn{1}{c}{} & $\ell_{100}$ &\bf{489.5} & 8.2& 155.1 \\
		\multicolumn{1}{c}{} & $\ell_{200}$ & \bf{1978.3} &63.3&  201.8 \\
		\multicolumn{1}{c}{} & $z_{100}$& \bf{6875.2}&3354.2& 5220.6  \\
		\multicolumn{1}{c}{} & $z_{200}$&\bf{4515.2}& 817.0& 910.6\\
		\multicolumn{1}{c}{} & $\lambda$ &\bf{193.4}& 18.4& 106.1 \\
		\midrule
		\multicolumn{1}{c}{\multirow{6}[1]{*}{SE}} & $\sigma_\varepsilon^2$  &2650.1& 5072.4& \bf{12442.0} \\
		\multicolumn{1}{c}{} & $\ell_{100}$ &  2.4& 70. & \bf{153.7} \\
		\multicolumn{1}{c}{} & $\ell_{200}$  & 2.5 &310.3	&  \bf{1339.5} \\
		\multicolumn{1}{c}{} & $z_{100}$ &3522.7& 49136.2 &\bf{6397.9}  \\
		\multicolumn{1}{c}{} & $z_{200}$ &2101.0  & 36809.1& \bf{4399.9}\\
		\multicolumn{1}{c}{} & $\lambda$ & \bf{93.4} & 2.5  & 27.2 \\
		\bottomrule
	\end{tabular}
	\caption{Results for Experiment 3: ESS after burnin period for both hyperprior and employing three different sampling algorithms. Highest values in boldface.}
	\label{tab:ESS_Bumps}%
\end{table}

\begin{table}[!t]
	\centering
	\scriptsize
	\begin{tabular}{lllrrrrrr}
		\toprule
		&   \textcolor[rgb]{ 1,  0,  0}{}& \multicolumn{3}{c}{AR(1)} & \multicolumn{3}{c}{SE} \\
		\cmidrule(lr){3-5}
		\cmidrule(lr){6-8}
		&         & \multicolumn{1}{l}{Burned} & \multicolumn{1}{l}{Non-burned} & \multicolumn{1}{l}{Total time} & \multicolumn{1}{l}{Burned} & \multicolumn{1}{l}{Non-burned} & \multicolumn{1}{l}{Total time} \\
		\cmidrule{1-9}
		\multirow{1}[1]{*}{MWG} & \(n=572\)  & 32.48 & \bf{297.78} & \bf{330.27}  &  1289.166  &  NA  & 1289.166  \\
		\multirow{1}[1]{*}{w-ELL-SS} &  \(n=572\)   & 106.43 &592.95 &699.38& 6.02& 1246.85& 1252.87\\
		\multicolumn{1}{c}{\multirow{1}[1]{*}{m-ELL-SS}} &  \(n=572\)  & \textbf{20.70} & 814.10& 834.79& \bf{85.17}  & \bf{810.19} & \bf{895.36} \\
		\bottomrule
	\end{tabular}%
	\caption{Experiment 3: CPU time (minutes) for $100,000$ iterations. NA denotes that MWG  for SE hyperprior did not converge. Best values in boldface.}
	\label{tab:Comp_time3}%
\end{table}%

\begin{table}[!htbp]
	\centering
	\scriptsize
	\begin{tabular}{lllrrrrrr}
		\toprule
		&   \textcolor[rgb]{ 1,  0,  0}{}& \multicolumn{3}{c}{AR(1)} & \multicolumn{3}{c}{SE} \\
		\cmidrule(lr){3-5}
		\cmidrule(lr){6-8}
		&         & \multicolumn{1}{l}{Burned} & \multicolumn{1}{l}{Non-burned} & \multicolumn{1}{l}{Total time} & \multicolumn{1}{l}{Burned} & \multicolumn{1}{l}{Non-burned} & \multicolumn{1}{l}{Total time} \\
		\cmidrule{1-9}
		\multirow{1}[1]{*}{MWG} & \(n=572\)  & 27.86 & \bf{249.14} & \bf{277.00}  &  956.78   &  NA      & 956.78   \\
		\multirow{1}[1]{*}{w-ELL-SS} &  \(n=572\)   & 45.91 &258.61 &304.52& 402.77& NA&\bf{402.77} \\
		\multicolumn{1}{c}{\multirow{1}[1]{*}{m-ELL-SS}} &  \(n=572\)  & \textbf{9.39} & 375.90& 385.29& \bf{42.98}  & \bf{397.12} & 440.10 \\
		\bottomrule
	\end{tabular}%
	\caption{Computational time for Experiment 3 in a High Performance Computer. Algorithms were run for $100,000$ iterations. m-ELL-SS and w-ELL-SS speed up by a factor of approximately $2.1$, while MWG by $1.1$.}
	\label{tab:Comp_time3_HPC}%
\end{table}%

\subsubsection{Prior elicitation} \label{Prior_elicitation_EX3}
As opposed to Experiment 1 and 2, where vague priors for covariance parameters sufficed, here we employ informative prior distributions for $\log (\lambda)$ and $\mathbf{u}$.
Knowledge about the parameters comes from the fact that the length-scales, for both stationary and non-stationary processes, are only identifiable between the minimum and maximum covariate distance. In this experiment, the maximum distance is 1 and the minimum is .0019; thus, the $\mathcal{N}(0,1)$ prior for each $u_j$ is inappropriate. Instead, we solve the system of equations in Section \ref{FixingHyperparameters} to fix the hyperparameters. 
Indeed, arbitrarily fixing the hyperparameters can greatly affect the inferences. See for instance the estimated length-scale process with MWG and AR hyperprior in Figure~\ref{fig:Bumps_ARprior}, where we set the prior of $\mathbf{u}$ to be a zero-centred GP with unit variance.
\begin{figure}[htbp]    
	\begin{center}
		{\includegraphics[scale=.42]{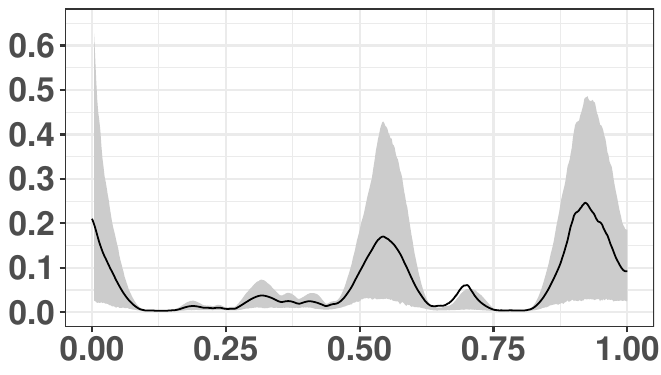}} 	
	\end{center}
	\caption{Posterior mean of lengh-scale for Experiment 3 with MGW and AR hyperprior with $\mu_\ell=0$ and $\tau_{\ell}^2=1$.}
	\label{fig:Bumps_ARprior}
\end{figure}

\subsection{Two-dimensional synthetic data}
\begin{figure}[H]   
	\begin{center}
	\subfloat[{{True}}]	{\includegraphics[scale=.2]{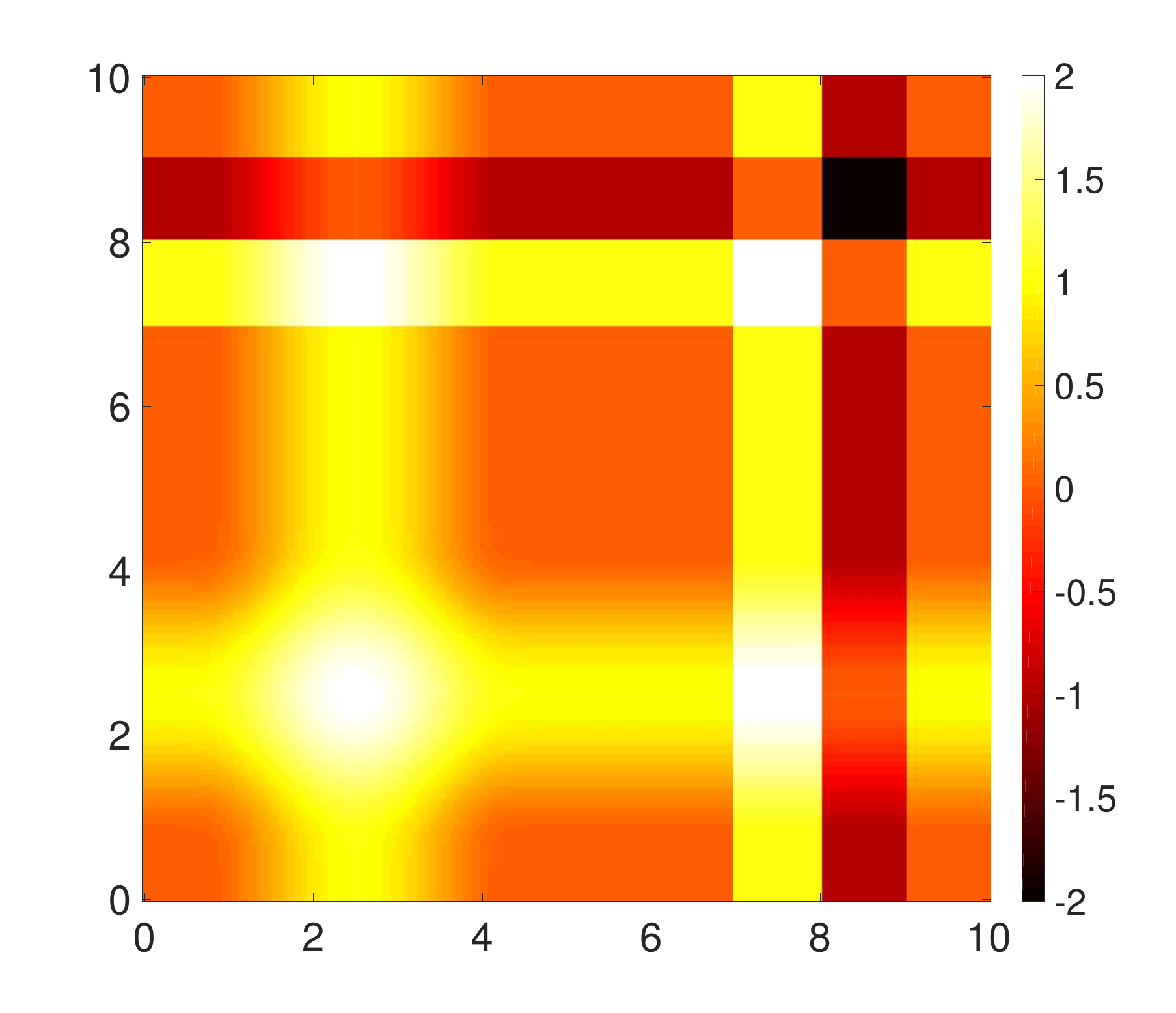}} 
\subfloat[Posterior mean ]{\includegraphics[scale=.2]{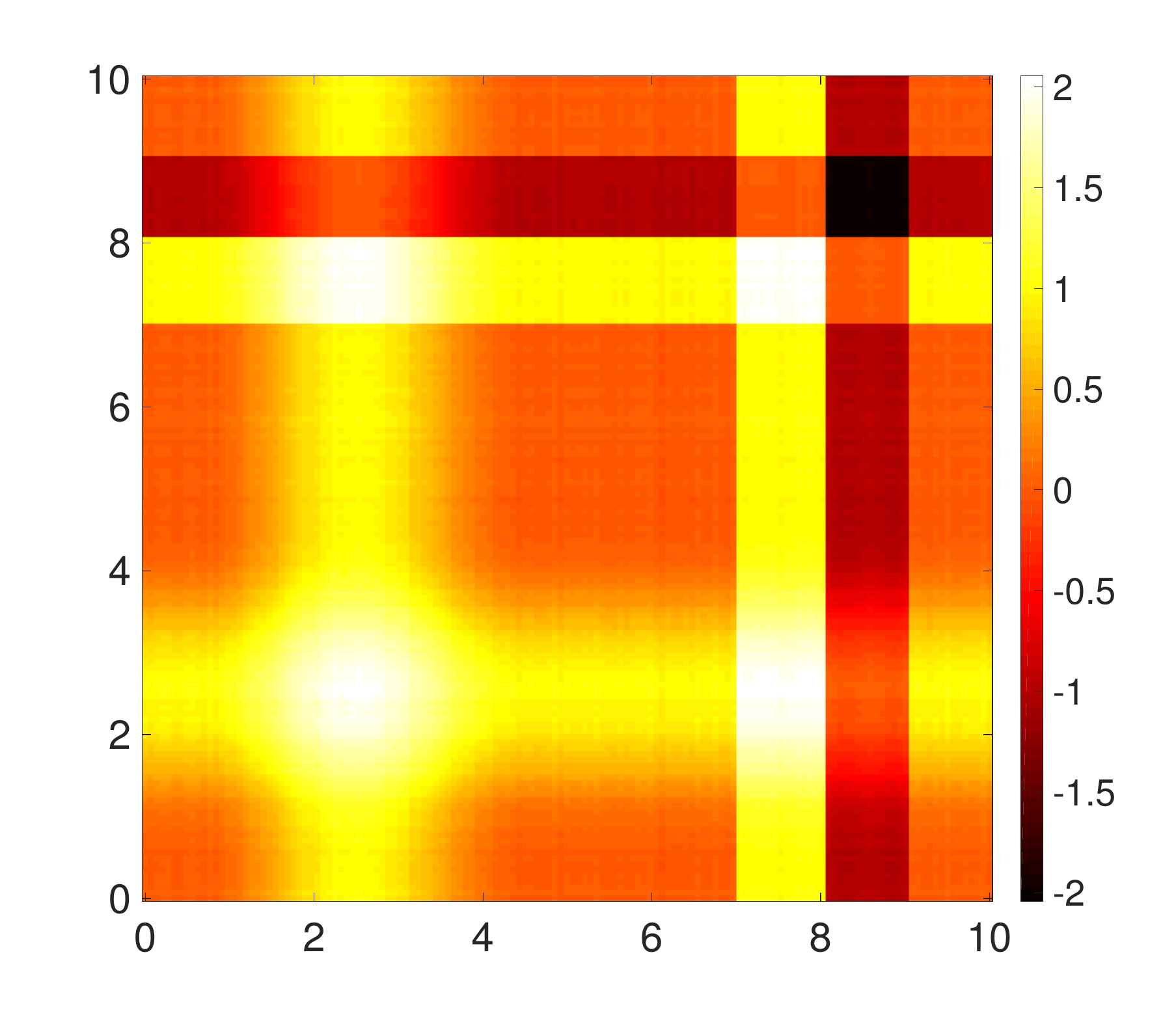}} 
	\end{center}
	\caption{Results for two-dimensional simulated dataset.} \label{fig:2dsimulated_data}
\end{figure}

\section{Comparative Evaluation}\label{Sup_Comparative_eval}
\begin{figure}[H]    
	\begin{center}
	\subfloat[{\tiny{MAE=$0.056$}}]{\includegraphics[scale=.42]{TGP_EX1-eps-converted-to.pdf}} 	
\subfloat[{\tiny{MAE=$.057$}}]{\includegraphics[scale=.42]{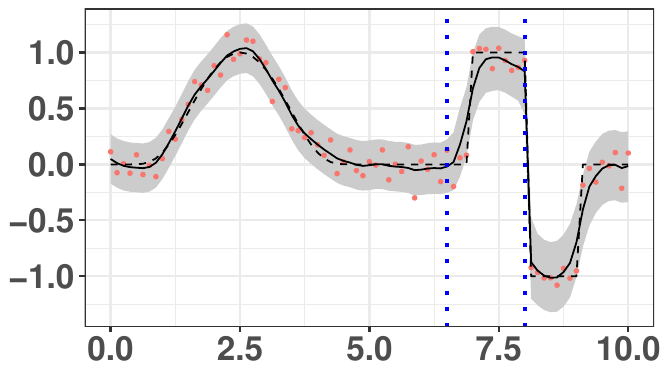}}
\subfloat[{\tiny{MAE=$.057$}}]{\includegraphics[scale=.42]{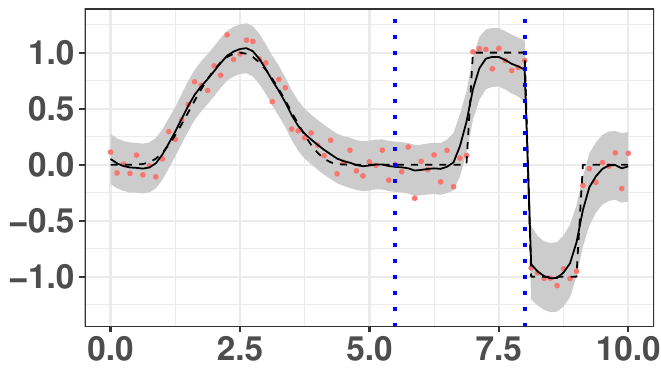}}
	\end{center}
	\caption{TGP model results for Experiment 1 with different chain lengths. (a):$100,000$ iterations with $20,000$ burn-in. (b): $200,000$ iterations with $50,000$ burn-in. (c): $500,000$ iterations with $100,000$ burn-in.}
	\label{fig:TGP_EX1}
\end{figure}
\begin{figure}[H]    
	\begin{center}
	\subfloat[{\tiny{MAE=$0.043$}}]{\includegraphics[scale=.42]{TGP_EX2-eps-converted-to.pdf}} 	
\subfloat[{\tiny{MAE=$0.043$}}]{\includegraphics[scale=.42]{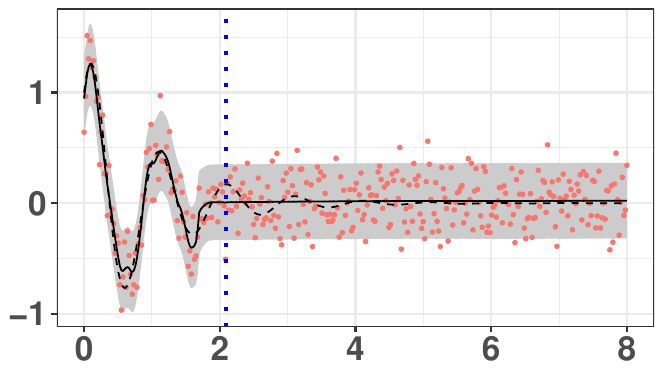}}
\subfloat[{\tiny{MAE=$.043$}}]{\includegraphics[scale=.42]{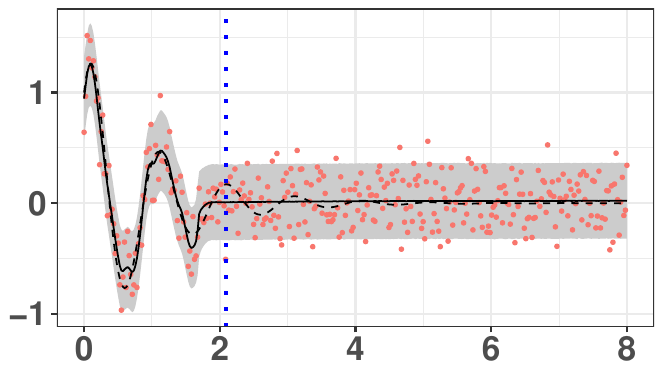}}
	\end{center}
	\caption{TGP model results for Experiment 2 with different chain lengths. (a):$100,000$ iterations with $20,000$ burn-in. (b): $200,000$ iterations with $50,000$ burn-in. (c): $500,000$ iterations with $100,000$ burn-in.}
	\label{fig:TGP_EX2}
\end{figure}

\begin{figure}[H]    
	\begin{center}
	\subfloat[{\tiny{MAE=$0.079$}}]{\includegraphics[scale=.42]{TGP_EX3-eps-converted-to.pdf}} 	
\subfloat[{\tiny{MAE=$.067$}}]{\includegraphics[scale=.42]{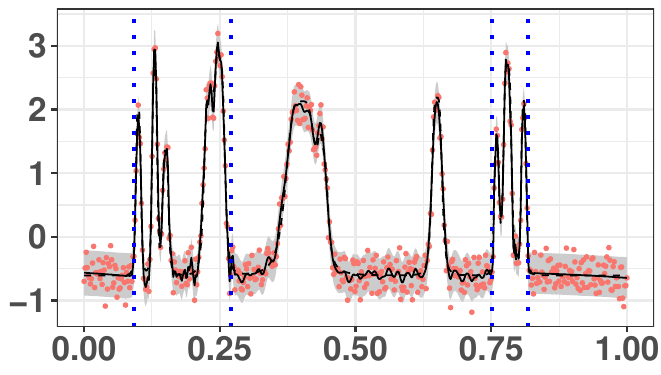}}
\subfloat[{\tiny{MAE=$.065$}}]{\includegraphics[scale=.42]{EX3_2-eps-converted-to.pdf}}
	\end{center}
	\caption{TGP model results for Experiment 3 with different chain lengths. (a):$100,000$ iterations with $20,000$ burn-in. (b): $200,000$ iterations with $50,000$ burn-in. (c): $500,000$ iterations with $100,000$ burn-in and thinning of $5$. Increasing the number of iterations has a positive effect on the number of partitions found. However, without knowing the ground truth, it is hard to know beforehand if the algorithm has been run for long enough to find the appropriate number of partitions.}
	\label{fig:TGP_EX3}
\end{figure}

\begin{figure}[H]    
	\begin{center}
		\hspace{-2mm}
			\subfloat[{\tiny{MAE=$0.122$}}]{\includegraphics[scale=.24]{TGP_EX4_subset-eps-converted-to.pdf}} 
	\subfloat[{\tiny{MAE=$.131$}}]{\includegraphics[scale=.24]{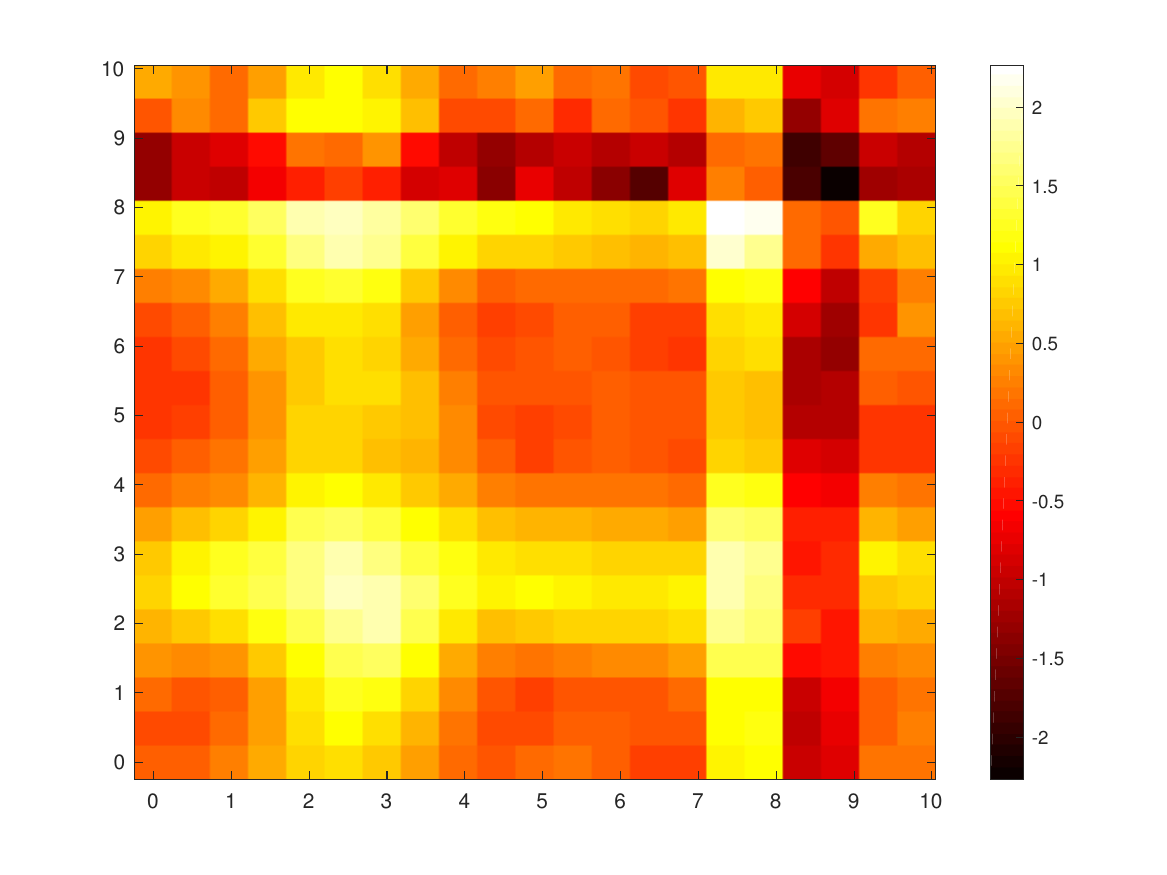}} 
	\subfloat[{\tiny{MAE=$.123$}}]{\includegraphics[scale=.24]{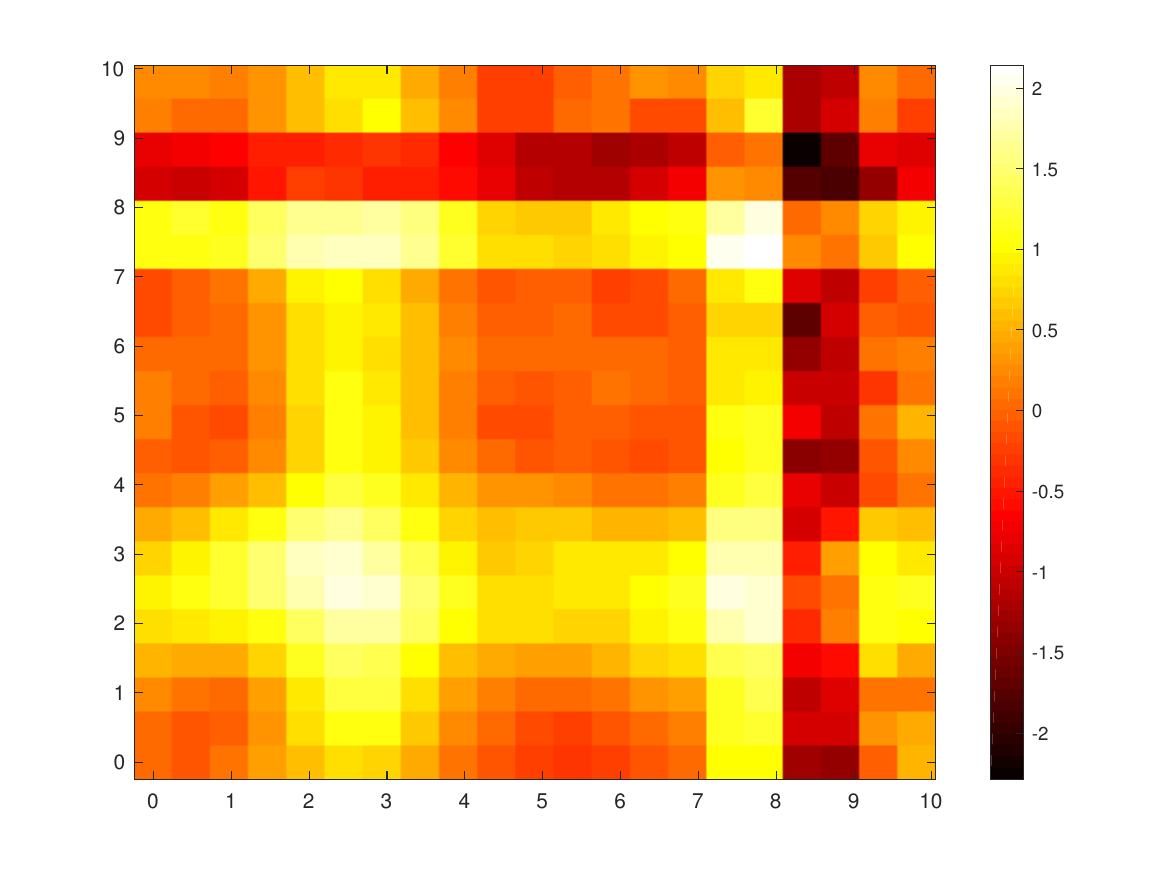}} 
	\end{center}
	\caption{TGP model results for Experiment 4 (subset) with different chain lengths. (a):$100,000$ iterations with $20,000$ burn-in. (b): $200,000$ iterations with $50,000$ burn-in. (c): $500,000$ iterations with $100,000$ burn-in and thinning of $5$.}
	\label{fig:TGP_EX3}
\end{figure}

\section{Real data: NASA rocket booster vehicle}\label{Sup_NASA_data}
\begin{figure}[H]    
	\begin{center}
		\hspace{-3mm}
		{\includegraphics[scale=.45]{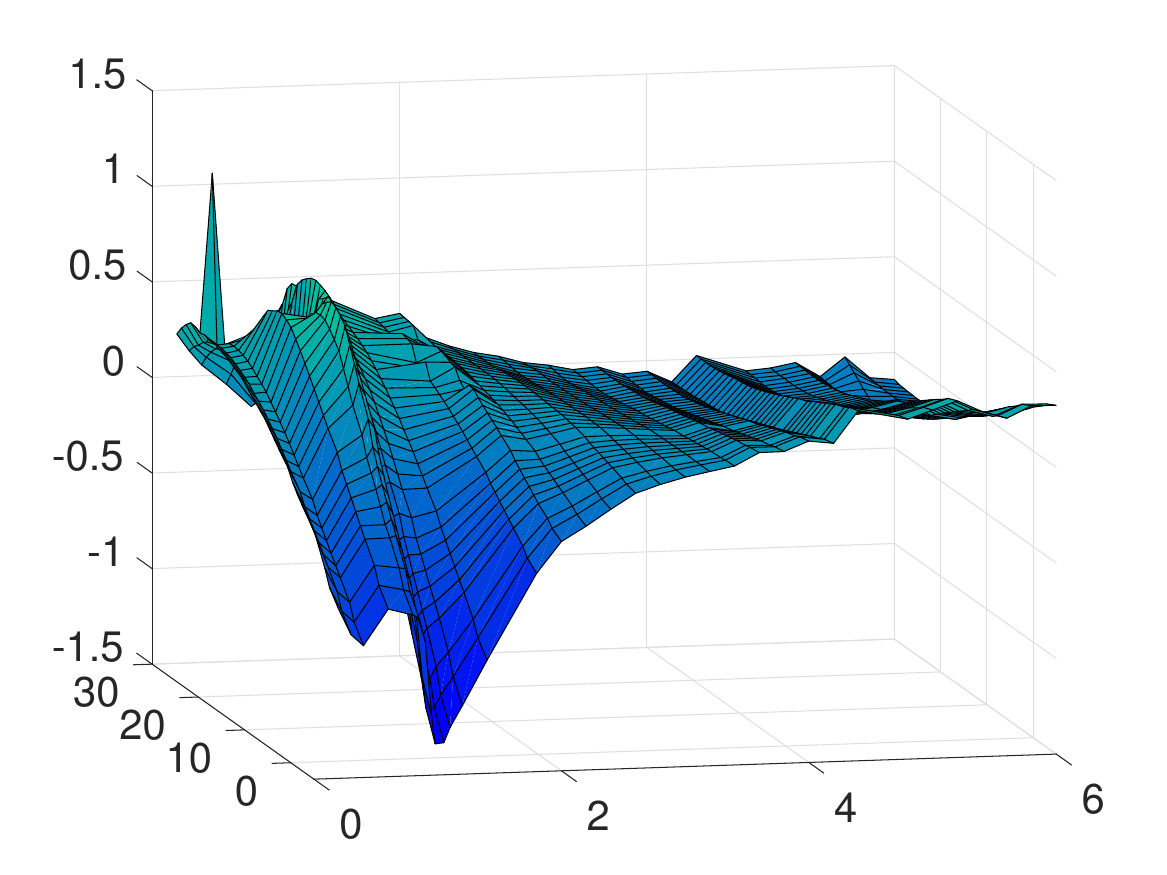}} 
	\end{center}
	\caption{Results for NASA rocket booster vehicle experiment. Posterior mean of non-stationary interaction term.}
	\label{fig:z3_Nasa}
\end{figure}

\begin{figure}[H]    
	\begin{center}
		\hspace{-3mm}
	\subfloat[{\tiny{$\boldsymbol{\ell}_1$}}]{\includegraphics[scale=.55]{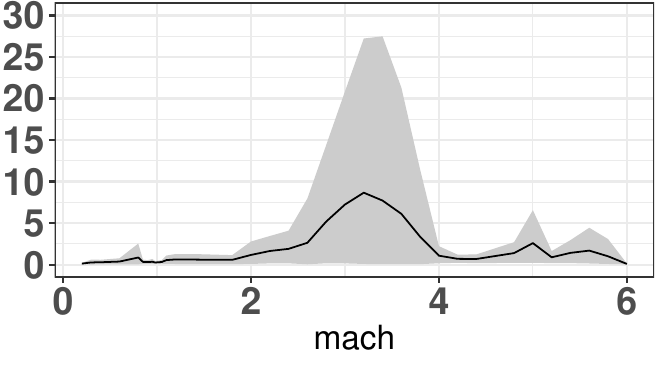}} 
\subfloat[{\tiny{$\boldsymbol{\ell}_2$}}]{\includegraphics[scale=.55]{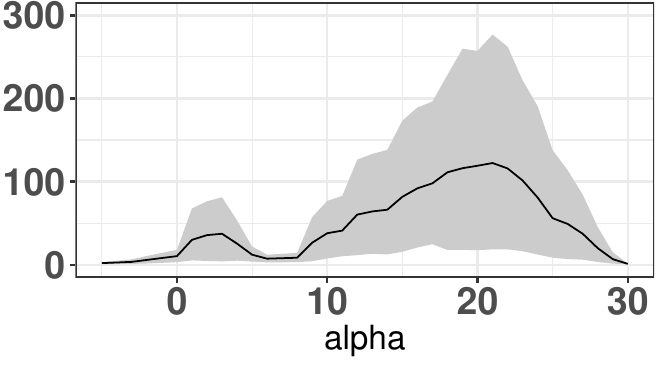}} \\
\subfloat[{\tiny{$\boldsymbol{\ell}_3$}}]{\includegraphics[scale=.55]{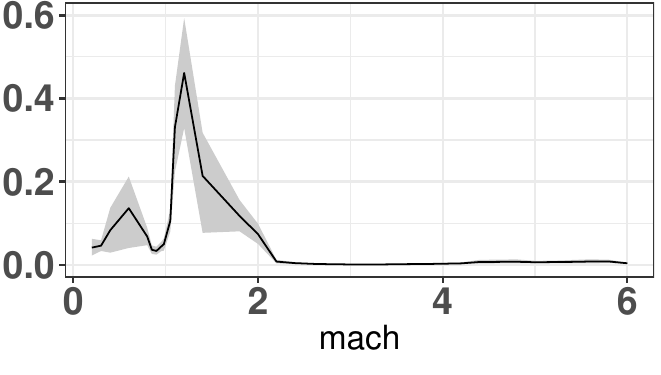}} 
\subfloat[{\tiny{$\boldsymbol{\ell}_4$}}]{\includegraphics[scale=.55]{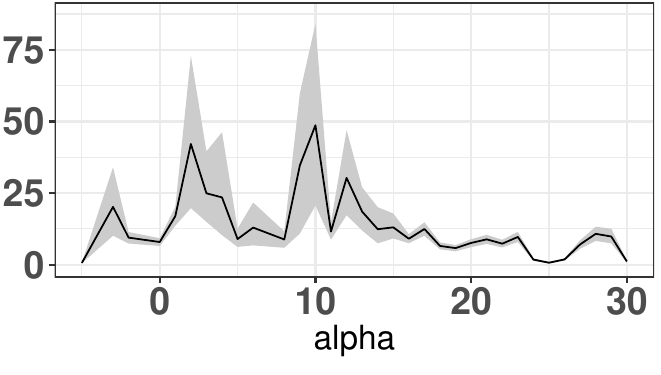}} 
	\end{center}
	\caption{ Posterior mean estimates of the stationary, one-dimensional length-scale processes with 95\% credible intervals. (a): Length-scale process for $\mathbf{z}_1$. (b): Length-scale process for $\mathbf{z}_2$. (c)-(d): Length-scale processes for the interaction term, $\mathbf{z}_3$. Notice a dip $\ell_4$ at alpha=25 to recover the peak, and the small values of $\ell_3$ around mach=1. }
	\label{fig:Nasa_ells}
\end{figure}

\end{changemargin}	


\end{document}